\documentclass[aps,jcp,superscriptaddress,amsmath,amssymb]{revtex4-2}

\usepackage{natbib}
\setcitestyle{super}

\usepackage{graphicx}
\usepackage{dcolumn}
\usepackage{bm}

\newcolumntype{.}{D{.}{.}{8}}

\usepackage[utf8]{inputenc}
\usepackage{physics}
\usepackage{amsmath, amssymb}
\usepackage{tabularx,booktabs,array,dcolumn}
\usepackage{setspace}
\usepackage{vmargin}
\usepackage{xcolor}
\usepackage{graphicx,psfrag,subfigure}

\usepackage{float}
\usepackage[all]{xy}
\usepackage{multirow}

\usepackage[round-mode=places,round-precision=4]{siunitx} 

\usepackage{bm}
\usepackage{mathtools}
\usepackage{physics}

\newcommand{\iim}{\mathrm{i}}

\usepackage[unicode]{hyperref}
\hypersetup{
   unicode=true,          
   plainpages=false,
   colorlinks=true,       
   citecolor=blue,        
}

\def\metmet{(CH$_4$)$_2$}
\def\metwat{CH$_4\cdot$H$_2$O}

\usepackage{soul}
\usepackage{color}

\def\epsi{\varepsilon}

\def\metmet{(CH$_4$)$_2$}
\def\metwat{CH$_4\cdot$H$_2$O}

\def\dd{\text{d}}

\newcommand{\bos}[1]{\pmb{#1}}
\newcommand{\mx}[1]{\pmb{#1}}
\newcommand{\pd}[2]{\frac{\partial #1}{\partial #2}}
\def\ncm{\text{CM}}
\def\transp{\text{T}}
\def\omone{\omega_1}
\def\omtwo{\omega_2}
\def\omthree{\omega_3}
\def\bKa{\bos{K}}

\def\detg{\tilde{g}}
\def\tbR{\tilde{\bos{R}}}
\def\bxi{\bos{\xi}}
\def\grad{\text{grad}}
\def\div{\text{div}}
\def\nco{D_\text{fixed}}

\def\vphi{\varphi}

\newcommand{\nb}[1]{n^\text{b}_{#1}}
\newcommand{\nq}[1]{k^\text{q}_{#1}}

\def\mcQ{Q}

\def\max{\text{max}}

\begin{document}

\title{%
Exact quantum dynamics developments for floppy molecular systems and complexes
}

\author{Edit M\'atyus}
\email{edit.matyus@ttk.elte.hu}
\author{Alberto Mart\'in Santa Dar\'ia}
\author{Gustavo Avila}

\affiliation{
ELTE, E\"otv\"os Lor\'and University, 
Institute of Chemistry, 
P\'azm\'any P\'eter s\'et\'any 1/A,
1117 Budapest, Hungary}

\date{\today}
\begin{abstract}
  \noindent 
Molecular rotation, vibration, internal rotation, isomerization, tunneling, intermolecular dynamics of weakly and strongly interacting systems, intra-to-inter-molecular energy transfer, hindered rotation and hindered translation over surfaces are important types of molecular motions. 
Their fundamentally correct and detailed description can be obtained by solving the nuclear Schrödinger equation on a potential energy surface. Many of the chemically interesting processes involve quantum nuclear motions which are `delocalized' over multiple potential energy wells.
These `large-amplitude' motions in addition to the high dimensionality of the vibrational problem represent challenges to the current (ro)vibrational methodology.
A review of the quantum nuclear motion methodology is provided, current bottlenecks of solving the nuclear Schrödinger equation are identified, and solution strategies are reviewed. 
Technical details, computational results, and analysis of these results in terms of limiting models and spectroscopically relevant concepts are highlighted for selected numerical examples.
\end{abstract}

\maketitle 

\clearpage


\section{Introduction \label{ch:intro}}
Molecules are never at rest, they constantly vibrate and rotate, and most interesting molecular phenomena involve nuclear motions extending over multiple potential energy wells. These multi-well motions are called large-amplitude motions that is in contrast to small-amplitude motions that correspond to small nuclear displacements localized to the bottom of a single potential energy well. Although atomic nuclei are ‘heavy’ particles and several types of molecular motion exhibit (semi-)classical features, the fundamentally correct theoretical description of molecules in motion\cite{Qu01} is based on quantum mechanics.\cite{HRSBook}

The basic quantum mechanical models of the most common types of nuclear motion, \emph{i.e.,} rotation and small-amplitude vibration, are the rigid rotor and the harmonic oscillator approximations that are almost as old as quantum mechanics \cite{BoJo25} and they are taught in established chapters of the undergraduate curriculum.\cite{WiDeCr55} The equilibrium rotational constants and harmonic frequencies are built-in features in most quantum chemical program packages and are often computed to obtain theoretical guidance in relation with experimentally recorded spectra. Perturbative corrections \cite{PaAlBook82} to these equilibrium quantities can be computed in electronic structure packages, but the simplest approach has limitations. As to the corrections of the harmonic frequencies, if there are several zeroth-order vibrational states close in energy, which is a common situation in polyatomic molecules, `resonance effects' introduce erroneous shifts in the perturbative correction. Furthermore, the quantum harmonic oscillator approximation is qualitatively wrong for large-amplitude motions that are common and important types of motions in molecular systems.

A straightforward solution to the beyond-harmonic-oscillator-problem accounting for large-amplitude motions, anharmonicities, mode coupling, etc., which is free of resonances, is offered by variational-type approaches. A direct variational solution of the rovibrational Schrödinger equation provides the direct or numerically `exact' solution corresponding to a given potential energy surface (PES). 
Hence, the common name, `exact quantum dynamics' is used to distinguish this approach from other techniques trying to `simulate' quantum `effects' by extending classical mechanical simulation of the nuclear motion. Quantum effect simulations by (imaginary-time) path-integral techniques have a favorable scaling property with the system size, but---apart from very special exceptions---they can have large errors on the vibrational band origins.

On the contrary, exact quantum dynamics methods can be used to systematically approach the numerically exact solutions of the nuclear Schrödinger equation. The severe shortcoming of the exact quantum approach is connected with an unfavorable, exponential scaling of the computational cost with the vibrational dimensionality.
	
This article starts with a general theoretical framework for the (ro)vibrational methodology, the Hamiltonian and curvilinear coordinates (Secs.~\ref{sec:coord} and \ref{sec:reddim}), basis functions and matrix representation (Secs.~\ref{sec:matrixH}) and lists some common algorithmic elements. 
This introduction leads us to identify the main bottlenecks (Sec.~\ref{sec:limitations}), and to review possible strategies to attenuate the computational cost increasing rapidly with the dimensionality (Secs.~\ref{sec:smolyak} and \ref{sec:other}).
The review is continued with analysis tools and spectroscopic limiting models for the computed rovibrational states (Sec.~\ref{sec:assign}), simulation of (ro)vibrational infrared and Raman spectra (Sec.~\ref{sec:tensint}).
To respect the typical length of Feature Articles and to avoid unnecessary repetition of already documented material (cited in the reference list), 
computational results from our own recent work are showcased during the course of the presentation of the theoretical and algorithmic background.

\section{Theoretical framework for quantum nuclear motion theory\label{sec:theo}}

\subsection{Nuclear Schrödinger equation and coordinate transformation \label{sec:coord}}
The time-independent Schrödinger equation is considered 
for the motion of the atomic nuclei on a potential energy surface (PES, $V$), 
\begin{align}
  (\hat{T} + V) \Psi = E \Psi \; .
  \label{eq:tise}
\end{align}
The direct solution of this differential (eigenvalue) equation can provide all stationary states of the system.
The $\hat{T}$ nuclear kinetic energy operator for $N$ nuclei with $m_i$ masses (in Hartree atomic units)  is
\begin{align}
  \hat{T}
  =
  -\sum_{i=1}^{N} \frac{1}{2m_i} 
  \left[%
    \hat{P}_{iX}^2 + \hat{P}_{iY}^2 + \hat{P}_{iZ}^2
  \right] \; ,
  \label{eq:keo}
\end{align}
where $\hat{P}_{i\alpha}=-\iim \partial / \partial R_{i\alpha}$ are the nuclear momenta conjugate to the $R_{i\alpha}$, $\alpha=X,Y,Z$ laboratory-frame (LF) Cartesian coordinates.

For an isolated system, the PES is invariant to the molecule's translation (and rotation), so it is convenient to separate the overall translation from the internal motion by defining $\bos{r}_i$ translationally invariant, body-fixed Cartesian coordinates,\cite{Su02}
\begin{align}
  \bos{R}_i
  =
  \bos{O}(\Omega) \bos{r}_i + \bos{R}_\ncm \; ,\quad i = 1,\ldots,N \;,
  \label{eq:defbfc}
\end{align}
where translational invariance means that $\bos{r}_i$ is a function of the laboratory coordinates which remain invariant to an overall translation of the system,
\begin{align}
  \bos{r}_i = \bos{f}_i(\bos{R}) = \bos{f}_i(\bos{R}') \; \quad\text{with}\quad
  \bos{R}_j'=\bos{R}_j+\bos{d},\quad\forall \bos{d}\in\mathbb{R}^3 \; .
  \label{eq:ti}
\end{align}
The overall translation of the system is described by the center of mass coordinates of the nuclei, 
\begin{align}
  \bos{R}_\ncm 
  = 
  \sum_{i=1}^{N} \frac{m_i}{m_{12\ldots N}}  \bos{R}_i \; 
  \quad\text{with}\quad
  m_{12\ldots N} = \sum_{i=1}^N m_i \; ,
\end{align}
and the center of the body-fixed frame can be fixed at the origin,
\begin{align}
  \sum_{i=1}^N m_i \bos{r}_i = 0 \; .\label{eq:BForiging}
\end{align}
%
In Eq.~(\ref{eq:defbfc}), $\bos{O}(\Omega)$ with $\Omega=(\omone,\omtwo,\omthree)$ describes the 3-dimensional rotation which connects the orientation of the body-fixed (BF) frame\cite{LiRe97} and the laboratory frame.

The commonly used normal coordinates (coordinates underlying the harmonic oscillator approximation) are defined as the linear combination of the $\bos{r}_i$ Eckart's\cite{Ec35} body-fixed Cartesian coordinates  with respect to a reference structure fixed at a PES minimum, and minimize the kinetic and potential couplings at the reference structure (Sec.~\ref{sec:optframe}).\cite{WiDeCr55}

\begin{figure}
  \begin{center}
    \includegraphics[width=8.cm]{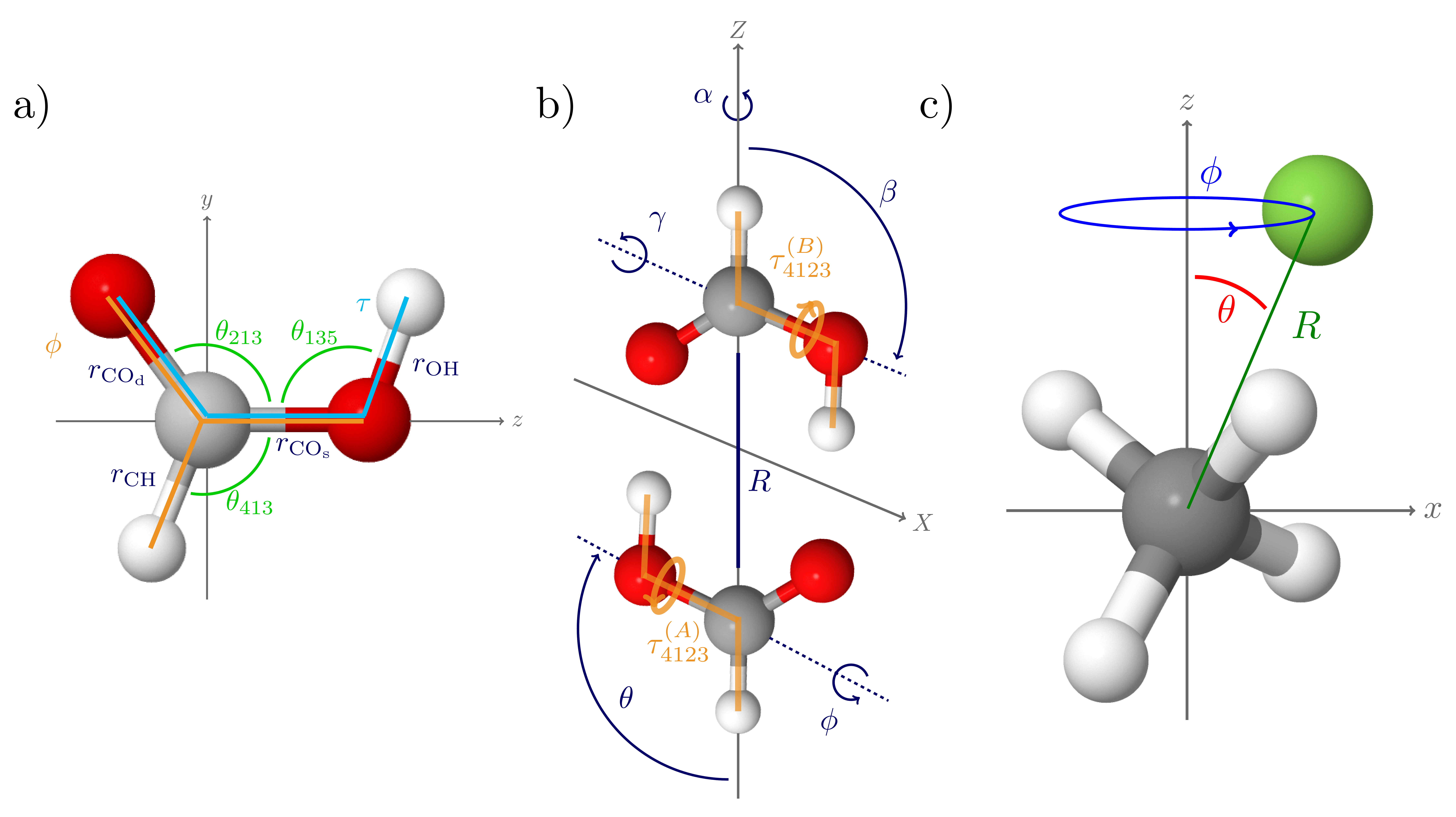}
  \end{center}    
  \caption{%
    Examples for internal coordinates: 
    (a) full 9D model for the formic acid monomer;\cite{DaAvMa22}
    (b) intermolecular-plus-torsional, 8D model for the formic acid dimer;\cite{DaAvMa21b}
    (c) full 12D model of the CH$_4\cdot$F$^-$ complex using the $q_1,\ldots,q_9$ normal coordinates for the methane fragment and     $(R,\theta,\phi)$ spherical polar coordinates for the relative ion-methane motion.\cite{AvMa19,AvMa19b,AvPaCzMa20,PaTaAvMa22}
    \label{fig:coords}
    }
\end{figure}

For an efficient description of quantum nuclear motion beyond small-amplitude vibrations about a local minimum, non-linear functions of $\bos{r}_i$ are commonly defined, 
\begin{align}
  q_k=g_k(\bos{r})\;, \quad k=1,\ldots,3N-6 \;
\end{align}
to efficiently describe the (extended) internal motion.
$g_k(\bos{r})$ is typically a function of scalar products of the $\bos{r}_i$ body-fixed Cartesian coordinates (hence independent of the frame definition), distance- and angle-type variables (Fig.~\ref{fig:coords}). It has been shown that beyond four-particle systems, $3N-6$ distance- and angle-type variables are not sufficient to uniquely characterize all molecular structures. \cite{ThJoCo98} So, beyond $N=4$, torsion-like coordinates (including some vector-type product of $\bos{r}_i$'s) must be also included 
for a unique description with $3N-6$ variables. 

The variable change
\begin{align}
  &(X_1,Y_1,Z_1,X_2,Y_2,Z_2,\ldots,X_{3N},Y_{3N},Z_{3N})\Rightarrow \nonumber \\ &(\xi_1,\ldots,\xi_{3N})
  =(q_1,q_2,\ldots,q_{3N-6},\omone,\omtwo,\omthree,X_\ncm,Y_\ncm,Z_\ncm)
  \label{eq:coortrans}
\end{align}
corresponds to a curvilinear coordinate `transformation', which 
can be characterized by the Jacobi matrix. The Jacobi matrix collects derivatives of the `old' coordinates with respect to the `new' ones, $J_{ik}=\partial \bos{R}_i/\partial \xi_k$.
For a vibrational degree of freedom, $\xi_k=q_k$ ($k=1,\ldots,3N-6$),
the Jacobian elements can be written as
\begin{align}
  J_{ik}=
  \pd{\bos{R}_{i}}{\xi_k}
  &=
  \pd{%
    [\bos{O} \bos{r}_i + \bos{R}_\ncm]
  }{q_k}
  =
  \bos{O}
  \pd{%
    \bos{r}_i 
  }{q_k}
  =
  \bos{O}\ \bos{t}_{ik} \; ,
  \label{eq:Tvib}
\end{align}
where we defined the `vibrational $t$-vector' as 
\begin{align}
  \bos{t}_{ik}
  =
  \pd{%
    \bos{r}_i 
  }{q_k}  \; , \quad k=1,2,\ldots,3N-6 \; .
  \label{eq:tvib}
\end{align}
For a rotational degree of freedom, $\xi_{3N-6+a}=\omega_{a}$, which is a rotation angle about the $a=1(x),2(y),3(z)$ axis of the body-fixed frame, we can write
\begin{align}
  J_{i,{3N-6+a}}
  &=
  \pd{\bos{R}_i}{\xi_{3N-6+a}}
  =
  \pd{%
    [\bos{O}(\Omega) \bos{r}_i + \bos{R}_\ncm]
  }{\omega_a}
  \nonumber \\
  &=
  \pd{%
    \bos{O}(\omone,\omtwo,\omthree)   
  }{\omega_a} \bos{r}_i 
  =
  \bos{O} [\bos{e}_a \times \bos{r}_i]
  =
  \bos{O}\ \bos{t}_{i,3N-6+a}
  \; , 
  \label{eq:Trot}
\end{align}
where the `rotational $t$-vector' was defined as
\begin{align}
  \bos{t}_{i,3N-6+a}
  =
  \bos{e}_a \times \bos{r}_i \;, \quad a=1(x),2(y),3(z) \; ,
  \label{eq:trot}
\end{align}
and $\bos{e}_a$ is the unit vector pointing along the (positive direction of the) coordinate axis, $(\bos{e}_a)_n=\delta_{an}\ (n=1,2,3)$. 

To obtain Eq.~(\ref{eq:Trot}), 
we used the following relation regarding the derivative of an abstract three-dimensional rotational operation, written in the Euler--Rodrigues' form, with respect to the rotation angle $\omega$ about the axis pointing in the direction of the unit vector
$\bos{k}=(k_x,k_y,k_z)$ with $|\bos{k}|=1$:
\begin{align}
  \pd{}{\omega^{(\bos{k})}}\bos{O}
  =
  \pd{}{\omega}
  \left[%
  \bos{I}
  + 
  \bKa\sin \omega 
  +
  \bKa (1-\cos\omega) 
  \right]
  =
  \bKa \cos\omega  + \bKa^2 \sin\omega  \; ,
\end{align}
where $\bos{I}$ is the three-dimensional unit matrix and
$\bKa$ is the `cross product matrix', $\bKa\bos{r}_i=\bos{k}\times \bos{r}_i$,
\begin{align}
  \bKa
  =
  \left(%
    \begin{array}{@{}ccc@{}}
       0     & -k_z &  k_y \\
       k_z &  0     & -k_x \\
      -k_y &  k_x & 0 \\
    \end{array}
  \right) \; .
\end{align}
Then, by using the $\bKa^3=-\bKa$ identity,
we can write
\begin{align}
  \bos{O}\bKa
  &=
  \left[%
  \bos{I}
  + 
  \bKa \sin \omega 
  +
  \bKa (1-\cos\omega) 
  \right] \bKa
  \nonumber \\
  &=
  \bKa + \bKa^2 \sin\omega  + \bKa^3 (1-\cos\omega)
  \nonumber \\
  &=
  \bKa \cos\omega  + \bKa^2 \sin\omega \; ,
\end{align}
and hence, the relation, used in Eq.~(\ref{eq:Trot}), is obtained as
\begin{align}
  \pd{}{\omega^{(\bos{k})}} \bos{O} \bos{r}_i
  =
  \bos{O} \bKa \bos{r}_i 
  =
  \bos{O} (\bos{k} \times \bos{r}_i)
  \; 
\end{align}
for $\bos{k}=\bos{e}_a$.
Finally, for a translational-type degree of freedom in the laboratory frame, 
$\xi_{3N-3+\alpha}=R_{\ncm,\alpha}$, $\alpha=1(X),2(Y),3(Z)$, we can write
\begin{align}
  J_{i,{3N-3+\alpha}}=&
  \pd{\bos{R}_i}{\xi_{3N-3+\alpha}}
  =
  \pd{%
    [\bos{O}(\omone,\omtwo,\omthree) \bos{r}_i + \bos{R}_\ncm]
  }{R_{\ncm,\alpha}} 
  =
  \bos{e}_{\alpha} \; .
  \label{eq:Ttrans}
\end{align}

Having the Jacobi-matrix elements at hand, Eqs.~(\ref{eq:Tvib}), (\ref{eq:Trot}), and (\ref{eq:Ttrans}), we can calculate the mass-weighted metric tensor elements, 
\begin{align}
  g_{kl} 
  = 
  \sum_{i=1}^{N} 
    m_{i} \pd{\bos{R}^\transp_i}{\xi_k} \pd{\bos{R}_i}{\xi_l} \; ,
  \label{eq:gmx}
\end{align}
where the introduction of the mass weights corresponds to mass-scaled Cartesian coordinates, $\tbR_i=\sqrt{m_i}\bos{R}_i$. 

\begin{figure}
    \centering
    \includegraphics[width=7.5cm]{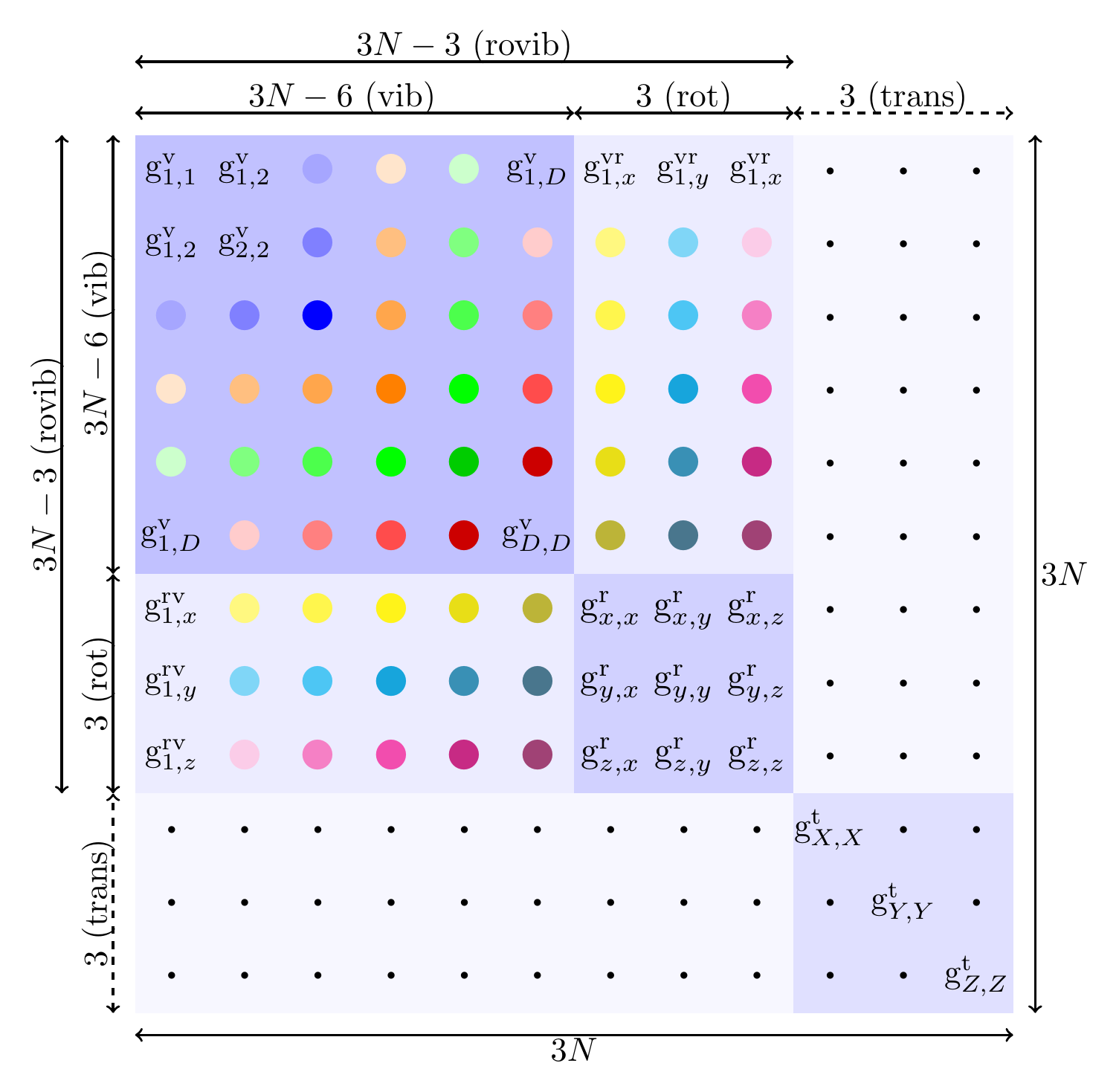}
    \caption{%
    Mass-weighted metric tensor: 
    vibrational (v), rotational (r), rovibrational (rv, vr), and translational (t) blocks of $\pmb{g}$. 
    The dots stand for `0' entries, \emph{i.e.,} the kinetic coupling is exactly zero between the rovibrational and the translational degrees of freedom.
    }
    \label{fig:g_matrix}
\end{figure}

For the rotation-vibration block of $\bos{g}$, $k,l=1,2,\ldots,3N-3$, is
\begin{align}
  g_{kl} 
  &=
  \sum_{i=1}^{N}
    m_{i} \pd{\bos{R}^\transp_i}{\xi_k} \pd{\bos{R}_i}{\xi_l} 
  \nonumber \\
  &=
  \sum_{i=1}^{N}
    m_{i} \bos{t}_{ik}^\transp \bos{O}^\transp \bos{O} \bos{t}_{il}
  \nonumber \\
  &=
  \sum_{i=1}^{N}
    m_{i} \bos{t}_{ik}^\transp \bos{t}_{il} \; , 
  \label{eq:grovib}
\end{align}
where orthogonality of the rotation matrix, $\bos{O}^\transp \bos{O}=\bos{I}$, allowed us to express
the rovibrational $g_{kl}$ elements solely in terms of body-fixed quantities, \emph{i.e.,} using the vibrational and rotational $t$-vectors, Eqs.~(\ref{eq:tvib}) and (\ref{eq:trot}), respectively.
Regarding the translation-vibration and translation-rotation blocks, 
we use the fact that the translational Jacobi matrix elements are coordinate independent (constant), and we fixed the origin of the body-fixed frame at the nuclear center of mass, Eq.~(\ref{eq:BForiging}), $\alpha=1(X),2(Y),3(Z)$ and $l=1,2,\ldots,3N-3$,
\begin{align}
  g_{3N-3+\alpha,l} 
  &=
  \sum_{i=1}^N 
    m_i \pd{\bos{R}^\transp_i}{\xi_{3N-3+\alpha}} \pd{\bos{R}_i}{\xi_l} 
  \nonumber \\
  &=
  \sum_{i=1}^N 
    m_i \bos{e}_{\alpha}^\transp \pd{\bos{O}\bos{r}_i}{\xi_l} 
  \nonumber \\
  &=
  \bos{e}_{\alpha}^\transp \pd{}{\xi_l} \bos{O} \sum_{i=1}^N m_i\bos{r}_i
  = 0 \; .
  \label{eq:grovibtrans}
\end{align}
The translational matrix elements are
\begin{align}
  g_{3N-3+\alpha,3N-3+\beta} 
  =
  \sum_{i=1}^N m_i \bos{e}_\alpha^\transp \bos{e}_\beta
  =
  m_{12\ldots N} \delta_{\alpha\beta} \; .
  \label{eq:gtrans}
\end{align}
The resulting structure of the mass-weighted metric tensor, $\bos{g}$, is highlighted in Fig.~\ref{fig:g_matrix}.

Next, let us define the $3N$-dimensional gradient operator, $(\grad_{\tbR})_{i\alpha}=\partial_{i\alpha}=\partial / \partial \tilde{R}_{i\alpha}$, and write the nuclear Schrödinger equation, Eq.~(\ref{eq:tise}), as
\begin{align}
  2(V-E) \Psi
  =
  \div_{\tbR} \grad_{\tbR} \Psi 
  =
  \div_{\bxi} \grad_{\bxi}\Psi \; ,
\end{align}
which is understood with the normalization of $\Psi$ is written as
\begin{align}  
  1 
  &=
  \int 
    \Psi^\ast \Psi\ \dd \tbR
  =
  \int 
    \Psi^\ast \Psi\ 
    \tilde{g}^{1/2}\ \dd \bos{\xi}  \; .
\end{align}
In the second equation,  due to the coordinate change, the determinant of the  Jacobi matrix, $\det\bos{J}$, appears, which equals the square root of the determinant of the metric tensor, 
$\det \bos{J}=\tilde{g}^{1/2}$ with $\tilde g = \det \bos{g}$. We can say in short that the `volume element' corresponding to integration in the new coordinates is
\begin{align}
  \dd V' = \detg^{1/2}\ \dd \xi_1 \dd \xi_2 \ldots \dd \xi_{3N}  \; .
  \label{eq:volel}
\end{align}
In curvilinear coordinates, the divergence of an $\bos{F}$ vector field is written 
(using Einstein's summation convention and the covariant and contravariant labelling) as 
\begin{align}
  \div_{\bxi} \bos{F}
  =
  \tilde{g}^{-1/2}
  \partial_k
  \tilde{g}^{1/2}
  F^k \; ,
\end{align}
and the gradient of a $\phi$ scalar field is 
\begin{align}
  \grad_{\bxi} \phi 
  =
  \partial^k \phi
  =
  g^{kl} \partial_l \phi \; ,
\end{align}
where $g^{kl}$ is the contravariant metric tensor, which is the inverse of the (covariant) metric tensor, Eq.~(\ref{eq:gmx}).
As a result, we can write the `$\div\grad$' operator in curvilinear coordinates as
\begin{align}
  \div_{\bxi} \grad_{\bxi}\Psi 
  &=  
  \detg^{-1/2} \partial_k \detg^{1/2} g^{kl} \partial_l \Psi  \; ,
\end{align}
and the integrals are calculated `with' the $\dd V'$ volume element, Eq.~(\ref{eq:volel}).
A more symmetric form of the differential operator appears, if we `merge' (the square root of) the Jacobi determinant in the wave function, so that we can use the 
simple normalization condition
\begin{align}
  1 = \int \psi^\ast \psi\ \dd \xi_1 \ldots\ \dd\xi_{3N} \; .
\end{align}
This implies that $\Psi = \detg^{-1/4} \psi$, and we obtain
the Schrödinger equation for $\psi$ as
\begin{align}
  2(V-E) \Psi 
  &= 
  \detg^{-1/2} \partial_k \detg^{1/2} g^{kl} \partial_l \Psi \\
  2(V-E) \detg^{-1/4} \psi 
  &= 
  \detg^{-1/2} \partial_k \detg^{1/2} g^{kl} \partial_l \detg^{-1/4} \psi \\  
  2(V-E) \psi 
  &= 
  \detg^{-1/4} \partial_k \detg^{1/2} g^{kl} \partial_l \detg^{-1/4} \psi  \; ,
\end{align}
which can be rearranged to
\begin{align}
  \left[%
    -\frac{1}{2}\detg^{-1/4} \partial_k \detg^{1/2} g^{kl} \partial_l \detg^{-1/4}
    +V
  \right] \psi
  =
  E \psi
\end{align}
or written in the more traditional form, 
\begin{align}
  \left[%
    \frac{1}{2}
    \sum_{k=1}^{3N}
    \sum_{l=1}^{3N}
      \detg^{-1/4} \hat{p}_k \detg^{1/2} G_{kl} \hat{p}_l  \detg^{-1/4}
    +V
  \right] \psi
  =
  E \psi 
  \label{eq:podol}
\end{align}
with the $\hat{p}_k=-\iim \partial / \partial \xi_k$ `generalized momenta' and the `big' $\bos{G}$ matrix
\begin{align}
  \bos{G}=\bos{g}^{-1} \; .
  \label{eq:bigGmx}
\end{align}
In the physical chemistry literature, the Eq.~(\ref{eq:podol}) form of the kinetic energy operator is commonly referred to as the `Podolsky form'.\cite{Po28} In 1928, Boris Podolsky drew Paul Dirac's attention to the correct form of the Laplace (`divgrad') operator in curvilinear coordinates (that had been known in mathematics for decades). Podolsky's choice for the normalization of the wave function resulted in a symmetric form of the operator, which is convenient for variational-type applications of the Schrödinger equation in curvilinear coordinates.

It is important to note that the curvilinear form of the operator can be expressed with the metric tensor, and it is not necessary to use the Jacobian matrix. Eqs.~(\ref{eq:grovib}),
(\ref{eq:grovibtrans}), and (\ref{eq:gtrans}) (see also Fig.~\ref{fig:g_matrix})
show that the metric tensor can be calculated from body-fixed quantities, \emph{i.e.,} rotational and vibrational $t$-vectors and the constant masses associated to the nuclei, and it depends neither on the rotation angles, nor the translational coordinates. The metric tensor can be written as a function of the vibrational coordinates and of the body-fixed frame definition ($\bos{e}_a$ vectors).

Due to the block-diagonal structure of $\bos{g}$ in terms of the rovibrational and the translational coordinates (Fig.~\ref{fig:g_matrix}) and since $\detg$ does not depend on the translational coordinates, the kinetic energy operator is the sum of a rovibrational and a  translational part, 
\begin{align}
  \left[%
    \frac{1}{2}
    \sum_{k=1}^{3N-3}
    \sum_{l=1}^{3N-3}
      \detg^{-1/4} \hat{p}_k \detg^{1/2} G_{kl} \hat{p}_l  \detg^{-1/4}
    +
    \frac{1}{2 m_{12\ldots N}} \bos{P}_{\ncm}^2
    +V
  \right] \psi
  =
  E \psi \; 
  \label{eq:podoltrv}
\end{align}
with $P_{\ncm,\alpha}=-\iim\partial / \partial R_{\ncm,\alpha}$.
The complete Eq.~(\ref{eq:podoltrv}) must be used to describe 
molecules in interaction with solid materials or their motion in cages.\cite{LaFeScBeBa19,FeLaScBeBa19}

At the same time, for an isolated molecular system, $V$ is independent of the translational degrees of freedom, and then, 
it is convenient to subtract the center-of-mass kinetic energy, $\bos{P}^2_{\ncm}/(2m_{12\ldots N})$ from the operator and separate the continuum spectrum of free translation. Thereby, we obtain the rovibrational Hamiltonian for an isolated molecule as
\begin{align}
  \hat{H}^\text{rv,P}
  =&\frac{1}{2}
  \sum_{k=1}^{D+3} \sum_{l=1}^{D+3}
    \detg^{-1/4} \hat{p}_k \detg^{1/2} G_{kl} \hat{p}_l  \detg^{-1/4}
    \label{eq:hampod}
  \\
  =
  &\frac{1}{2} 
  \sum_{k=1}^{D} \sum_{l=1}^{D} 
    \detg^{-1/4} \hat{p}_k G_{kl} \detg^{1/2} \hat{p}_l \detg^{-1/4} \nonumber \\
  & 
  + \frac{1}{2} 
  \sum_{k=1}^{D} \sum_{a=1}^3
  (\hat{p}_k G_{k,D+a} 
  + 
  G_{k,D+a}\hat{p}_k)\hat{J}_a \nonumber \\
  & + \frac{1}{2} \sum_{a=1}^3 G_{D+a,D+a} \hat{J}_a^2 \nonumber \\
  & + \frac{1}{2} \sum_{a=1}^3 \sum_{b>a}^3 G_{D+a,D+b}[\hat{J}_a,\hat{J}_b]_+ 
  + V \; ,  
  \label{eq:detPodrv}
\end{align}
where $D=3N-6$ applies. 
Eq.~(\ref{eq:detPodrv}) was obtained from Eq.~(\ref{eq:hampod}) by exploiting the fact that $\detg$ does not depend on the rotational coordinates.\cite{MaSzCs14} Furthermore, we used that the derivative with respect to the $a$th rotational angle, is the body-fixed angular momentum component, $\hat{J}_a=-\iim\partial / \partial \omega_a$. $[\hat{J}_a,\hat{J}_b]_+$ labels the anti-commutator of $\hat{J}_a$ and $\hat{J}_b$.\cite{Za98}
We also note that the body-fixed angular momentum operators satisfy the (anomalous) commutation relations, expressed with the $\epsilon_{abc}$ Levi-Civita symbol,
\begin{align}
  \label{eq:commutation}
  [\hat{J}_a,\hat{J}_b] = -i \epsilon_{abc} \hat{J}_c,
  \quad\quad a,b,c = 1(x),2(y),3(z)
\end{align}

The $\hat{H}^\text{P}$ Podolsky form of the rovibrational Hamiltonian was found to be advantageous in some computer implementation,\cite{MaCzCs09} because it was possible to arrive at a stable numerical code by using only first-order coordinate derivatives (vibrational $t$-vectors computed by finite differences).
Efficient use of the Podolsky form assumes that the truncated resolution of identity in the finite vibrational basis provides an accurate matrix representation (Mx) for
the operator identity $\text{Mx}(\hat{p}_k^2)\approx \text{Mx}(\hat{p}_k)\cdot \text{Mx}(\hat{p}_k)$ for the vibrational derivative operators ($k=1,\ldots,D$).

If this approximation is inaccurate in the vibrational basis, then, it is better to use the `fully-rearranged' form of the Hamiltonian,\cite{Lu00,LaNa02,AvMa19,AvMa19b} which is obtained from the vibrational Podolsky form by mathematically equivalent manipulations,
\begin{align}
  \hat{H}^\text{v} 
  = 
  -\frac{1}{2} 
  \sum_{k=1}^{D} \sum_{l=1}^{D} 
    G_{kl} \frac{\partial}{\partial q_{k}}\frac{\partial}{\partial q_{l}}
  -\frac{1}{2} 
  \sum_{l=1}^{D} B_{l}
    \frac{\partial}{\partial q_{l}}
  +U
  +V \; ,
  \label{eq:hamiltonian_rear}
\end{align}
where 
\begin{align}
  B_{l} 
  = 
  \sum_{k=1}^{D} 
  \frac{\partial}{\partial q_{k}} G_{kl} \; ,
\end{align}
and the potential-like term,\cite{LaNa02,MaCzCs09}
\begin{align}
  U= \frac{1}{32} \sum_{k=1}^{D} \sum_{l=1}^{D}
    \Bigg[
    \frac{G_{kl}}{\tilde{g}^2} \frac{\partial\tilde{g}}{\partial q_k}
    \frac{\partial\tilde{g}}{\partial q_l}
    +4\frac{\partial}{\partial q_k} 
    \Bigg(
    \frac{G_{kl}}{\tilde{g}}
    \frac{\partial\tilde{g}}{\partial q_l}
    \Bigg)
    \Bigg] \; 
  \label{eq:pseudopot}
\end{align}
are functions of the vibrational coordinates only (similarly to $G_{kl}$ and $V$).

Evaluation of $B_l$ and $U$ assumes the evaluation of higher-order coordinate derivatives (vibrational coordinate derivatives of the $t$-vectors, up to third order, \emph{i.e.,} $\partial^3 \bos{r}_i/\partial q_k \partial q_l \partial q_m$). In this case, evaluation of the derivatives by finite differences requires the use of increased precision arithmetic (quadruple precision in Fortran),\cite{MaCzCs09} alternatively analytic derivatives, \emph{e.g.,} for $Z$-matrix coordinates,\cite{LaNa02} or automated differentiation\cite{YaYu15} can be used.

The idea of developing a general and unifying approach to the rovibrational kinetic energy operator has an at least half-a-century history, and we only cite here some of the prominent examples.\cite{MeGu69,MeJMS79,Lu00,Lu03,LaNa02,YuThJe07}
During the rest of this work, developments will be reviewed in relation with 
the numerical kinetic energy operator approach of Ref.~\citenum{MaCzCs09} and its applications within the GENIUSH (GENeral Internal-coordinate USer-defined Hamiltonians) program.\cite{MaCzCs09,FaMaCs11,SaCs16,DaAvMa21a}

\subsection{Geometrical constraints and reduced-dimensionality models\label{sec:reddim}}
The number of vibrational degrees of freedom increase linearly with the number of atoms in the system and result in a rapid scale-up of the computational cost of variational applications (Secs.~\ref{sec:matrixH}--\ref{sec:smolyak}). To attenuate this growth, it is possible to introduce constraints on the motion of the nuclei. 

Geometrical constraints, \emph{e.g.,} fixing certain bond lengths or angles, can be implemented\cite{MaCzCs09}  in the outlined formalism by `deleting' the corresponding rows and columns of the $\bos{g}$ matrix corresponding to the constrained coordinate.\cite{WiDeCr55}
Zeroing rows and columns in the $\bos{G}$ matrix does not correspond to imposing a rigorous geometrical constraint, but it corresponds to fixing the `generalized' momenta corresponding to the selected coordinate. Although `zeroing' in the $\bos{G}$ matrix (within some specific computational setup) may result in a better numerical approximation to the full-dimensional result, the numerical values depend on the coordinate representation used for the \emph{constrained fragment,} which is conceptually problematic.\cite{MaCzCs09}

The (ro)vibrational Hamiltonians corresponding to rigorous geometrical constraints take the same from as in Eqs.~(\ref{eq:hampod}), (\ref{eq:detPodrv}) and (\ref{eq:hamiltonian_rear}), but  $D<3N-6$  and it labels the number of the active vibrational degrees of freedom. The $\detg$ and $\bos{G}$ functions in the Hamiltonian are obtained by 
(a) constructing the full $\bos{g}\in \mathbb{R}^{(3N-3)\times(3N-3)}$ matrix (the block-diagonal translational block can be inverted separately, if needed);
(b) deleting the last $\nco$ rows and columns of the full $\bos{g}$ corresponding to the constrained degrees of freedom;
(c) the remaining $D$-dimensional `reduced' $\bos{g}$ matrix is used to calculate $\detg$ and to obtain $\bos{G}$ by inversion.

Experience shows these reduced dimensionality-models (with fixed geometrical parameters) can be useful in weakly interacting systems,\cite{SaCsAlWaMa16,SaCsMa17,FeMa19,AvMa19,AvMa19b,DaAvMa21a} but they can introduce large errors within bound molecules.\cite{DaAvMa22} 
It remains a question to be explored whether the error of (a qualitatively meaningful) reduced-dimensionality model is introduced due to the lack of `quantum coherence' of the discarded modes with the active degrees of freedom or it is rather a structural effect due to the dissection of a certain $D$-dimensional cut of the full-dimensional PES corresponding to some fixed coordinate values.

All in all, ideally we can aim for a systematically improvable (series of) approximate solution(s) of the nuclear Schrödinger equation by including all vibrational degrees of freedom as dynamical variable.

\subsection{A variational rovibrational approach: basis set and integration grid for the matrix representation of the Hamiltonian\label{sec:matrixH}}

In bound-state quantum mechanics, a common and powerful approach for solving the wave equation, a differential equation with no known analytic solution, is provided by the variational method. This method can be straightforwardly applied to the rovibrational problem, since the rovibrational Hamiltonian is bounded from below.
According to Walther Ritz' \cite{Ri1909} linear variational procedure from 1909, the variationally best linear combination coefficients of a fixed (orthonormal) basis set are obtained by diagonalization of the matrix-eigenvalue problem 
\begin{align}
  \bos{H} \bos{c}_n = E_n \bos{c}_n \; ,
\end{align}
which provides the best approximation to the exact wave function over the space spanned by the $\Phi_I$ basis functions,
\begin{align}
  \psi_n = \sum_{I=1}^{N_\text{b}^\text{(rv)}} c_{n,I} \Phi_I \;  
\end{align}
and $E_n$ approaches the $n$th exact eigenvalue from above.

These rigorous variational properties apply if the Hamiltonian matrix is constructed exactly, which assumes evaluation of integrals for the Hamiltonian with the pre-defined basis functions
\begin{align}
  H_{IJ} 
  = 
  \langle 
    \Phi_I | \hat{H} \Phi_J
  \rangle \; .
\end{align}
In all applications reviewed in this work, the $\langle \Phi_I| \Phi_J\rangle =\delta_{IJ}$ orthonormality applies. 

The $\Phi_I$ multi-dimensional basis functions are often defined as
\begin{align}
  \Phi_I
  &=
  \Phi_{i K\tau}^{(JM)}(q_1,\ldots,q_D,\Omega)
  =
  \Phi_i(q_1,\ldots,q_k) \Theta^{(JM)}_{K\tau}(\Omega)  \; ,
  \label{eq:rovib}
\end{align}
where $\Theta^{(JM)}_{K\tau}(\Omega)$ and $\Phi_i(q_1,\ldots,q_k)$
label rotational and vibrational basis functions, respectively.
The choice of the rotational and vibrational basis functions is motivated by analytically solvable quantum mechanical models, and most importantly, those of the rigid rotor and the harmonic oscillator approximations.\cite{WiDeCr55,BuJe98}

\subsubsection{Rotational basis functions and matrix elements}
Basis functions for the rotational part are obtained from eigenfunctions of the symmetric top problem, $\ket{J\bar{K}M}$ $(\bar{K}=-J,\ldots,J)$.\cite{BuJe98} The $J=0,1,\ldots$  and $M=-J,\ldots,J$ quantum numbers correspond to the total rotational angular momentum square and its laboratory-frame projection, which are exact quantum numbers for isolated molecules. 
For every $JM$ pair, the $2J+1$ symmetric top eigenfunctions corresponding to different $\bar{K}=-J,\ldots,J$ values span the rotational subspace. 

For a variational-like computation, the matrix representation of the Hamiltonian, Eq.~(\ref{eq:detPodrv}), must be constructed over the basis set. Matrix elements including the $\hat{J}_a$ angular momentum operators and the $|J\bar{K}M\rangle$ rigid-rotor functions are\cite{Za98,BuJe98,FaMaCs11}
\begin{align}
  \langle J\bar{K}M | \hat{J}_x | J(\bar{K}\pm 1)M \rangle 
  &= \frac{1}{2} \sqrt{J(J+1)-\bar{K}(\bar{K}\pm 1)} \\
  \langle J\bar{K}M | \hat{J}_y | J(\bar{K}\pm 1)M \rangle 
  &= \mp \frac{\iim}{2} \sqrt{J(J+1)-\bar{K}(\bar{K}\pm 1)} \\
  \langle J\bar{K}M | \hat{J}_z | J\bar{K}M \rangle 
  &= \bar{K} \; .
\end{align}
To avoid complex-valued Hamiltonian matrix elements, instead of $\ket{J\bar{K}M}$, their
linear combination, the so-called called Wang-functions\cite{Wa29} are used, for  $K=|\bar{K}|=1,\ldots,J$,
\begin{align}
\label{eq:wang}
\Theta^{(JM)}_{K\tau}
=
\left\lbrace 
\begin{array}{@{}c@{\ }c@{}}
    \frac{1}{\sqrt{2}} \left[ \ket{J\bar{K}M} + \ket{J-\bar{K}M} \right], & \text{for even}~\bar{K}, \tau=0  \\
    \frac{\iim}{\sqrt{2}} \left[ \ket{J\bar{K}M} - \ket{J-\bar{K}M} \right], & \text{for even}~\bar{K}, \tau=1 \\
    \frac{1}{\sqrt{2}} \left[ \ket{J\bar{K}M} - \ket{J-\bar{K}M} \right], & \text{for odd}~\bar{K},  ~\tau=0\\
    \frac{\iim}{\sqrt{2}} \left[ \ket{J\bar{K}M} + \ket{J-\bar{K}M} \right], & \text{for odd}~\bar{K},  ~\tau=1\\
\end{array}
\right.
\end{align}
For $K=0$, $\Theta^{(JM)}_{00}=|J 0M \rangle$ (and $\Theta^{(JM)}_{01}=0$).

\subsubsection{Vibrational basis functions and matrix elements}
A common and straightforward choice for a multi-dimensional vibrational basis function is 
a product form,
\begin{align}
  \Phi_i(q_1,\ldots,q_D)
  =
  \prod_{k=1}^D 
    \vphi^{(k)}_{i_k}(q_k) \; ,
  \label{eq:vibprodbasis}
\end{align} 
where $\vphi^{(k)}_{i_k}\ (i_k=1,\ldots,\nb{k})$ are 1-dimensional (1D) functions of the vibrational coordinates forming an orthonormal basis set, \emph{e.g.,} Hermite functions, obtained from the analytic solution of the harmonic oscillator eigenproblem.\cite{WiDeCr55,BuJe98}

The matrix representation for the $\hat{p}_k$ differential operator is known in analytic form for all commonly used 1D basis function types. At the same time, the potential energy is, in general, a complicated $(3N-6)$-dimensional function, and its matrix elements can be computed only by numerical integration, unless some special form is assumed for the function representing the potential energy (Sec.~\ref{sec:other}). 
For special coordinate choices, the curvilinear kinetic energy operator can be written in a closed analytic form, which can be integrated by analytic expressions for the vibrational basis. But, if we aim for a `black-box-type' vibrational procedure, we consider $G_{kl}$, $\detg$, $B_l$, and $U$ as general, vibrational-coordinate dependent functions, similarly to the $V$ potential energy, which is integrated by numerical techniques. Furthermore, the coordinates which are most convenient for fitting the potential, most commonly, interatomic distances, do not represent a good choice for doing the rovibrational computation.

A 1D numerical integral, 
\begin{align}
  \int_a^b
    w(q) F(q)\ \dd q
  =
  \sum_{i=1}^{k^\text{q}} w_i\ F (q_i) \; 
  \label{eq:oneDnumint}
\end{align}
can be most efficiently calculated by some Gaussian quadrature. If an appropriate Gaussian quadrature rule, $w_i$ weights and the $q_i$ points, exists for the $a$ and $b$ boundaries and the $w$ weight function, then, the numerical result is exact if $F$ is at most a $(2k^\text{q}-1)$-order polynomial of $q$, and $2k^\text{q}-1$ is called the (1D) accuracy of the Gaussian quadrature.\cite{golub}
A straightforward generalization of 1D quadratures to multi-dimensional integration relies on forming a multi-dimensional, direct-product quadrature of the 1D rules, 
\begin{align}
  &\int_{a_1}^{b_1} \ldots \int_{a_D}^{b_D} w^{1}(q_1)\ldots w^{D}(q_D)
    F(q_1,\ldots,q_D)\ \dd q_1 \ldots \dd q_D
  \nonumber \\
  &\approx
  \sum_{i_1=1}^{\nq{1}} \ldots \sum_{i_D=1}^{\nq{D}}
    w_{1,i_1} \ldots w_{D,i_D}\ 
    F(q_{1,i_1},\ldots,q_{D,i_D}) \; 
  \label{eq:multiquad}
\end{align}
with the $w_{k,i_k}$ weights and $q_{k,i_k}$ points ($k=1,\ldots,D$) corresponding to the 1D quadrature rule of the $k$th degree of freedom. 
We note, that in some cases (for periodic coordinates), we do not use orthogonal polynomial functions, but sine and cosine functions. Integrals of periodic functions can be very efficiently integrated by a trapezoidal (equidistant) quadrature that converges exponentially fast with respect to the number of grid points.

In the vibrational methodology, a finite basis and finite grid representation used to construct the Hamiltonian matrix is called the finite basis representation (FBR). Since we assume a general $V$ function representing the PES, the integration is in general non-exact, but the exact value of the integrals for the basis set can be approached (arbitrarily close) by increasing the number of quadrature points.

\begin{figure}
    \centering
    \includegraphics[width=6.5cm]{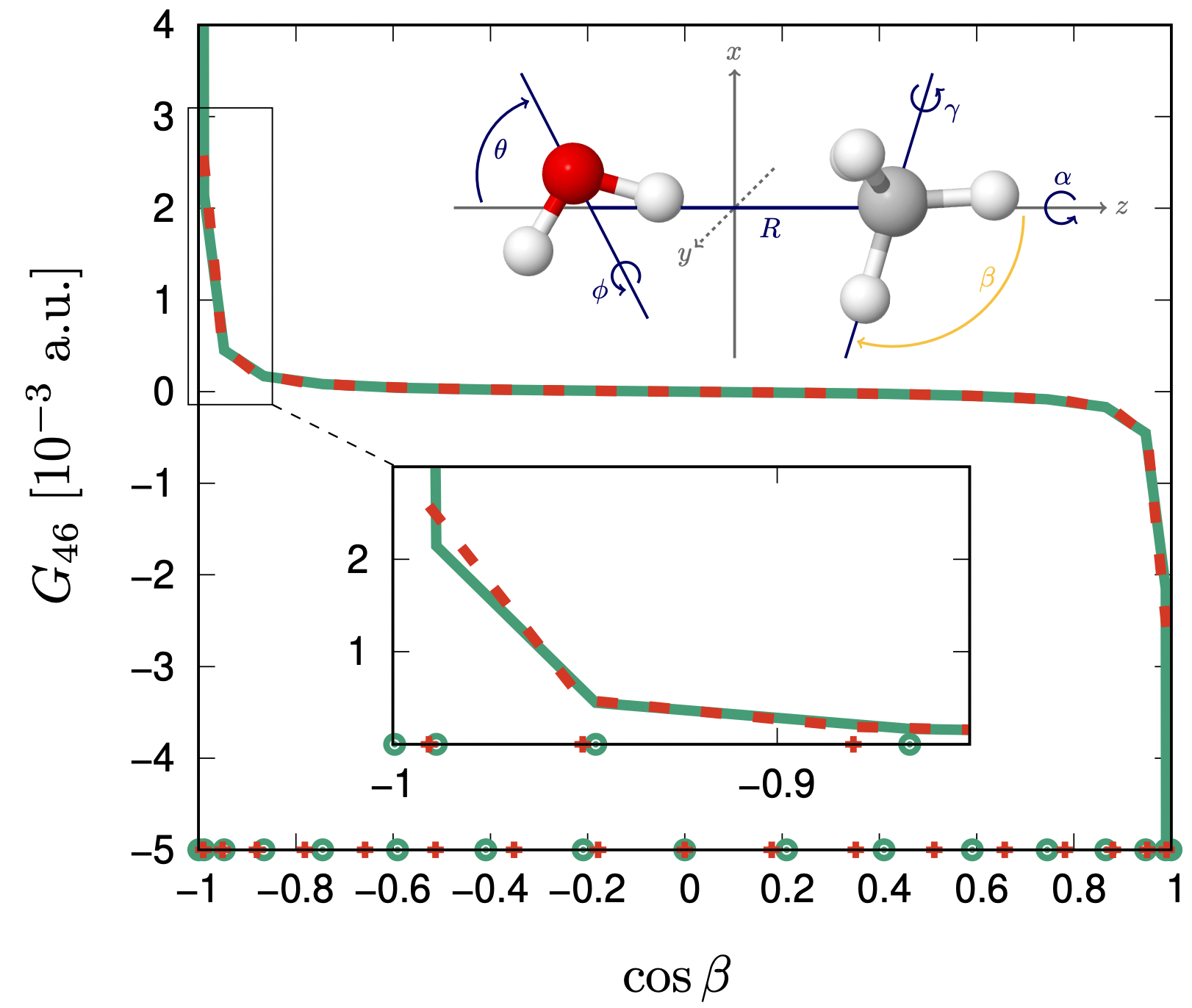}
    \caption{%
      Singular behaviour of a $\bos{G}$ matrix element, `$G_{46}$', for the example of the \metwat\ dimer along the $\cos\beta$ coordinate. The definition of the intermolecular coordinates, $(R,\theta,\phi,\alpha,\beta,\gamma)$, is shown in the inset.
      The cot-DVR points\cite{ScMa10} (green) have a higher density than Legendre DVR (red) near the singularities, $\cos\beta=-1$ and $+1$. 
     \label{fig:Gsingular}
      }
\end{figure}

For the special case of an equal number of basis functions and quadrature points for the $k$th degree of freedom ($\nb{k}=\nq{k}$), it is possible to define a similarity transform of the finite basis with the prescription that the transformed basis functions diagonalize the $q_k$ coordinate operator matrix. This choice results in the discrete variable representation (DVR) of the basis, and the eigenvalues of the coordinate operator provide us with the DVR grid points.\cite{Szalay96,BaLi86,LiCa00} This representation became popular in the vibrational community in conjunction with the approximation
\begin{align}
  (\bos{q^n}_k)_{ij} \approx (\bos{q}_k\cdot \ldots \cdot \bos{q}_k)_{ij} = q_{k,i}^n \delta_{ij} \; ,
  \label{eq:diagonal}
\end{align}  
which relies on multiple ($n-1$) insertion of the truncated resolution of identity (in the finite basis), and allows us to approximate any function of the coordinates by a diagonal matrix. 
For a finite grid (and basis), this is certainly an approximation, which spoils the rigorous variational property (and convergence of the eigenvalues from `above') of the computations. In practice, beyond a certain minimal number of grid points, the eigenvalues properly converge to their `numerically' exact value upon increase of the number of grid points (equal to the number of basis functions).
Eq.~(\ref{eq:diagonal}) allows to straightforwardly include any complicated function of the coordinates in a `black-box' fashion. Complicated functions of the coordinates,
\emph{e.g.,} $V$, $G_{kl}$, $\detg$, $B_l$, $U$ (Sec.~\ref{sec:coord}), 
are ubiquitous in the vibrational theory. These general coordinate-dependent functions do not need to be explicitly represented with polynomials of the coordinates, but it is assumed that they are smooth functions of the coordinates and their change with the coordinates is characterized by not too high-order polynomials (the maximal polynomial order is not too high), and hence, the numerically exact results can be systematically approached by increasing the number of quadrature points. This is generally the case for the $V$ PES corresponding to an isolated electronic state over the dynamically relevant coordinate range.  Certain terms in the kinetic energy operator ($G_{kl}$, ...) may become singular over the dynamically relevant coordinate range, which commonly happens in floppy molecular systems (singularity of some elements of the $\bos{G}$ matrix is highlighted for the example of the methane-water dimer in Fig.~\ref{fig:Gsingular}). 
In practice, this singular behaviour can be determined (either from the analytic formulation of the kinetic energy or) by numerically `measuring' the behaviour upon the change of the coordinates.\cite{AvMa19} Most often an appropriate (Gaussian-)quadrature can be found, in which the weight function, Eq.~(\ref{eq:oneDnumint}), corresponds to this singular behaviour. Ref.~\citenum{AvMa19} presents an example, when this is not the case, and the DVR-FBR approach is used. 
Another option would be to deal with singularities using special functions of multiple coordinates. Most importantly, singularities due to spherical motion can be treated by using spherical harmonics\cite{WaCa03HF} or Wigner $D$-functions\cite{WaCa21}, but this alternative would mean losing a product basis set (of one-particle basis functions), which allows a general implementation.

The size of the direct-product basis and the direct-product grid grows exponentially with the number of active vibrational degrees of freedom. 
The corresponding computational cost can be mitigated by efficient algorithmic and implementation techniques, most importantly by 
(a) evaluating nested sums, \emph{e.g.,} Eq.~(\ref{eq:multiquad}), sequentially;\cite{BraCa93,BraCa94,RoCa96,WaCa01,WaCa03,CaWa11,WaCa21}
(b) using an iterative (Lanczos) eigensolver,\cite{lanczos,BrCa94} which requires only multiplication of a trial vector with the Hamiltonian matrix without storage or even explicit construction of the full matrix.\cite{BrCa94,MaSiCs09,MaCzCs09} Nevertheless, even in the most efficient implementation, a few vectors of the total size of the basis (and the grid) must be stored, which grows exponentially with the vibrational dimensionality. The matrix-vector multiplication can be parallelized with the OpenMP protocol (as it is described in Ref.~\citenum{MaSiCs09}, for example).

During the course of the Lanczos eigensolver iterations, one eigenpair is converged after the other. The computational effort, which is determined by the number of matrix-vector products is determined by the number of required eigenvalues, which means that, in practice, up to a few hundred (a thousand) states can be efficiently computed. 
The smallest (or largest) eigenvalues of the Hamiltonian matrix can be efficiently computed. To compute a spectral window (from an arbitrarily chosen range of the entire spectrum), there is currently no known methodology that would be more efficient than computing all states up to the desired range.\cite{MaSiCs09} A truly efficient computation of solely a spectral window is still an open problem. 

An efficient implementation and available computing power typically allows us to use this simple, `direct'(-product) methodology up to 6--10 fully coupled vibrational degrees of freedom.\cite{FaSaCs14,SaCs16,SaCsAlWaMa16,SaCsMa17,DaAvMa21a,DaAvMa21b}

To be able to solve the vibrational Schrödinger equation beyond this system size (or for challenging floppy systems up to a higher energy range already with up to ca.~10 degrees of freedom), it is necessary to develop vibrational methodologies which attenuate the rapid increase of the computational cost with the dimensionality.

\subsection{Computational bottleneck of the variational vibrational methodology \label{sec:limitations}} 
\begin{figure}
  \centering
  \includegraphics[width=6.5cm]{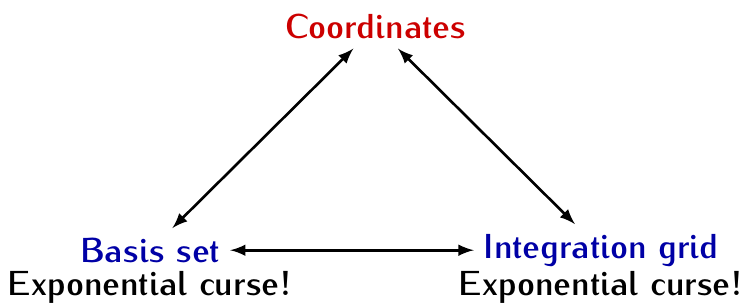}
  \caption{%
    Complexity of the vibrational problem. In general, the PES is a multi-dimensional function.
    \label{fig:complexity}
  }
\end{figure}

While the electronic structure problem scales exponentially with respect to the basis size, the general vibrational problem suffers from a \emph{double} exponential scale-up: the computational effort grows exponentially fast with the dimensionality with respect to the basis set size \emph{and} also with respect to the integration grid size (Fig.~\ref{fig:complexity}). 

This double exponential scale-up is due to the fact that the potential energy surface (PES) is in general a multi-($3N-6$-)dimensional function, for which the matrix representation must be constructed by means of numerical integration (hence, the exponentially growing integration grid).

Over the past decade, there have been important developments in the quantum nuclear motion methodology that make it possible to attenuate the exponential growth of the vibrational problem in a systematic manner. 
In the following subsection (Sec.~\ref{sec:smolyak}), we present a strategy that we used and developed during the past few years, and it is followed by a short overview of other 
possible strategies in Sec.~\ref{sec:other}.

\subsection{A possible strategy for efficient vibrational computations: Smolyak quadrature and pruning \label{sec:smolyak}} 
A possible strategy to attenuate the rapidly increasing computational cost of solving the vibrational Schrödinger equation relies on a systematically improvable reduction of both the basis set size and the integration grid size. 

The foundations of this direction were laid down by Avila and Carrington about a decade ago. \cite{AvCa09,AvCa11,AvCa11b} More recently, we have `embedded' this approach, originally used for semi-rigid systems, in computations involving both floppy and semi-rigid parts.\cite{AvMa19,AvMa19b,AvPaCzMa20,DaAvMa22}

\begin{figure}
  \centering
  \includegraphics[width=8cm]{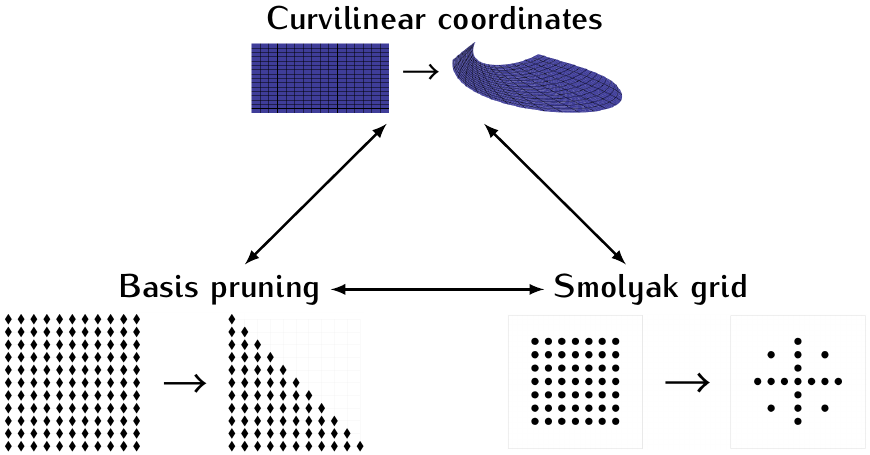}
  \caption{%
    Illustration of a possible strategy to attenuate the double-exponential scale-up of the vibrational problem: system-adapted curvilinear coordinates, basis pruning, and grid pruning.
    \label{fig:strategy}
  }
\end{figure}

The strategy is highlighted in Fig.~\ref{fig:strategy}.
An initial pre-requisite for practical usefulness of this approach is to find good curvilinear coordinates, which make the kinetic and potential coupling small.

\subsubsection{Towards optimal internal coordinates}
To be able to efficiently truncate the product basis set, Eq.~(\ref{eq:vibprodbasis}), we need good coordinates.
For good coordinates, both the kinetic and the potential energy coupling is small. For semi-rigid molecules, like methane, CH$_4$, and ethene, C$_2$H$_4$, a reasonably good coordinate representation is immediately provided by the commonly used  rectilinear normal coordinates. By construction \emph{(vide infra)} both the kinetic and the potential coupling is zero at the equilibrium structure, and since the vibrations are of small amplitude, the coupling remains small over the dynamically relevant coordinate range. Hence efficient basis pruning is possible, and grid pruning was introduced for these systems used as examples in Refs.~\citenum{AvCa09,AvCa11,AvCa11b}.

\begin{figure}
    \centering
    \includegraphics[scale=0.4]{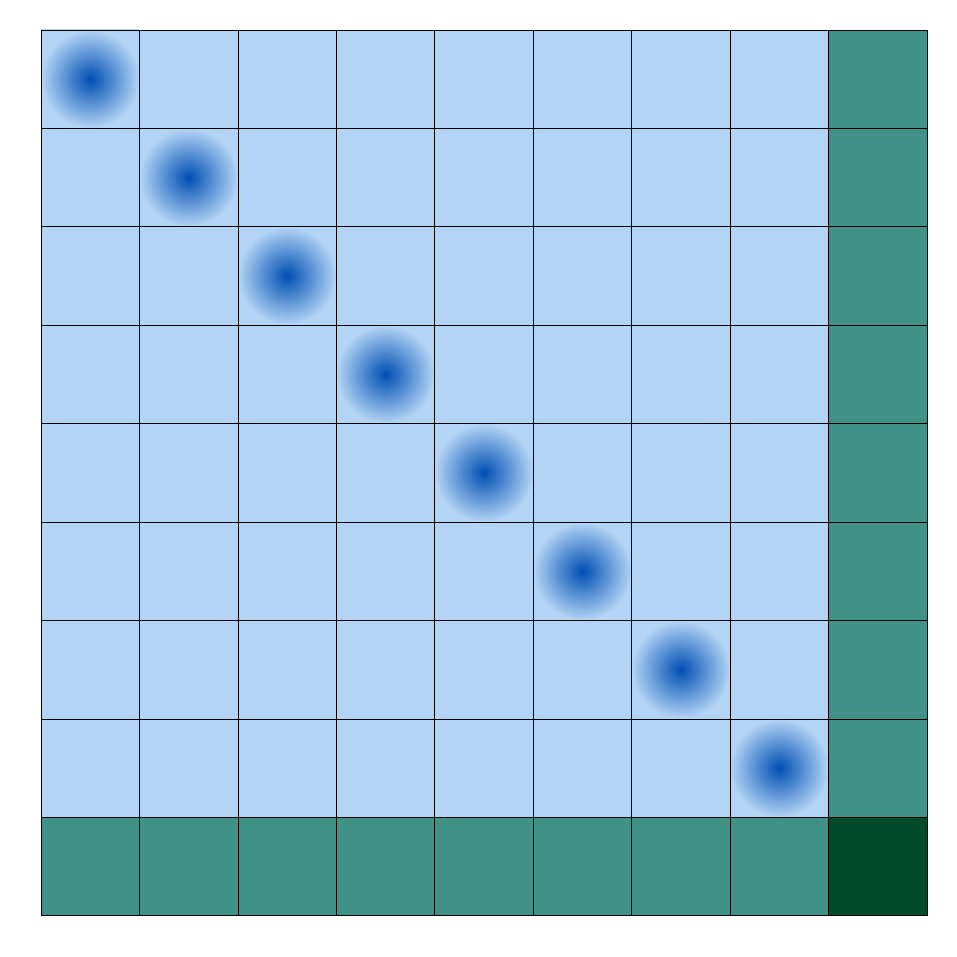}
    \caption{Optimal coordinates ensure small coupling (light blue).}
    \label{fig:smalcpl}
\end{figure}

At the same time, finding optimal coordinates for floppy systems is non-trivial.
Fortunately, a very broad and chemically important class of floppy molecular systems have just a few large-amplitude motions (LAMs) and many small-amplitude motions (SAMs), examples include, NH$_{3}$, methanol, CH$_{3}$OH,\cite{LaNa14} ethane, CH$_3$CH$_3$, ethanol, CH$_3$CH$_2$OH, or glycine, NH$_2$CH$_2$COOH. 
For these types of systems, a simple strategy is to accept that a small number of vibrational degrees of freedom, \emph{i.e.,} the LAMs, are strongly coupled (among themselves and with the small-amplitude ones), and hence basis and grid truncation is inefficient. At the same time, the remaining motions are of small amplitude for which good coordinates can be constructed (Fig.~\ref{fig:smalcpl}). For doing this, we need to abandon the equilibrium structure as reference, and use instead a reference path (surface, volume, hyper-volume\ldots) which depends on the LAMs. With respect to this (multi-dimensional) reference `path', we can define linear combination of the small-amplitude coordinates, which minimize both the kinetic and the potential energy coupling among the small-amplitude modes. These `good' small-amplitude coordinates are constructed similarly to the (curvilinear) normal coordinates.\cite{PaAlBook82,ZhKlMa91,CaAvMaAgScCa16}

Papu\v{s}ek and co-workers pioneered this direction of research starting from the 1970s with first applications for the ammonia molecule described by one large-amplitude and many (five) small-amplitude degrees of freedom.\cite{PaStSp73,PaAlBook82}
For the case of using rectilinear normal coordinates for representing the small-amplitude vibrations and a one-dimensional path, the `so-called' reaction-path Hamiltonian\cite{MiHaAd80} has been derived and used in many applications.\cite{MiHaAd80,BoHuHaNoCa07} For two large-amplitude degrees of freedom, the reaction-surface Hamiltonian\cite{CaMi84} has been formulated, and it was used for a quantum dynamical model of hydrogen (proton) transfer in malonaldehyde\cite{CaMi86} and in the formic acid dimer.\cite{ShBaAl91}

Anharmonicities of one-dimensional bond-stretching and bending-type motions are better represented by the curvilinear bond stretching- and bending-type coordinates, than by rectilinear `analogues',\cite{DaAvMa22} and this motivates the definition of \emph{curvilinear} normal (c-normal) coordinates, $\bos{\mcQ}$. Although the analytic kinetic energy operator is not immediately available for these coordinates, the kinetic energy coefficients can be straightforwardly computed over a grid using ideas of Sec.~\ref{sec:coord}.

So, let us consider some physically motivated `primitive' internal coordinates, $\bos{\rho}_i$ (bond distances, angles, etc.),
to describe the small-amplitude vibrations. Then, the curvilinear normal coordinates, $\mcQ_k$, are defined with respect to the reference structure, $\bos{\rho}^{\text{(0)}}$, obtained as the minimum of the PES for a selected value of the LAM coordinates, $\tau$, 
\begin{align}
  \bos{\rho}(\tau)
  = 
  \bos{\rho}^{(\text{0})}(\tau) + 
  \bos{L}(\tau) \bos{\mcQ} \; ,
\end{align}
where
the linear combination coefficient matrix, $\bos{L}$, solves the $\bos{L}^{-1} \bos{\mathcal{GF}} \bos{L} = \bos{\Lambda}$ equation with normalization $\bos{L}^\text{T}\bos{\mathcal{G}}^{-1}\bos{L}=\bos{I}$, and thereby minimize the coupling both in the potential and in the kinetic energy near the $\bos{\rho}^{(0)}$ reference structure.\cite{PaAlBook82} $\bos{\mathcal{G}}(\bos{\rho}^{(0)},\tau)$ is the small-amplitude block of the `big' $\bos{G}$ matrix, Eq.~(\ref{eq:bigGmx}), corresponding to the reference structure for $(\bos{\rho}^{(0)},\tau)$, and $(\bos{\mathcal{F}})_{ij}=\partial^2 V/\partial \rho_i \partial \rho_j|_{(\bos{\rho}^{(0)},\tau)}$ is the force-constant matrix at the same structure.

Whether it is necessary to determine $\bos{L}(\tau)$ for `every' $\tau$ (in practice,  carried out over a grid and interpolated or fitted to have a functional representation)\cite{DaAvMa22} or some ($\tau$ independent) averaged $\bar{\bos{L}}$ is sufficient for the relevant $\tau$ range of the dynamics, depends on the system and energy range, and, in principle, should be always checked (numerically).

The more common rectilinear normal coordinates,\cite{WiDeCr55} $\bar{Q}_i$ could be obtained as
\begin{align}
  \bos{r} = \bos{r}^\text{(0)} + \bos{l} \bos{Q} \;  ,
    \label{eq:normalcoor1}  
\end{align}
where $\bos{l}=\bos{T}^\text{(0)} \bos{L}$, and $(\bos{T}^{(\text{0})})_{ik}=t_{ik}^{(0)}$ collects the vibrational t-vectors, Eq.~(\ref{eq:tvib}), corresponding to the semi-rigid degrees of freedom at structure $(\bos{\rho}^{(0)},\tau)$.

For practical computations with c-normal coordinates, it is necessary to pay attention to the domain of the various coordinates. The primitive internal coordinates are typically angle- and distance-type coordinates, and they are calculated as linear combination of the c-normal coordinates, defined on the entire real axis, $Q_k\in\mathbb{R}$. 
`Mapping functions' were used in Ref.~\citenum{DaAvMa22} to ensure that the calculated coordinate values are in the mathematically correct range.

\subsubsection{Basis pruning}
If good coordinates are found, which ensure that the coupling of the different degrees of freedom is small (off-diagonal Hamiltonian matrix elements less than ca.~30~\% of the corresponding diagonal elements), then, some product basis functions are not required, can be discarded, in order to converge the lowest energy levels,\cite{WhHa75,BoCaHu03} assuming that reasonably good one-dimensional basis functions can be found.
We note that for basis (and grid) pruning, we use FBR, since (unfortunately), there is no known efficient way to prune a DVR, in which the basis and grid representations are inherently coupled. 

Discarding basis functions from a direct-product basis means that instead of writing the vibrational ansatz as
\begin{align}
  \Psi_{i}(q_1,\ldots,q_D) 
  = 
  \sum_{n_1=1}^{n_1^\max}
  \ldots
  \sum_{n_D=1}^{n_D^\max}
    c_{i,n_1,\ldots,n_D} 
    \prod_{k=1}^D 
    \varphi_{n_k}(q_k) \; ,
  \label{eq:productbasis}
\end{align}
we choose an $f(n_1,\ldots,n_D)$ function of the basis function `excitation' indices to define basis combinations of one-dimensional basis functions that are important, 
\begin{align}
  \Psi_{i}(q_1,\ldots,q_D) 
  = 
  \sum_{f(n_1,\ldots,n_D) \leq b}  
    c_{i,n_1,\ldots,n_D}
    \prod_{k=1}^D 
      \vphi_{n_k}(q_k) \; ,
  \label{eq:prunedbasis}    
\end{align}
and discard the rest without jeopardizing the accuracy of the results.
The simplest possible pruning function is the sum of the basis excitation indices,
\begin{align}
  f(n_{q_1},...,n_{q_{D}}) = \sum_{i=1}^D n_{q_i} = n_{q_1} + \ldots + n_{q_{D}} \; .
\end{align}
This function can be improved\cite{HaPo15,HaPo15b,SaPo21} by weighting `down' the contribution from the $n_{q_i}$ excitation `quanta' of the lowest-energy $D$th and $(D-1)$th, $\ldots$, vibrations, \emph{e.g.,} 
\begin{align}
  f(n_{q_1},...,n_{q_{D}}) = n_{q_1} + \ldots + \frac{1}{2} n_{q_{D-1}}+ \frac{1}{3} n_{q_{D}} \; .
\end{align}
For an \emph{a priori} assessment of the importance of a basis function, the following considerations can be made.
Regarding the (un)importance of an $|\bos{n}'\rangle$ basis state ($\bos{n}'$ collects the basis indexes) in a wave function dominated by the $|\bos{n}\rangle$ basis state, the magnitude of the ratio, 
\begin{align}
\frac{%
  \langle  
    \bos{n}
    | \hat{H} |
    \bos{n}'
  \rangle
}{%
  E^{(0)}_{\bos{n}}
  -
  E^{(0)}_{\bos{n}'}
}
\label{eq:coup}
\end{align}
can be indicative. 
The ratio is small, \emph{i.e.,} the $|\bos{n}'\rangle$ basis function can be neglected, 
either if (a) the coupling through the Hamiltonian matrix (nominator) is small, or (b) the zeroth-order energy difference (denominator) is large.
The order of magnitude of the Hamiltonian matrix element can be estimated by considering the Taylor expansion of the potential and the kinetic energy written in good coordinates.
Pruning is almost a separate field within theoretical spectroscopy, since there are infinitely many possible choices (and many possible physically motivated arguments). For example, there are pruning choices that are effective only to compute fundamental vibrations and other pruning choices are useful to compute thousands of states.\cite{HaPo15,HaPo15b,SaPo21}

\subsubsection{Grid pruning}
Reducing the basis size allows us to tackle part of the problem (Figs.~\ref{fig:complexity}, \ref{fig:strategy}), the exponential scale-up of the overall variational procedure can be attenuated only if the integration grid is also pruned.
A general approach to the variational grid-size problem is provided by the non-product Smolyak quadrature grid approach first introduced by Avila and Carrington \cite{AvCa09} in vibrational computations, and, since then, it has been used to efficiently describe the semi-rigid skeleton of a variety of systems \cite{AvCa11,AvCa11b,LaNa14,AvMa19,AvMa19b,ChLa21,DaAvMa22}.

In an abstract manner, we can write the direct-product quadrature rule  as
\begin{align}
  \hat{Q}(D,k_1^\max,\ldots,k_D^\max)
  &=
  \sum_{i_1=1}^{k_1^\max}
  \ldots
\sum_{i_D=1}^{k_D^\max}
  \hat{Q}^{i_{q_1}}_{q_1}   \otimes  \cdots \otimes  \hat{Q}^{i_{q_D}}_{q_D} \; ,
\end{align}
to generate the sum in Eq.~(\ref{eq:multiquad}), where every $\hat{Q}_{q_i}^{i_{q_i}}$ operator generates a sum to approximate that 1D integral, Eq.~\eqref{eq:oneDnumint}.
The $\hat{Q}(D,k_1^\max,\ldots,k_D^\max)$ direct-product quadrature could be used to integrate
all matrix elements of the full, direct-product basis, Eq.~(\ref{eq:productbasis}), but it is `too good'
to integrate the matrix elements of a pruned basis. 
It is also important to note that the basis pruning, Eq.~(\ref{eq:prunedbasis}), has some structure, determined by the $f(n_1,\ldots,n_D)$ function, and typically, products of multiple high-order polynomials are discarded. So, in the pruned basis, it is sufficient to integrate functions including high-order polynomials for only one (or a few) degrees of freedom at a time. The products of the highest-order polynomials for multiple (all) degrees of freedom are discarded in the pruned basis, and thus, the corresponding part of the grid, which would be necessary to integrate these high-order polynomial products can also be discarded.

A systematic formulation and implementation of this idea was provided by Avila and Carrington\cite{AvCa09} by using a non-product Smolyak quadrature, which can be defined as a linear combination of direct-product quadratures rules as 
\begin{align}
\label{eq:smolgrid2}
  \hat{Q}(D,H)
  &=
  \sum_{{{\pmb{\sigma}}_{\pmb{s}}}(i)\le H} 
    C_{i_{q_1},\ldots,i_{q_D}}
    \hat{Q}^{i_{q_1}}_{q_1}   \otimes  \cdots \otimes  \hat{Q}^{i_{q_D}}_{q_D} \; ,
  \nonumber \\ 
  & \quad \quad \quad 
  i_{\chi}=1,2,3,4,\ldots \quad \text{and} \quad \chi=1,\ldots,D,
\end{align}
where $H$ is a grid-pruning parameter and ${\pmb{\sigma}_{\pmb{s}}}(i)$ is the grid-pruning function, for which the simplest form is\cite{AvCa12}
\begin{align}
   {{\pmb{\sigma}}_{\pmb{s}}}(i) = s^{(q_{1})}(i_{q_{1}})+\ldots+s^{(q_{D})}(i_{q_{D}}) \leq H.
  \label{eq:smolcondi}
\end{align}
With this non-product grid, the number of points kept for the accurate integration of the potential and kinetic integrals is (much) smaller than the direct product grid that we would need to evaluate the same integrals with the same accuracy. 
There are three factors that can be tuned to modify and improve the accuracy of the Smolyak integration grid:
    (a) the grid-pruning functions,  $s^{(q_k)}(i_{k})$, which must be monotonic increasing functions, \emph{i.e.}, $s^{(q_k)}(i_{k})\ge s^{(q_k)}(i_{k}-1)$; 
    (b) the grid-pruning parameter, $H$ (the larger $H$, the better the convergence, but the grid is  larger); and
    (c) the underlying grid sizes of the nested quadrature rules. For a nested quadrature, all points of the $j$th quadrature rule are within the $(j+1)$th quadrature rule. These quadratures do not have a Gaussian accuracy.
    For harmonic basis functions, we use a sequence of quadrature rules explained by Heiss and Winschel.\cite{HeWi08} Nested quadratures are necessary to have a structure for the Smolyak grid, and thereby, to be able to calculate the Hamiltonian matrix-vector product in a sequential way.

So, following Ref.~\citenum{AvCa09}, we can write a multi-dimensional integral of an $F(x_{1},\ldots,x_{D})$  multivariate function as
\begin{align}
&  \int \cdots \int F(x_{1},\ldots,x_{D})\ \dd x_{1}\ldots \dd x_{D} \approx \nonumber\\ 
&  \sum_{k_{1}=1}^{k_{1}^{\rm max}}\ldots\sum_{k_{D}=1}^{k_{D}^{\rm max}} 
    W^{\rm Smol}(k_{1},\ldots,k_{D}) 
    F(x_{1}^{k_{1}},\ldots, x_{D}^{k_{D}}) \; ,
\end{align}
where $W^{\rm Smol}(k_{1},\ldots,k_{D})$ is the Smolyak weight for every $(k_1,\ldots,k_D)$ point. The Smolyak grid structure appears in the summation indexes $k_{c}^{\rm max}$ as follows: 
$k_{1}$ depends on $H$, 
$k_{2}^{\rm max}$ depends on $H$ and $k_{1}^{\rm max}$, 
$k_{3}^{\rm max}$ depends on $H$, $k_{1}^{\rm max}$ and $k_{2}^{\rm max}$, etc. 
Due to this special structure, the matrix-vector products can be computed using sequential summations. \cite{AvCa09,AvCa11b,AvMa19,AvMa19b,AvPaCzMa20}. 
The implementation details of the matrix-vector multiplication in relation with a pruned Smolyak integration was described in Refs.~\cite{AvCa09,AvCa11b,AvMa19}.
Finally, we note that Lauvergnat and co-workers use a somewhat different grid-pruning approach.\cite{LaNa14,NaLa18,ChLa21}

\subsection{Other possible strategies for efficient vibrational computations \label{sec:other}} 

\subsubsection{Other efficient variational strategies}
Instead of pruning the multi-dimensional direct product basis, it is possible to
reduce the number of multi-dimensional functions by 
ensuring that the number of one-particle (1D) basis functions remains very small, which is possible by using optimized 1D functions in the direct-product expansion. 
This idea is realized in the multi-configurational time-dependent Hartree approach (MCTDH), \cite{BeMiJa2000} which ensures that the 1D functions are optimal by using time-dependent basis functions.
Alternatively, one can build a compact multi-dimensional basis set by using contracted basis functions obtained from solving the (reduced-dimensionality) subsystems' Schrödinger equation.\cite{BaLi89,HeTe90,WaCa04,WaCa18,FeBa19,FeBa20,LiLiFeBa21}

Similarly to the basis-size problem, there are different strategies to tackle the grid-size problem of multi-dimensional numerical integration, including: 
(a) a Sum-Of-Products (SOP) expansion\cite{JaMe96,ThoCarr17,ThoCarr18} of the Hamiltonian, which allows to construct the multi-dimensional integrals from 1D integrals;
and 
(b) a truncated $n$-mode expansion\cite{CaCuBo97,CaBoHa98,BoCaHu03,ZiRa19} of the Hamiltonian, which replaces the $D$-dimensional integrals with a combination of maximum $n$-dimensional integrals $(n\ll D)$.

Aiming for a precise representation of the PES, the number of terms in a sum-of-products expansion may grow rapidly, and the overall computational cost increases significantly.
Regarding the $n$-mode expansion, the variational computations can be made very efficient up to 3-4-modes, but (going beyond) 4-mode coupling has been found to be important for spectroscopic precision.\cite{WaCaBo15}
Furthermore, the multi-mode expansion of the PES is carried out about a single reference structure (typically the equilibrium structure). For floppy systems this expansion is not efficient, but the more general high-dimensional model representation (HDMR),\cite{RaAl99} which has been successfully used for PES development,\cite{MaCa06,MaCa07} can be used, in principle. The $n$-mode expansion about a single equilibrium structure can be considered as a special case of HDMR (`cut-HDMR').

\subsubsection{A non-variational approach}
A fundamentally different and promising direction is about fully abandoning the variational approach and aiming to solve the eigenvalue equation as a differential equation using numerical (finite element) methods. This approach, named `collocation', has been introduced and pursued by Carrington and co-workers for solving the vibrational Schrödinger equation over the past decade.\cite{MaYaCa11,AvCa13,AvCa15,AvCa17,WoCa19,WoCa21,Ca21}
A great advantage of collocation is that it does not require the computation of any integrals, so the computational burden (and unfavorable scale-up with the vibrational dimensionality) of the evaluation of multi-dimensional integrals is completely avoided. 
It can be efficiently used if a very good basis set is available.
Its current technical `difficulty' is connected with an efficient computation of many eigenvectors for a (generalized) non-symmetric (real) matrix eigenvalue problem. 

\subsubsection{Prospects for quantum hardware}
As to further possible strategies, we mention the popular direction of considering alternative hardware, \emph{i.e.,} use of future quantum computers (QC) for quantum dynamics.\cite{OlBaReTa20} For these types of applications, it is convenient to construct the second-quantized form of the vibrational Hamiltonian following Christiansen,\cite{Chr04,OlBaReTa20}
or more recently, to use first-quantized operators.\cite{SuBeWiRuBa21}
For a QC implementation, it appears to be the most advantageous if we have a very sparse matrix over the multi-dimensional basis constructed from one-dimensional basis functions, for which the entries can be `assembled' from lower-dimensional objects. The multi-dimensional (product) basis size does not enter the problem, only the sum of the number of the one-dimensional basis functions (instead of their product, which builds up the exponential growth of the `classical' computational effort). If these conditions are realized, then, we may speculate that a DVR representation will be advantageous (which we abandoned due the exponential scale-up of the matrix size over the one-particle basis and grid). 
Furthermore, if a sufficiently accurate $d<3N-6$ dimensional HDMR representation can be found for the system, then it is possible to cut the exponential growth of the storage (memory) requirement. 
So, based on this speculation, a DVR-HDMR($d$), where $d$ is ideally not (much) larger than 4, may turn out to be an efficient representation of the vibrational problem on a quantum computer.

There are also arguments for performing directly a `pre-Born--Oppenheimer'-type  computation, \emph{i.e.,} by completely avoiding the Born--Oppenheimer approximation \cite{MaHuMuRe11a,MaHuMuRe11b,MaRe12} that may be the most efficient implementation of the molecular quantum mechanics problem in a quantum computer.\cite{KaWhPeYuAs11}

\begin{figure}
  \centering
  \includegraphics[width=8cm]{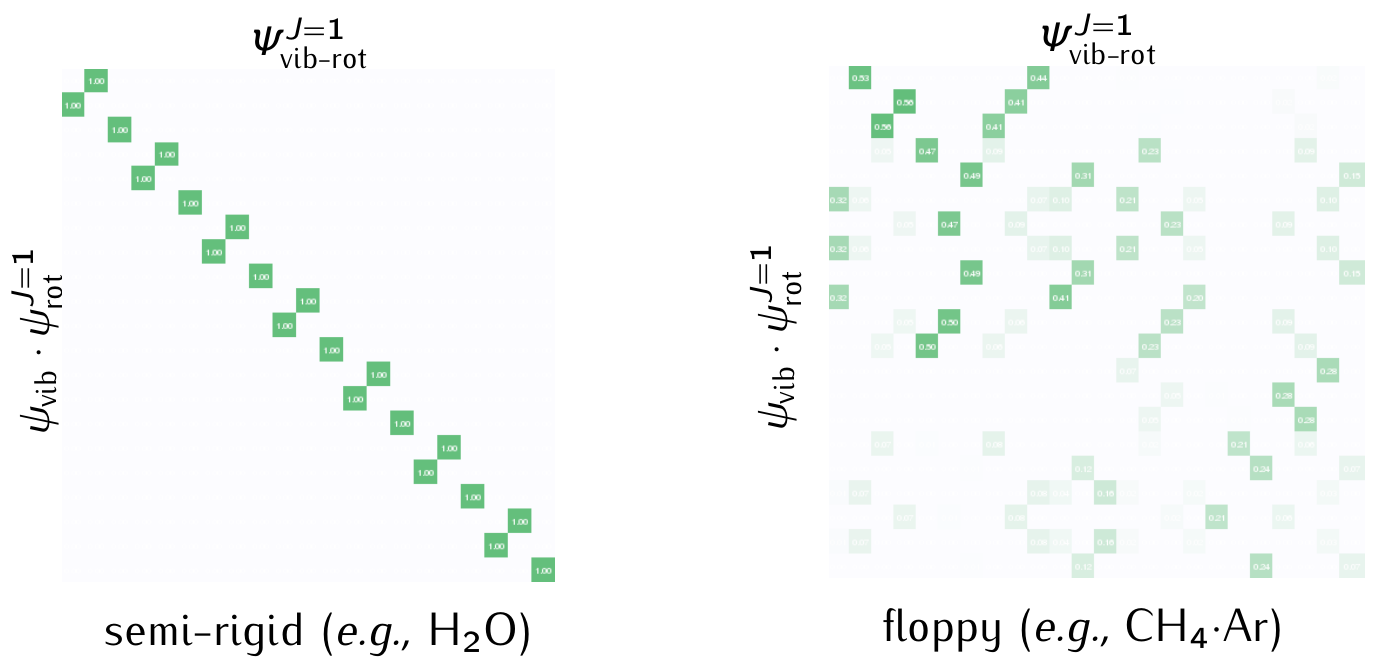}
  \caption{%
    Vibrational parent assignment for a semi-rigid \emph{vs.} floppy system.
  \label{fig:floppyrot}}
\end{figure}

\subsection{Efficient rotational-vibrational computations \label{sec:optframe}}

Let us assume that all vibrational states have been computed for the dynamically relevant energy range. 
If the rovibrational coupling is small, then the rovibrational states with rotational quantum number $J$ can be well approximated with a linear combination of products of the vibrational eigenfunction and the rigid rotor functions,
\begin{align}
  \Psi^{(JM)}_i 
  = 
  \sum_{n=1}^{N_\text{vib}} \sum_{K=0}^J \sum_{\tau=0,1} c^{(JM,i)}_{nK\tau} \psi^\text{v}_n \Theta^{(JM)}_{K\tau} \; ,
  \label{eq:rovibprod}
\end{align}
and precise variational results can be obtained by diagonalizing a (small) rovibrational Hamiltonian matrix constructed over the $JM$ rotational basis (consisting of $2J+1$ basis functions) and the vibrational eigenstates up to the energetically relevant region.

The rovibrational coupling can be made small for semi-rigid systems by using the Eckart frame as a body-fixed frame, \emph{i.e.,} by choosing the $\bos{r}_i$ coordinates (molecular orientation) so that they satisfy, in addition to the center of mass condition, Eq.~(\ref{eq:BForiging}), also the rotational (Eckart) condition,\cite{Ec35}
\begin{align}
  \sum_{i=1}^N m_i \bos{c}_i \times\bos{r}_i = 0    \; ,
\end{align}
where $\bos{c}_i$ is some reference structure, typically the relevant equilibrium structure on the PES. Fulfillment of these conditions ensures that the rovibrational coupling remains small near $\bos{c}_i$, which is sufficient for an efficient rovibrational computation of not too high rotational and vibrational excitation of semi-rigid systems (see, for example, Ref.~\citenum{FaMaFuNeMiZoCs11}), \emph{i.e.,} for which the dynamically relevant coordinate range remains in the proximity of the equilibrium structure.

Let us choose the $\mx{c}_i$ Cartesian reference configuration 
and solve the orientational Eckart condition at Cartesian
structures which belong to the $\mx{\delta}_i^{(\pm k)}$ Cartesian displacements by making
small changes in the $q_k\pm\epsi$ internal coordinates about the reference structure, $\mx{c}_i$.
So, we solve the Eckart condition at 
$\mx{r}_i^{(\pm k)}=\mx{c}_i \pm \mx{\delta}_k$:
\begin{align}
  \sum_{i=1}^N 
    m_i \mx{c}_i\times \mx{r}_i^{(+ k)} & = 0 
    \label{eq:eckartp}
    \\
  \sum_{i=1}^N 
    m_i \mx{c}_i\times \mx{r}_i^{(- k)} & = 0 \; .
    \label{eq:eckartm}
\end{align}
Then, the rovibrational block of the $g$ matrix reads as
\begin{align}
  g_{D+a,k} 
  &=
  \sum_{i=1}^N
    m_i
    [\mx{e}_a \times \mx{r}_i]
    \cdot 
      \pd{\mx{r}_i}{q_k}
    \nonumber \\
  &=
  \mx{e}_a \cdot
  \sum_{i=1}^N
    m_i
    \left[\mx{r}_i
    \times 
      \pd{\mx{r}_i}{q_k}\right]
    \nonumber \\
  &=
  \lim_{\epsi\rightarrow 0}\ 
  \mx{e}_a \cdot
  \sum_{i=1}^N
    m_i
    \left[\mx{r}_i
    \times 
      \frac{\mx{r}_i^{(+k)}+\mx{r}_i^{(-k)}-2\mx{r}_i}{\epsi}\right]
    \nonumber \\  
  &=
  \lim_{\epsi\rightarrow 0}\ 
  \frac{1}{\epsi}  
  \left(%
  \mx{e}_a \cdot
  \sum_{i=1}^N
    m_i
    \mx{r}_i
    \times 
    \mx{r}_i^{(+k)}
  +
  \mx{e}_a \cdot
  \sum_{i=1}^N
    m_i
    \mx{r}_i
    \times 
    \mx{r}_i^{(-k)}  
  \right) \; ,
\end{align}
which shows the known result that the Eckart condition ensures that the rovibrational coupling vanishes at the reference point with $\bos{r}_i=\bos{c}_i$.
Unfortunately, it is impossible to make the rovibrational coupling vanish over an extended part (beyond finite many points) of the configuration space.\cite{LiRe97} Nevertheless, it may be possible to reduce the rovibrational coupling by numerical methods.

In order to make the rovibrational coupling small, $\mx{G}^\text{rv}$ ($=(\mx{G}^\text{vr})^\text{T}$ must be small, where the `rv' and `vr' blocks are understood similarly to the blocks of  $\bos{g}$ in Fig.~\ref{fig:g_matrix}. $\mx{G}^\text{rv}$ depends on the choice of the $\bos{r}_i$ body-fixed Cartesian coordinates, and in principle, the coupling can be modified by defining an optimal shape-dependent body-fixed frame by rotating the system from some initial orientation, $\bos{r}^{(0)}_i$,  by a 3D rotation, described by a rotation matrix $\bos{C}$ and parameterized by three angles, which depend on the $\bos{q}$ shape coordinates, $\mx{\alpha}(\bos{q})=(\alpha(\bos{q}),\beta(\bos{q}),\gamma(\bos{q}))$:
\begin{align}
  \mx{r}_i = \mx{C}[\bos{\alpha}(\bos{q})] \mx{r}^{(0)}_i \; .
\end{align}


\begin{figure}
  \begin{center}
  \includegraphics[width=8cm]{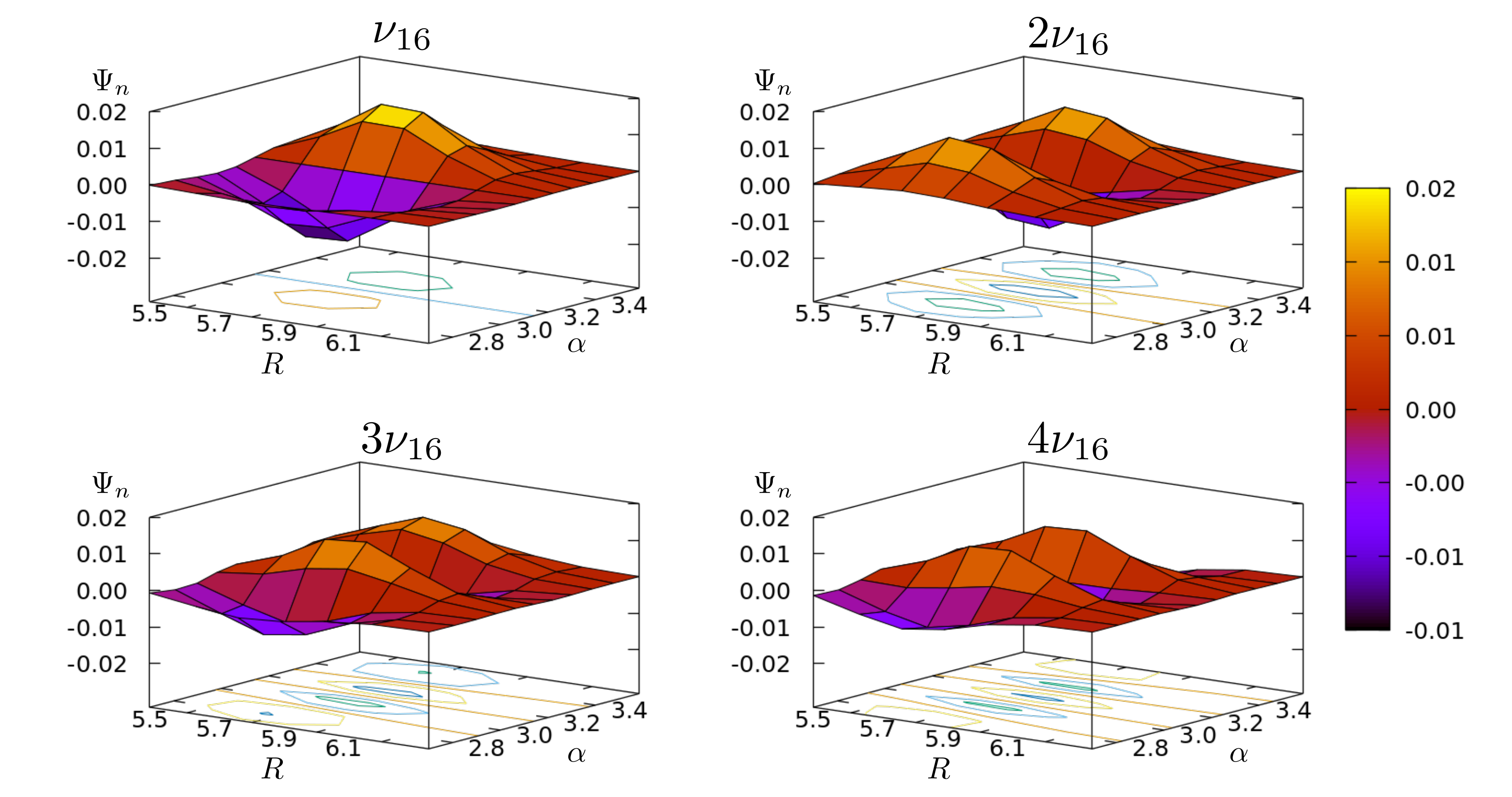}
  \end{center}
  \caption{%
    Assignment of vibrational excitation based on counting the nodes of the wave function for the example of the formic acid dimer. 
    Reproduced from Ref.~\citenum{DaAvMa21b} with permission from the PCCP Owner Societies. 
  \label{fig:nodecounting}}
\end{figure}
\section{Assignment of the rovibrational states\label{sec:assign}}
Variational (ro)vibrational computations provide list of energies and corresponding wave functions (stationary states) that can be used to simulate spectral patterns (Sec.~\ref{sec:ir_tran}) or time-dependent phenomena.\cite{OwZaChYuTeYa17,SzYa17}
It is nevertheless a relevant aim to characterize the computed (hundreds or thousands of) stationary states.

A simple characterization of a state is possible by visual inspection of (1D and 2D cuts of) its wave function and counting the nodes (nodal surfaces), which carry information about the vibrational excitation along the relevant degrees of freedom. As an example, we show 2D cuts of the formic acid dimer vibrational states corresponding to the fundamental vibration and overtones of the twisting mode (Fig.~\ref{fig:nodecounting}).\cite{DaAvMa21b} Node counting is (a) least model dependent way of characterizing a wave function.

Further assignment options include `measuring' the similarity of the variational wave function with simple models (of rotation and vibration).
In the most common representation, a rovibrational wave function can be considered as a linear combination of rigid-rotor (Wang) functions and the lowest $N_\text{vib}$ vibrational eigenfunctions used as a vibrational basis, Eq.~(\ref{eq:rovibprod}).

\subsection{Vibrational parent analysis}
If there is a single, dominant $\Psi^\text{v}_n$ vibrational state in the expansion, Eq.~(\ref{eq:rovibprod}), then we may say that it is the `vibrational parent', from which we can `derive' the rovibrational state by `rotational excitation'.
The term `vibrational parent' was first defined by Wang and Carrington in Ref.~\citenum{WaCa11}

A related analysis tool, the rigid-rotor decomposition (RRD)\cite{MaFaSzCzAlCs10} scheme was defined, which amounts to computing the overlap between the $m$th rovibrational wave function $\Psi_m^{(J>0)}$, with the product of a vibrational wave function and a rigid-rotor function, 
\begin{align}
  S_{nK\tau,m} 
  = 
  \langle%
    \psi_n^\text{v} \Theta^{(JM)}_{K\tau}  | 
    \Psi_m^{(J>0)}
  \rangle \; .
  \label{eq:rrd}
\end{align}
Figure~\ref{fig:floppyrot}
exemplifies that the low-energy states shown in the figure for the (semi-rigid) water molecule have clear, dominant RRD coefficients, whereas the floppy methane-argon dimer (from a 3D computation\cite{FeMa19}) strong mixing is visible already in the low-energy range. We note that the RRD coefficients depend on the body-fixed frame, which is a computational `parameter' (Sec.~\ref{sec:optframe}).

\begin{figure}
  \centering
  \includegraphics[width=9cm]{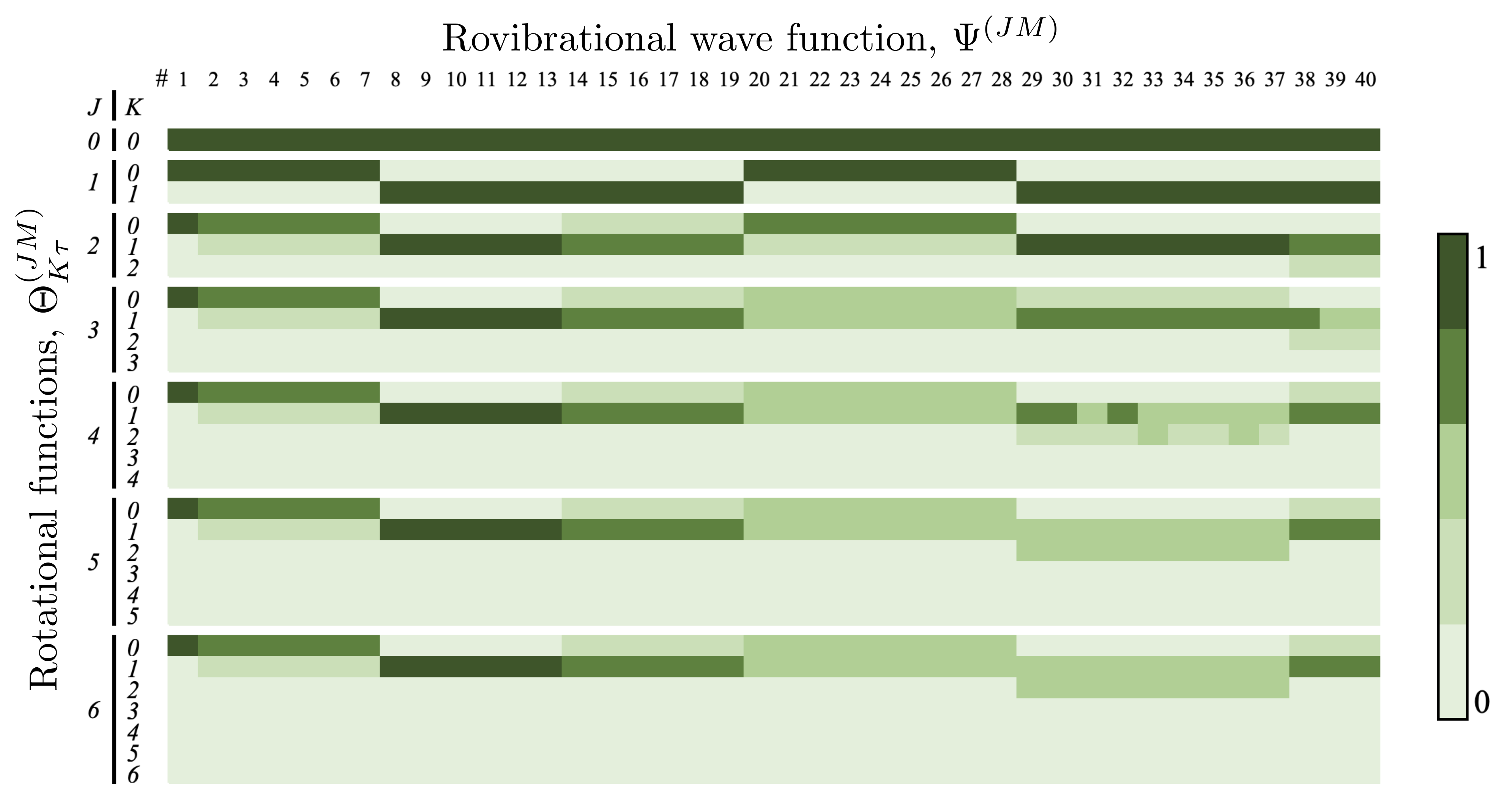}
  \caption{Assignment of the $K$ label, Eqs.~\eqref{eq:rotpar}--\eqref{eq:kappaK} of the rotational functions to the intermolecular-rotational states of $(\text{CH}_4)_2$ computed with GENIUSH using the 6D intermolecular coordinates (similar to methane-water in Fig.~\ref{fig:Gsingular}) and the PES of Ref.~\citenum{dimers}. }
  \label{fig:Klabel}
\end{figure}
\subsection{Rotational parent analysis}
Similarly to the vibrational parent assignment, we can ask whether there is any dominant $\Theta^{(JM)}_{K\tau}$ Wang function in the Eq.~(\ref{eq:rovibprod}) expansion of a rovibrational state. This question may be relevant even if there is no dominant vibrational parent, which is often the case for floppy systems.

As a measure of the importance of $\Theta^{(JM)}_{K\tau}$ in the rovibrational wave function,
we use the fact that the rovibrational $\psi^{(JM)}_{i}$ wave function is normalized and the basis functions in its basis set expansion, Eq.~\eqref{eq:rovibprod}, are orthonormalized,
\begin{align}
  1
  &=
  \langle 
    \psi^{(JM)}_{i}|\psi^{(JM)}_{i} 
  \rangle
  \nonumber \\
  &=
  \int \dd \bos{q}
  \int \dd \Omega  
  \left[%
     \sum_{n'K'\tau'}
     c^{(JM,i)\ast}_{n'K'\tau'} 
     \psi^{\text{v}\ast}_{n'}(\bos{q}) 
     \Theta^{(JM)\ast}_{K'\tau'}(\Omega)
  \right] \nonumber \\
  &\quad\quad\quad\quad\quad\quad
  \left[%
     \sum_{nK\tau}
     c^{(JM,i)}_{nK\tau} 
     \psi^{\text{v}}_{n}(\bos{q}) 
     \Theta^{(JM)}_{K\tau}(\Omega)
  \right]  
  \nonumber \\
  &=
  \sum_{K\tau}
  \sum_{n}
     |c^{(JM,i)}_{nK\tau}|^2 
  \nonumber \\
  &=:
  \sum_{K\tau}
    \tilde\kappa^{(JM,i)}_{K\tau}     
    \label{eq:rotpar}
\end{align}
and
\begin{align}
  \tilde\kappa^{(JM,i)}_{K\tau}     
  =
  \sum_{n}
     |c^{(JM,i)}_{nK\tau}|^2 
  \label{eq:kappaKtau}
\end{align}
is a measure for the `weight' of the $\Theta_{K\tau}^{(JM)}$ rotational basis state with the $K\tau$ label in the rovibrational wave function.
We can also have a measure only for the $K$ label by summing for $\tau=0,1$
\begin{align}
  \kappa^{(JM,i)}_K = \sum_{\tau=0,1} \tilde\kappa^{(JM,i)}_{K\tau} \; .
  \label{eq:kappaK}
\end{align}
Due to the normalization of the rovibrational wave function $\sum_{K=0}^J\kappa^{(JM)}_K=1$.

Figure~\ref{fig:Klabel} shows for the example of the lowest-energy states of the $(\text{CH}_4)_2$ dimer, for which the equilibrium structure on the used PES\cite{dimers} is a symmetric top. The $K\tau$ labels can be unambiguously assigned up to $J=1$(2), but beyond $J>2$ this is not always possible due to the  strong rovibrational coupling.

\subsection{Coupled-rotor decomposition}
\begin{figure}
    \centering
    \includegraphics[width=8.5cm]{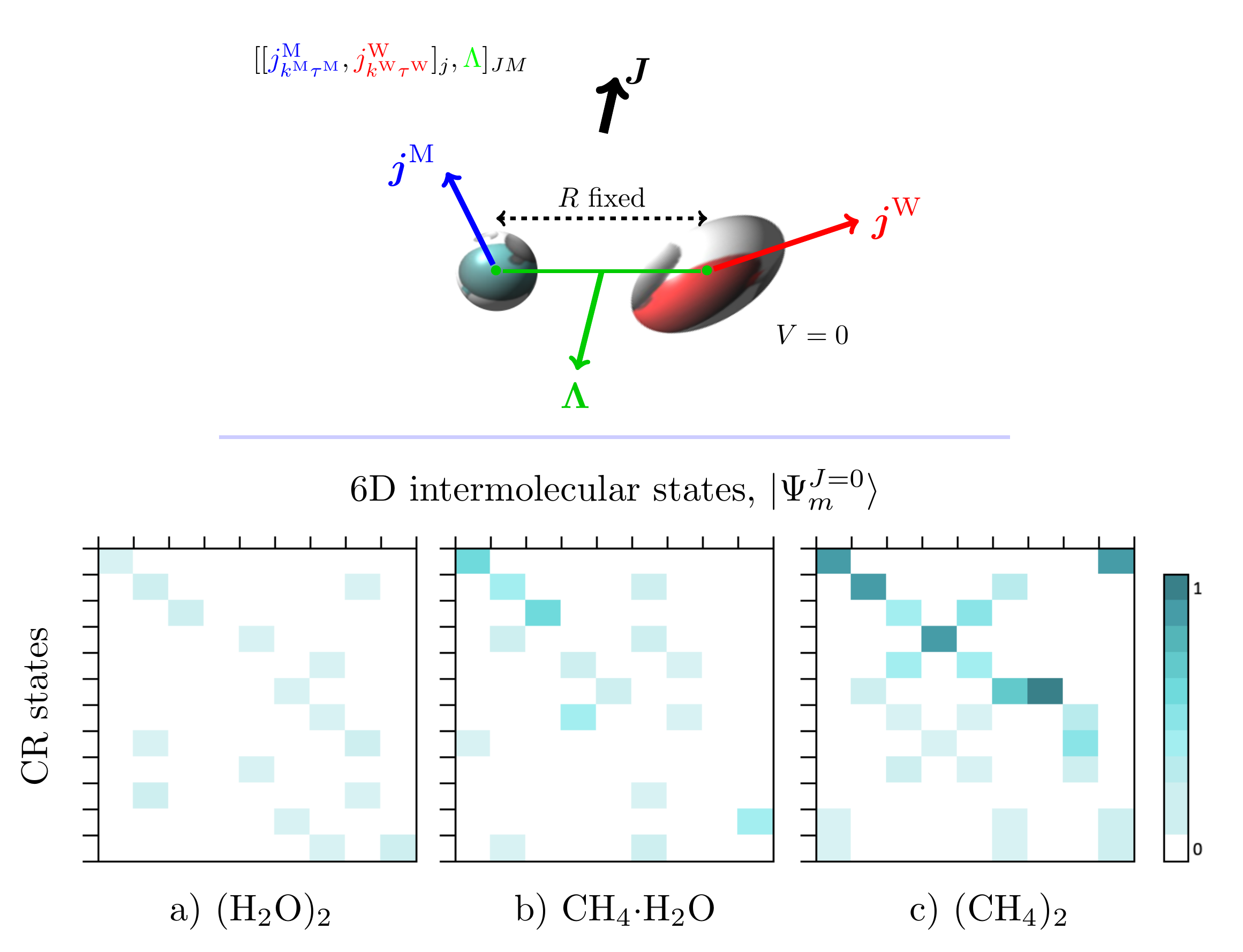}
    \caption{%
    Coupled-rotor decomposition coefficients computed for the $J=0$ states of the \metmet\ dimer.
    }
    \label{fig:crd}
\end{figure}

For the analysis of rovibrational states of floppy molecular dimers, it is useful to assign subsystems' angular momenta to the dimer state according to the coupling of the angular momenta of monomer $A$,  ($\bos{j}^A$), monomer $B$, ($\bos{j}^B$), and the effective diatom, ($\bos{\Lambda}$) to a dimer state with $JM$ angular momentum quantum numbers corresponding to the overall rotation, $\bos{J}^2$, and its projection to a space-fixed axis, $J_Z$, (Fig.~\ref{fig:crd}a)
\begin{align}
  [[j^A_{k^A\tau^A}, j^B_{k^B\tau^B}]_j , \Lambda]_{JM} \; .
\end{align}
Since during the computations performed with GENIUSH\cite{SaCsAlWaMa16,SaCsMa17,FeMa19,DaAvMa21a} (with DVR), we did not use monomer rotational basis functions, we computed the overlap, named coupled rotor decomposition (CRD),\cite{SaCsMa17} of the intermolecular rovibrational state, $\Psi_{m,K\tau}^{(JM)}$ and the coupled rotor (CR) function with a fixed monomer distance and without the PES (Fig.~\ref{fig:crd}a), 
\begin{align}
  \text{CRD}^{(JM)}_{nm} = \sum_{r=1}^{N_R} \Bigg| 
  \sum_{K\tau}
  \sum_{o =1}^{N_\Omega}\tilde{\Psi}^{(JM)}_{m,K\tau}
    (\rho_r,\omega_o) \, \cdot \, \tilde{\varphi}^{(J)}_{n,K\tau}(\omega_o) \Bigg|^2  \; ,
  \label{eq:crd}
\end{align}
where $\bar{\Psi}^{(J)}_m$ is the $m$th rovibrational state
and $\bar{\varphi}^{(J)}_n$ is the $n$th CR function in the DVR representation used in the cited references.

The CRD overlaps were used to assign the rovibrational states to irreducible representations of the molecular symmetry (MS) group, by using the formal symmetry properties of the CR functions.\cite{SaCsMa17,FeMa19}
%

\section{Simulating rovibrational infrared and Raman spectra \label{sec:tensint}}
We can consider interaction of (non-relativistic) molecular matter with external electromagnetic fields, not accounted for in the description, by using the rovibrational eigenstates as a basis. To account for the effect of `external fields', weak-field spectroscopies, strong-field interactions, or for corrections `within the system' beyond the Coulomb interactions, \emph{i.e.,} hyperfine,\cite{CoOwKuYa18,YaYaZaYuKu22} we have implemented the evaluation of 
the transition moments connecting the rovibrational states $\Psi^{(J'M')}_{i'}$ and $\Psi^{(JM)}_{i}$, for a $T^{(\rm LF)}_A$ general tensorial property of rank $\Omega$, where $A$ collects the LF Cartesian indexes, according to\cite{OwYa18} %
 \begin{equation}
    \bra{\Psi^{(J'M')}_{i'}} T^{(\rm LF)}_A 
    \ket{\Psi^{(JM)}_{i}} 
    = 
    \sum_{\omega=0}^\Omega
    \mathcal{M}_{A\omega}^{(J'M',JM)}
    \mathcal{K}_{\omega}^{(J'i',Ji)}
    \label{eq:tensmatrix}
 \end{equation}
with
 \begin{equation}
 \begin{split}
&    \mathcal{M}_{A\omega}^{(J'M',JM)} = (-1)^{M'}
     \sqrt{(2J'+1)(2J+1)} \\ 
&    \times\quad\quad
     \sum_{\sigma=-\omega}^\omega 
     [U^{(\Omega)}]^{-1}_{A,\omega\sigma}
      \begin{pmatrix}
          J & \omega & J'\\
          M & \sigma & -M'
      \end{pmatrix}
 \end{split}
 \label{eq:tensmatrixM}
 \end{equation}
and
 \begin{equation}
 \begin{split}
    & \mathcal{K}_{\omega}^{(J'i',Ji)} = 
    \sum_{\substack{n,K,\tau\\n',K',\tau'}}
    [c_{n'K'\tau'}^{(J',i')}]^\ast c_{nK\tau}^{(J,i)}
     \sum_{\pm K', \pm K}[d_{K'}^{(\tau')}]^\ast
     d_K^{(\tau)}\\
     & \times (-1)^{K'}
     \sum_{\sigma=-\omega}^\omega \sum_{a} 
      \begin{pmatrix}
          J & \omega & J'\\
          K & \sigma & -K'
      \end{pmatrix}
      U^{(\Omega)}_{\omega\sigma, a}
     \bra{\psi^\text{v}_{n'}}T_a^{(\rm BF)}\ket{\psi^\text{v}_{n}} \; ,
 \end{split}
  \label{eq:tensmatrixK}
\end{equation}
where $c_{nK\tau}^{(J,i)}$ are the linear combination coefficients of Eq.~(\ref{eq:rovibprod}) and $d_K^{(\tau)}$ correspond to the Wang combination coefficients, Eq.~(\ref{eq:wang}).
The vibrational integrals $\bra{\psi^\text{v}_{n'}}T_a^{(\rm BF)}\ket{\psi^\text{v}_{n}}$ are computed in the body-fixed (BF) frame using (the DVR or FBR) vibrational wave functions $\psi^\text{v}_{n}$ and the $T_a^{(\rm BF)}$ `property' hypersurface available from \emph{ab initio} electronic structure theory (in a fitted or interpolated functional form).
The $\mathcal{K}_{\omega}$ coefficients correspond to the rank-$\Omega$ tensor in the body-fixed frame and the corresponding rovibrational wave functions, so they can be evaluated separately and stored for every $\omega$ value. 
 
This implementation has been reported and was tested for the example of electric dipole transitions (for $\Omega=1$) of the methane-water complex\cite{DaAvMa21a} in comparison with the transition moments reported by Wang and Carrington.\cite{WaCa21}
 
In what follows, the expressions are summarized for rovibrational infrared ($\Omega=1$) and Raman ($\Omega=2$) experiments.
%
\subsection{Infrared transition moments \label{sec:ir_tran}}
Intensities of infrared spectroscopy experiments can be simulated by using the electric dipole moment, which is an $\Omega=1$ rank property, including the
$U^{(1)}_{\omega\sigma, \alpha}$ matrix elements.\cite{OwYa18,DaAvMa21a}
The working formula for dipole moment transitions is obtained after some manipulation from the general expressions, Eqs.~(\ref{eq:tensmatrix})--(\ref{eq:tensmatrixK}),\cite{BuJe98,OwYa18,ErTsRa20} 
\begin{equation}
 \begin{split}
    \mathcal{R}_1 
    &= 
    g_\text{ns}
    \sum_{M,M'}\sum_{A=X,Y,Z}| 
      \bra{\Psi^{(J'M')}_{i'}} T^{(\rm LF)}_A \ket{\Psi^{(JM)}_{i}}|^2    
    \nonumber \\
    &=
    g_{\text{ns}}
    (2J'+1)(2J+1) \\
    & \times \Bigg|
    \sum_{\substack{n,K,\tau\\n',K',\tau'}}
    [c_{n'K'\tau'}^{(J',i')}]^* c_{nK\tau}^{(J,i)}
     \sum_{\pm K', \pm K}[d_{K'}^{(\tau')}]^\ast
     d_K^{(\tau)}\\
     & \times (-1)^{K'}
     \sum_{\sigma=-1}^1 \sum_{\alpha} 
      \begin{pmatrix}
          J & 1 & J'\\
          K & \sigma & -K'
      \end{pmatrix}
      U^{(1)}_{\omega\sigma, \alpha}
     \bra{\psi^\text{v}_{n'}}\mu_\alpha^{(\rm BF)}\ket{\psi^\text{v}_{n}}
    \Bigg|^2     \;.
 \end{split}
 \label{eq:transmoment_1}
\end{equation}

\subsection{Raman transition moments\label{sec:raman}}
Another important technique in rovibrational spectroscopy is Raman scattering. For computing Raman intensities, the relevant property is the rank $\Omega=2$ polarizability matrix, $\bos{\alpha}$.
It is convenient to distinguish to types of transition moments, the isotropic (independent of the molecular orientation) with $\omega=0$ and the anisotropic with $\omega=2$ components.
By using the general expressions in Eqs.~(\ref{eq:tensmatrix})--(\ref{eq:tensmatrixK}), the isotropic ($\omega=0$) component can be written as\cite{BuJe98,ErTsRa22}
\begin{equation}
 \begin{split}
    & \mathcal{R}_0 
    = 
    \delta_{JJ'} g_{\text{ns}}
    (2J'+1)(2J+1) \\
    & \times \Bigg| 
    \sum_{\substack{n,K,\tau\\n',K',\tau'}}
    [c_{n'K'\tau'}^{(J',i')}]^* c_{nK\tau}^{(J,i)}
     \sum_{\pm K', \pm K}[d_{K'}^{(\tau')}]^\ast
     d_K^{(\tau)}\\
     & \quad\quad \times (-1)^{K'}
     \sum_{ab} 
      \begin{pmatrix}
          J & 0 & J'\\
          K & 0 & -K'
      \end{pmatrix}
      U^{(2)}_{00,ab}
     \bra{\psi^\text{v}_{n'}}\alpha_{ab}^{(\rm BF)}\ket{\psi^\text{v}_{n}}
    \Bigg |^2  ,
 \end{split}
 \label{eq:Riso}
 \end{equation}
where the 3$J$-symbol vanishes unless $J=J'$, leading to the selection rule $\Delta J=0$ for the isotropic transition moments, incorporated in the equations with the Kroneker delta, $\delta_{JJ'}$.
For the anisotropic contribution ($\omega=2$),
the working equation is
 \begin{equation}
 \begin{split}
    & \mathcal{R}_2 
    = 
    g_{\text{ns}}
    (2J'+1)(2J+1) \\
    & \times \Bigg|
    \sum_{\substack{n,K,\tau\\n',K',\tau'}}
    [c_{n'K'\tau'}^{(J',i')}]^* c_{nK\tau}^{(J,i)}
     \sum_{\pm K', \pm K}[d_{K'}^{(\tau')}]^\ast
     d_K^{(\tau)}\\
     & \quad\quad \times (-1)^{K'}
     \sum_{\sigma=-2}^{2}
     \sum_{ab} 
      \begin{pmatrix}
          J & 2 & J'\\
          K & \sigma & -K'
      \end{pmatrix}
     U^{(2)}_{2\sigma, ab}
     \bra{\psi^\text{v}_{n'}}\alpha_{ab}^{(\rm BF)}\ket{\psi^\text{v}_{n}}
    \Bigg |^2  .
 \end{split}
 \label{eq:Raniso}
 \end{equation}
The non-zero matrix elements of $U^{(2)}_{\omega\sigma, \alpha\beta}$ have been collected in Ref.~\citenum{OwYa18}.
The non-vanishing value of 3$J$-symbols gives the $\Delta J = 0, 1, 2$ selection rules for the anisotropic component.
The current procedure makes it possible to compute polarizability transition moments connecting rovibrational states of floppy systems that are relevant for high-resolution Raman scattering experiments.

\section{Summary and conclusions}
We have reviewed the theoretical and methodological (ro)vibrational framework developed by ourselves or in close collaboration with colleagues over the past decade. The reviewed methodology is implemented in the GENIUSH program, which was used to carry out the highlighted computations.

The focus of this Feature Article was on (ro)vibrational spectroscopy of isolated floppy molecules and complexes, presenting the numerically exact solution of the (ro)vibrational Schrödinger equation on a potential energy surface, analysis of the computed wave function, assignment of the computed stationary states, and evaluation of electric dipole and polarizability transition moments for simulating (weak-field) high-resolution infrared and Raman spectra of (semi-rigid and) floppy molecular systems and complexes. 
The methodology can be straightforwardly extended to compute molecules in cages and at surfaces.

An important methodological challenge, which is currently at the focus of our main efforts, is the increase of the vibrational dimensionality of feasible computations.
Exact quantum dynamics computations of molecular systems with up to 20--30 fully coupled vibrational degrees of freedom, allowing for a few large-amplitude motions,  which carry lots of interesting chemistry,  will be an important next milestone to reach in the next couple of years.  
We aim to pursue these developments by developing general, $N$-atomic, $D$-dimensional approaches, \emph{i.e.,} pushing the frontier of a black-box-type rovibrational research program. Thereby, as available computational power increases, in parallel with the methodological developments, rovibrational computations of spectroscopic quality become possible for higher-dimensional ($>12$D) floppy molecular systems.

\section*{Conflicts of interest}
There are no conflicts to declare.

\section*{Acknowledgements}
We thank the financial support of the Swiss National Science Foundation 
(PROMYS Grant, No.~IZ11Z0\_166525). 
EM thanks a DFG Mercator Fellowship granted in the framework of the BENCh project (No.~389479699/GRK2455) and hospitality of the Institute of Physical Chemistry at the University of Göttingen.


\begin{thebibliography}{118}%
\makeatletter
\providecommand \@ifxundefined [1]{%
 \@ifx{#1\undefined}
}%
\providecommand \@ifnum [1]{%
 \ifnum #1\expandafter \@firstoftwo
 \else \expandafter \@secondoftwo
 \fi
}%
\providecommand \@ifx [1]{%
 \ifx #1\expandafter \@firstoftwo
 \else \expandafter \@secondoftwo
 \fi
}%
\providecommand \natexlab [1]{#1}%
\providecommand \enquote  [1]{``#1''}%
\providecommand \bibnamefont  [1]{#1}%
\providecommand \bibfnamefont [1]{#1}%
\providecommand \citenamefont [1]{#1}%
\providecommand \href@noop [0]{\@secondoftwo}%
\providecommand \href [0]{\begingroup \@sanitize@url \@href}%
\providecommand \@href[1]{\@@startlink{#1}\@@href}%
\providecommand \@@href[1]{\endgroup#1\@@endlink}%
\providecommand \@sanitize@url [0]{\catcode `\\12\catcode `\$12\catcode
  `\&12\catcode `\#12\catcode `\^12\catcode `\_12\catcode `\%12\relax}%
\providecommand \@@startlink[1]{}%
\providecommand \@@endlink[0]{}%
\providecommand \url  [0]{\begingroup\@sanitize@url \@url }%
\providecommand \@url [1]{\endgroup\@href {#1}{\urlprefix }}%
\providecommand \urlprefix  [0]{URL }%
\providecommand \Eprint [0]{\href }%
\providecommand \doibase [0]{https://doi.org/}%
\providecommand \selectlanguage [0]{\@gobble}%
\providecommand \bibinfo  [0]{\@secondoftwo}%
\providecommand \bibfield  [0]{\@secondoftwo}%
\providecommand \translation [1]{[#1]}%
\providecommand \BibitemOpen [0]{}%
\providecommand \bibitemStop [0]{}%
\providecommand \bibitemNoStop [0]{.\EOS\space}%
\providecommand \EOS [0]{\spacefactor3000\relax}%
\providecommand \BibitemShut  [1]{\csname bibitem#1\endcsname}%
\let\auto@bib@innerbib\@empty
\bibitem [{\citenamefont {Quack}(2001)}]{Qu01}%
  \BibitemOpen
  \bibfield  {author} {\bibinfo {author} {\bibfnamefont {M.}~\bibnamefont
  {Quack}},\ }\bibfield  {title} {\bibinfo {title} {Molecules in motion},\
  }\href {https://doi.org/10.2533/chimia.2001.753} {\bibfield  {journal}
  {\bibinfo  {journal} {Chimia}\ }\textbf {\bibinfo {volume} {55}},\ \bibinfo
  {pages} {753} (\bibinfo {year} {2001})}\BibitemShut {NoStop}%
\bibitem [{\citenamefont {Quack}\ and\ \citenamefont {Merkt}(2011)}]{HRSBook}%
  \BibitemOpen
  \bibinfo {editor} {\bibfnamefont {M.}~\bibnamefont {Quack}}\ and\ \bibinfo
  {editor} {\bibfnamefont {F.}~\bibnamefont {Merkt}},\ eds.,\ \href@noop {}
  {\emph {\bibinfo {title} {Handbook of {H}igh-resolution {S}pectroscopy}}}\
  (\bibinfo  {publisher} {John Wiley \& Sons},\ \bibinfo {address}
  {Chichester},\ \bibinfo {year} {2011})\BibitemShut {NoStop}%
\bibitem [{\citenamefont {Born}\ and\ \citenamefont {Jordan}(1925)}]{BoJo25}%
  \BibitemOpen
  \bibfield  {author} {\bibinfo {author} {\bibfnamefont {M.}~\bibnamefont
  {Born}}\ and\ \bibinfo {author} {\bibfnamefont {P.}~\bibnamefont {Jordan}},\
  }\bibfield  {title} {\bibinfo {title} {Zur {Q}uantenmechanik},\ }\href
  {https://doi.org/10.1007/BF01328531} {\bibfield  {journal} {\bibinfo
  {journal} {Z. Phys.}\ }\textbf {\bibinfo {volume} {34}},\ \bibinfo {pages}
  {858} (\bibinfo {year} {1925})}\BibitemShut {NoStop}%
\bibitem [{\citenamefont {{Wilson, Jr.}}\ \emph {et~al.}(1955)\citenamefont
  {{Wilson, Jr.}}, \citenamefont {Decius},\ and\ \citenamefont
  {Cross}}]{WiDeCr55}%
  \BibitemOpen
  \bibfield  {author} {\bibinfo {author} {\bibfnamefont {E.~B.}\ \bibnamefont
  {{Wilson, Jr.}}}, \bibinfo {author} {\bibfnamefont {J.~C.}\ \bibnamefont
  {Decius}},\ and\ \bibinfo {author} {\bibfnamefont {P.~C.}\ \bibnamefont
  {Cross}},\ }\href@noop {} {\emph {\bibinfo {title} {Molecular Vibrations: The
  Theory of Infrared and Raman Vibrational Spectra}}}\ (\bibinfo  {publisher}
  {McGraw-Hill Book Company, Inc.},\ \bibinfo {year} {1955})\BibitemShut
  {NoStop}%
\bibitem [{\citenamefont {Papousek}\ and\ \citenamefont
  {Aliev}(1982)}]{PaAlBook82}%
  \BibitemOpen
  \bibfield  {author} {\bibinfo {author} {\bibfnamefont {D.}~\bibnamefont
  {Papousek}}\ and\ \bibinfo {author} {\bibfnamefont {M.~R.}\ \bibnamefont
  {Aliev}},\ }\href@noop {} {\emph {\bibinfo {title} {Molecular
  {V}ibrational-rotational {S}pectra}}}\ (\bibinfo {year} {1982})\BibitemShut
  {NoStop}%
\bibitem [{\citenamefont {Sutcliffe}(2002)}]{Su02}%
  \BibitemOpen
  \bibfield  {author} {\bibinfo {author} {\bibfnamefont {B.}~\bibnamefont
  {Sutcliffe}},\ }\bibinfo {title} {Coordinate {S}ystems and
  {T}ransformations},\ in\ \href@noop {} {\emph {\bibinfo {booktitle}
  {{H}andbook of {M}olecular {P}hysics and {Q}uantum {C}hemistry, ed. {S}.
  {W}ilson. {V}ol. 1: {F}undamentals}}}\ (\bibinfo  {publisher} {John Wiley \&
  Sons, Ltd.},\ \bibinfo {year} {2002})\ Chap.~\bibinfo {chapter} {31}, pp.\
  \bibinfo {pages} {485--500}\BibitemShut {NoStop}%
\bibitem [{\citenamefont {Littlejohn}\ and\ \citenamefont
  {Reinsch}(1997)}]{LiRe97}%
  \BibitemOpen
  \bibfield  {author} {\bibinfo {author} {\bibfnamefont {R.~G.}\ \bibnamefont
  {Littlejohn}}\ and\ \bibinfo {author} {\bibfnamefont {M.}~\bibnamefont
  {Reinsch}},\ }\bibfield  {title} {\bibinfo {title} {Gauge fields in the
  separation of rotations and internal motions in the $n$-body problem},\
  }\href {https://doi.org/10.1103/RevModPhys.69.213} {\bibfield  {journal}
  {\bibinfo  {journal} {Rev. Mod. Phys.}\ }\textbf {\bibinfo {volume} {69}},\
  \bibinfo {pages} {213} (\bibinfo {year} {1997})}\BibitemShut {NoStop}%
\bibitem [{\citenamefont {Eckart}(1935)}]{Ec35}%
  \BibitemOpen
  \bibfield  {author} {\bibinfo {author} {\bibfnamefont {C.}~\bibnamefont
  {Eckart}},\ }\bibfield  {title} {\bibinfo {title} {Some studies concerning
  rotating axes and polyatomic molecules},\ }\href
  {https://doi.org/10.1103/PhysRev.47.552} {\bibfield  {journal} {\bibinfo
  {journal} {Phys. Rev.}\ }\textbf {\bibinfo {volume} {47}},\ \bibinfo {pages}
  {552} (\bibinfo {year} {1935})}\BibitemShut {NoStop}%
\bibitem [{\citenamefont {Mart{\'\i}n Santa~Dar{\'\i}a}\ \emph
  {et~al.}(2022)\citenamefont {Mart{\'\i}n Santa~Dar{\'\i}a}, \citenamefont
  {Avila},\ and\ \citenamefont {M{\'a}tyus}}]{DaAvMa22}%
  \BibitemOpen
  \bibfield  {author} {\bibinfo {author} {\bibfnamefont {A.}~\bibnamefont
  {Mart{\'\i}n Santa~Dar{\'\i}a}}, \bibinfo {author} {\bibfnamefont
  {G.}~\bibnamefont {Avila}},\ and\ \bibinfo {author} {\bibfnamefont
  {E.}~\bibnamefont {M{\'a}tyus}},\ }\bibfield  {title} {\bibinfo {title}
  {Variational vibrational states of {HCOOH}},\ }\href
  {https://doi.org/10.1016/j.jms.2022.111617} {\bibfield  {journal} {\bibinfo
  {journal} {J. Mol. Spectrosc.}\ }\textbf {\bibinfo {volume} {385}},\ \bibinfo
  {pages} {111617} (\bibinfo {year} {2022})}\BibitemShut {NoStop}%
\bibitem [{\citenamefont {Martín Santa~Daría}\ \emph
  {et~al.}(2021{\natexlab{a}})\citenamefont {Martín Santa~Daría},
  \citenamefont {Avila},\ and\ \citenamefont {Mátyus}}]{DaAvMa21b}%
  \BibitemOpen
  \bibfield  {author} {\bibinfo {author} {\bibfnamefont {A.}~\bibnamefont
  {Martín Santa~Daría}}, \bibinfo {author} {\bibfnamefont {G.}~\bibnamefont
  {Avila}},\ and\ \bibinfo {author} {\bibfnamefont {E.}~\bibnamefont
  {Mátyus}},\ }\bibfield  {title} {\bibinfo {title} {Fingerprint region of the
  formic acid dimer: variational vibrational computations in curvilinear
  coordinates},\ }\href {https://doi.org/10.1039/D0CP06289H} {\bibfield
  {journal} {\bibinfo  {journal} {Phys. Chem. Chem. Phys.}\ }\textbf {\bibinfo
  {volume} {23}},\ \bibinfo {pages} {6526} (\bibinfo {year}
  {2021}{\natexlab{a}})}\BibitemShut {NoStop}%
\bibitem [{\citenamefont {Avila}\ and\ \citenamefont
  {M\'atyus}(2019)}]{AvMa19}%
  \BibitemOpen
  \bibfield  {author} {\bibinfo {author} {\bibfnamefont {G.}~\bibnamefont
  {Avila}}\ and\ \bibinfo {author} {\bibfnamefont {E.}~\bibnamefont
  {M\'atyus}},\ }\bibfield  {title} {\bibinfo {title} {Toward breaking the
  curse of dimensionality in (ro)vibrational computations of molecular systems
  with multiple large-amplitude motions},\ }\href
  {https://doi.org/10.1063/1.5090846} {\bibfield  {journal} {\bibinfo
  {journal} {J. Chem. Phys.}\ }\textbf {\bibinfo {volume} {150}},\ \bibinfo
  {pages} {174107} (\bibinfo {year} {2019})}\BibitemShut {NoStop}%
\bibitem [{\citenamefont {Avila}\ and\ \citenamefont
  {Mátyus}(2019)}]{AvMa19b}%
  \BibitemOpen
  \bibfield  {author} {\bibinfo {author} {\bibfnamefont {G.}~\bibnamefont
  {Avila}}\ and\ \bibinfo {author} {\bibfnamefont {E.}~\bibnamefont
  {Mátyus}},\ }\bibfield  {title} {\bibinfo {title} {Full-dimensional ({12D})
  variational vibrational states of {CH}$_4\cdot${F}$^-$: {I}nterplay of
  anharmonicity and tunneling},\ }\href {https://doi.org/10.1063/1.5124532}
  {\bibfield  {journal} {\bibinfo  {journal} {J. Chem. Phys.}\ }\textbf
  {\bibinfo {volume} {151}},\ \bibinfo {pages} {154301} (\bibinfo {year}
  {2019})}\BibitemShut {NoStop}%
\bibitem [{\citenamefont {Avila}\ \emph {et~al.}(2020)\citenamefont {Avila},
  \citenamefont {Papp}, \citenamefont {Czakó},\ and\ \citenamefont
  {Mátyus}}]{AvPaCzMa20}%
  \BibitemOpen
  \bibfield  {author} {\bibinfo {author} {\bibfnamefont {G.}~\bibnamefont
  {Avila}}, \bibinfo {author} {\bibfnamefont {D.}~\bibnamefont {Papp}},
  \bibinfo {author} {\bibfnamefont {G.}~\bibnamefont {Czakó}},\ and\ \bibinfo
  {author} {\bibfnamefont {E.}~\bibnamefont {Mátyus}},\ }\bibfield  {title}
  {\bibinfo {title} {Exact quantum dynamics background of dispersion
  interactions: case study for {CH}$_4\cdot${A}r in full (12) dimensions},\
  }\href {https://doi.org/10.1039/C9CP04426D} {\bibfield  {journal} {\bibinfo
  {journal} {Phys. Chem. Chem. Phys.}\ }\textbf {\bibinfo {volume} {22}},\
  \bibinfo {pages} {2792} (\bibinfo {year} {2020})}\BibitemShut {NoStop}%
\bibitem [{\citenamefont {Papp}\ \emph {et~al.}(2022)\citenamefont {Papp},
  \citenamefont {Tajti}, \citenamefont {Avila}, \citenamefont {Mátyus},\ and\
  \citenamefont {Czakó}}]{PaTaAvMa22}%
  \BibitemOpen
  \bibfield  {author} {\bibinfo {author} {\bibfnamefont {D.}~\bibnamefont
  {Papp}}, \bibinfo {author} {\bibfnamefont {V.}~\bibnamefont {Tajti}},
  \bibinfo {author} {\bibfnamefont {G.}~\bibnamefont {Avila}}, \bibinfo
  {author} {\bibfnamefont {E.}~\bibnamefont {Mátyus}},\ and\ \bibinfo {author}
  {\bibfnamefont {G.}~\bibnamefont {Czakó}},\ }\bibfield  {title} {\bibinfo
  {title} {{CH}$_4\cdot${F}$^-$ {PES} revisited: {F}ull-dimensional ab initio
  potential energy surface and variational vibrational states},\ }\href
  {https://doi.org/10.1080/00268976.2022.2113565} {\bibfield  {journal}
  {\bibinfo  {journal} {Mol. Phys.}\ ,\ \bibinfo {pages} {e2113565}} (\bibinfo
  {year} {2022})}\BibitemShut {NoStop}%
\bibitem [{\citenamefont {Thompson}\ \emph {et~al.}(1998)\citenamefont
  {Thompson}, \citenamefont {Jordan},\ and\ \citenamefont
  {Collins}}]{ThJoCo98}%
  \BibitemOpen
  \bibfield  {author} {\bibinfo {author} {\bibfnamefont {K.~C.}\ \bibnamefont
  {Thompson}}, \bibinfo {author} {\bibfnamefont {M.~J.~T.}\ \bibnamefont
  {Jordan}},\ and\ \bibinfo {author} {\bibfnamefont {M.~A.}\ \bibnamefont
  {Collins}},\ }\bibfield  {title} {\bibinfo {title} {Polyatomic molecular
  potential energy surfaces by interpolation in local internal coordinates},\
  }\href {https://doi.org/10.1063/1.476259} {\bibfield  {journal} {\bibinfo
  {journal} {J. Chem. Phys.}\ }\textbf {\bibinfo {volume} {108}},\ \bibinfo
  {pages} {8302} (\bibinfo {year} {1998})}\BibitemShut {NoStop}%
\bibitem [{\citenamefont {Podolsky}(1928)}]{Po28}%
  \BibitemOpen
  \bibfield  {author} {\bibinfo {author} {\bibfnamefont {B.}~\bibnamefont
  {Podolsky}},\ }\bibfield  {title} {\bibinfo {title} {Quantum-mechanically
  correct form of {H}amiltonian function for conservative systems},\ }\href
  {https://doi.org/10.1103/PhysRev.32.812} {\bibfield  {journal} {\bibinfo
  {journal} {Phys. Rev.}\ }\textbf {\bibinfo {volume} {32}},\ \bibinfo {pages}
  {812} (\bibinfo {year} {1928})}\BibitemShut {NoStop}%
\bibitem [{\citenamefont {Lauvergnat}\ \emph {et~al.}(2019)\citenamefont
  {Lauvergnat}, \citenamefont {Felker}, \citenamefont {Scribano}, \citenamefont
  {Benoit},\ and\ \citenamefont {Bačić}}]{LaFeScBeBa19}%
  \BibitemOpen
  \bibfield  {author} {\bibinfo {author} {\bibfnamefont {D.}~\bibnamefont
  {Lauvergnat}}, \bibinfo {author} {\bibfnamefont {P.}~\bibnamefont {Felker}},
  \bibinfo {author} {\bibfnamefont {Y.}~\bibnamefont {Scribano}}, \bibinfo
  {author} {\bibfnamefont {D.~M.}\ \bibnamefont {Benoit}},\ and\ \bibinfo
  {author} {\bibfnamefont {Z.}~\bibnamefont {Bačić}},\ }\bibfield  {title}
  {\bibinfo {title} {{H}$_2$, {HD}, and {D}$_2$ in the small cage of structure
  {II} clathrate hydrate: Vibrational frequency shifts from fully coupled
  quantum six-dimensional calculations of the vibration-translation-rotation
  eigenstates},\ }\href {https://doi.org/10.1063/1.5090573} {\bibfield
  {journal} {\bibinfo  {journal} {J. Chem. Phys.}\ }\textbf {\bibinfo {volume}
  {150}},\ \bibinfo {pages} {154303} (\bibinfo {year} {2019})}\BibitemShut
  {NoStop}%
\bibitem [{\citenamefont {Felker}\ \emph {et~al.}(2019)\citenamefont {Felker},
  \citenamefont {Lauvergnat}, \citenamefont {Scribano}, \citenamefont
  {Benoit},\ and\ \citenamefont {Bačić}}]{FeLaScBeBa19}%
  \BibitemOpen
  \bibfield  {author} {\bibinfo {author} {\bibfnamefont {P.~M.}\ \bibnamefont
  {Felker}}, \bibinfo {author} {\bibfnamefont {D.}~\bibnamefont {Lauvergnat}},
  \bibinfo {author} {\bibfnamefont {Y.}~\bibnamefont {Scribano}}, \bibinfo
  {author} {\bibfnamefont {D.~M.}\ \bibnamefont {Benoit}},\ and\ \bibinfo
  {author} {\bibfnamefont {Z.}~\bibnamefont {Bačić}},\ }\bibfield  {title}
  {\bibinfo {title} {Intramolecular stretching vibrational states and frequency
  shifts of ({H}$_2$)$_2$ confined inside the large cage of clathrate hydrate
  from an eight-dimensional quantum treatment using small basis sets},\ }\href
  {https://doi.org/10.1063/1.5124051} {\bibfield  {journal} {\bibinfo
  {journal} {J. Chem. Phys.}\ }\textbf {\bibinfo {volume} {151}},\ \bibinfo
  {pages} {124311} (\bibinfo {year} {2019})}\BibitemShut {NoStop}%
\bibitem [{\citenamefont {Mátyus}\ \emph {et~al.}(2014)\citenamefont
  {Mátyus}, \citenamefont {Szidarovszky},\ and\ \citenamefont
  {Császár}}]{MaSzCs14}%
  \BibitemOpen
  \bibfield  {author} {\bibinfo {author} {\bibfnamefont {E.}~\bibnamefont
  {Mátyus}}, \bibinfo {author} {\bibfnamefont {T.}~\bibnamefont
  {Szidarovszky}},\ and\ \bibinfo {author} {\bibfnamefont {A.~G.}\ \bibnamefont
  {Császár}},\ }\bibfield  {title} {\bibinfo {title} {Modelling non-adiabatic
  effects in {H}$_3^+$: {S}olution of the rovibrational {S}chrödinger equation
  with motion-dependent masses and mass surfaces},\ }\href
  {https://doi.org/10.1063/1.4897566} {\bibfield  {journal} {\bibinfo
  {journal} {J. Chem. Phys.}\ }\textbf {\bibinfo {volume} {141}},\ \bibinfo
  {pages} {154111} (\bibinfo {year} {2014})}\BibitemShut {NoStop}%
\bibitem [{\citenamefont {Zare}(1998)}]{Za98}%
  \BibitemOpen
  \bibfield  {author} {\bibinfo {author} {\bibfnamefont {R.~N.}\ \bibnamefont
  {Zare}},\ }\href@noop {} {\emph {\bibinfo {title} {Angular Momentum:
  Understanding Spatial Aspects in Chemistry and Physics}}}\ (\bibinfo
  {publisher} {Wiley-Interscience},\ \bibinfo {address} {New York},\ \bibinfo
  {year} {1998})\BibitemShut {NoStop}%
\bibitem [{\citenamefont {M\'atyus}\ \emph {et~al.}(2009)\citenamefont
  {M\'atyus}, \citenamefont {Czak\'o},\ and\ \citenamefont
  {Cs\'asz\'ar}}]{MaCzCs09}%
  \BibitemOpen
  \bibfield  {author} {\bibinfo {author} {\bibfnamefont {E.}~\bibnamefont
  {M\'atyus}}, \bibinfo {author} {\bibfnamefont {G.}~\bibnamefont {Czak\'o}},\
  and\ \bibinfo {author} {\bibfnamefont {A.~G.}\ \bibnamefont {Cs\'asz\'ar}},\
  }\bibfield  {title} {\bibinfo {title} {Toward black-box-type full- and
  reduced-dimensional variational (ro)vibrational computations},\ }\href
  {https://doi.org/10.1063/1.3076742} {\bibfield  {journal} {\bibinfo
  {journal} {J. Chem. Phys.}\ }\textbf {\bibinfo {volume} {130}},\ \bibinfo
  {pages} {134112} (\bibinfo {year} {2009})}\BibitemShut {NoStop}%
\bibitem [{\citenamefont {Luckhaus}(2000)}]{Lu00}%
  \BibitemOpen
  \bibfield  {author} {\bibinfo {author} {\bibfnamefont {D.}~\bibnamefont
  {Luckhaus}},\ }\bibfield  {title} {\bibinfo {title} {{6D} vibrational quantum
  dynamics: Generalized coordinate discrete variable representation and
  (a)diabatic contraction},\ }\href {https://doi.org/10.1063/1.481924}
  {\bibfield  {journal} {\bibinfo  {journal} {J. Chem. Phys.}\ }\textbf
  {\bibinfo {volume} {113}},\ \bibinfo {pages} {1329} (\bibinfo {year}
  {2000})}\BibitemShut {NoStop}%
\bibitem [{\citenamefont {Lauvergnat}\ and\ \citenamefont
  {Nauts}(2002)}]{LaNa02}%
  \BibitemOpen
  \bibfield  {author} {\bibinfo {author} {\bibfnamefont {D.}~\bibnamefont
  {Lauvergnat}}\ and\ \bibinfo {author} {\bibfnamefont {A.}~\bibnamefont
  {Nauts}},\ }\bibfield  {title} {\bibinfo {title} {Exact numerical computation
  of a kinetic energy operator in curvilinear coordinates},\ }\href
  {https://doi.org/10.1063/1.1469019} {\bibfield  {journal} {\bibinfo
  {journal} {J. Chem. Phys.}\ }\textbf {\bibinfo {volume} {116}},\ \bibinfo
  {pages} {8560} (\bibinfo {year} {2002})}\BibitemShut {NoStop}%
\bibitem [{\citenamefont {Yachmenev}\ and\ \citenamefont
  {Yurchenko}(2015)}]{YaYu15}%
  \BibitemOpen
  \bibfield  {author} {\bibinfo {author} {\bibfnamefont {A.}~\bibnamefont
  {Yachmenev}}\ and\ \bibinfo {author} {\bibfnamefont {S.~N.}\ \bibnamefont
  {Yurchenko}},\ }\bibfield  {title} {\bibinfo {title} {Automatic
  differentiation method for numerical construction of the
  rotational-vibrational {H}amiltonian as a power series in the curvilinear
  internal coordinates using the {E}ckart frame},\ }\href
  {https://doi.org/10.1063/1.4923039} {\bibfield  {journal} {\bibinfo
  {journal} {J. Chem. Phys.}\ }\textbf {\bibinfo {volume} {143}},\ \bibinfo
  {pages} {014105} (\bibinfo {year} {2015})}\BibitemShut {NoStop}%
\bibitem [{\citenamefont {Meyer}\ and\ \citenamefont
  {G{\"u}nthard}(1969)}]{MeGu69}%
  \BibitemOpen
  \bibfield  {author} {\bibinfo {author} {\bibfnamefont {R.}~\bibnamefont
  {Meyer}}\ and\ \bibinfo {author} {\bibfnamefont {H.~H.}\ \bibnamefont
  {G{\"u}nthard}},\ }\bibfield  {title} {\bibinfo {title} {Internal rotation
  and vibration in {CH}$_2$={CC}l–{CH}$_2${D}},\ }\href
  {https://doi.org/10.1063/1.1670803} {\bibfield  {journal} {\bibinfo
  {journal} {J. Chem. Phys.}\ }\textbf {\bibinfo {volume} {50}},\ \bibinfo
  {pages} {353} (\bibinfo {year} {1969})}\BibitemShut {NoStop}%
\bibitem [{\citenamefont {Meyer}(1979)}]{MeJMS79}%
  \BibitemOpen
  \bibfield  {author} {\bibinfo {author} {\bibfnamefont {R.}~\bibnamefont
  {Meyer}},\ }\bibfield  {title} {\bibinfo {title} {Flexible models for
  intramolecular motion, a versatile treatment and its application to
  glyoxal},\ }\href {https://doi.org/10.1016/0022-2852(79)90230-3} {\bibfield
  {journal} {\bibinfo  {journal} {J. Mol. Spectrosc.}\ }\textbf {\bibinfo
  {volume} {76}},\ \bibinfo {pages} {266} (\bibinfo {year} {1979})}\BibitemShut
  {NoStop}%
\bibitem [{\citenamefont {Luckhaus}(2003)}]{Lu03}%
  \BibitemOpen
  \bibfield  {author} {\bibinfo {author} {\bibfnamefont {D.}~\bibnamefont
  {Luckhaus}},\ }\bibfield  {title} {\bibinfo {title} {The vibrational spectrum
  of {HONO}: Fully coupled {6D} direct dynamics},\ }\href
  {https://doi.org/10.1063/1.1567713} {\bibfield  {journal} {\bibinfo
  {journal} {J. Chem. Phys.}\ }\textbf {\bibinfo {volume} {118}},\ \bibinfo
  {pages} {8797} (\bibinfo {year} {2003})}\BibitemShut {NoStop}%
\bibitem [{\citenamefont {Yurchenko}\ \emph {et~al.}(2007)\citenamefont
  {Yurchenko}, \citenamefont {Thiel},\ and\ \citenamefont {Jensen}}]{YuThJe07}%
  \BibitemOpen
  \bibfield  {author} {\bibinfo {author} {\bibfnamefont {S.~N.}\ \bibnamefont
  {Yurchenko}}, \bibinfo {author} {\bibfnamefont {W.}~\bibnamefont {Thiel}},\
  and\ \bibinfo {author} {\bibfnamefont {P.}~\bibnamefont {Jensen}},\
  }\bibfield  {title} {\bibinfo {title} {Theoretical {ROV}ibrational {E}nergies
  ({TROVE}): {A} robust numerical approach to the calculation of rovibrational
  energies for polyatomic molecules},\ }\href
  {https://doi.org/10.1016/j.jms.2007.07.009} {\bibfield  {journal} {\bibinfo
  {journal} {J. Mol. Spectrosc.}\ }\textbf {\bibinfo {volume} {245}},\ \bibinfo
  {pages} {126} (\bibinfo {year} {2007})}\BibitemShut {NoStop}%
\bibitem [{\citenamefont {F\'abri}\ \emph {et~al.}(2011)\citenamefont
  {F\'abri}, \citenamefont {M\'atyus},\ and\ \citenamefont
  {Cs\'asz\'ar}}]{FaMaCs11}%
  \BibitemOpen
  \bibfield  {author} {\bibinfo {author} {\bibfnamefont {C.}~\bibnamefont
  {F\'abri}}, \bibinfo {author} {\bibfnamefont {E.}~\bibnamefont {M\'atyus}},\
  and\ \bibinfo {author} {\bibfnamefont {A.~G.}\ \bibnamefont {Cs\'asz\'ar}},\
  }\bibfield  {title} {\bibinfo {title} {Rotating full- and reduced-dimensional
  quantum chemical models of molecules},\ }\href
  {https://doi.org/10.1063/1.3533950} {\bibfield  {journal} {\bibinfo
  {journal} {J. Chem. Phys.}\ }\textbf {\bibinfo {volume} {134}},\ \bibinfo
  {pages} {074105} (\bibinfo {year} {2011})}\BibitemShut {NoStop}%
\bibitem [{\citenamefont {Sarka}\ and\ \citenamefont
  {Császár}(2016)}]{SaCs16}%
  \BibitemOpen
  \bibfield  {author} {\bibinfo {author} {\bibfnamefont {J.}~\bibnamefont
  {Sarka}}\ and\ \bibinfo {author} {\bibfnamefont {A.~G.}\ \bibnamefont
  {Császár}},\ }\bibfield  {title} {\bibinfo {title} {Interpretation of the
  vibrational energy level structure of the astructural molecular ion {H}$_5^+$
  and all of its deuterated isotopomers},\ }\href
  {https://doi.org/https://doi.org/10.1063/1.4946808} {\bibfield  {journal}
  {\bibinfo  {journal} {J. Chem. Phys.}\ }\textbf {\bibinfo {volume} {144}},\
  \bibinfo {pages} {154309} (\bibinfo {year} {2016})}\BibitemShut {NoStop}%
\bibitem [{\citenamefont {Martín Santa~Daría}\ \emph
  {et~al.}(2021{\natexlab{b}})\citenamefont {Martín Santa~Daría},
  \citenamefont {Avila},\ and\ \citenamefont {Mátyus}}]{DaAvMa21a}%
  \BibitemOpen
  \bibfield  {author} {\bibinfo {author} {\bibfnamefont {A.}~\bibnamefont
  {Martín Santa~Daría}}, \bibinfo {author} {\bibfnamefont {G.}~\bibnamefont
  {Avila}},\ and\ \bibinfo {author} {\bibfnamefont {E.}~\bibnamefont
  {Mátyus}},\ }\bibfield  {title} {\bibinfo {title} {Performance of a
  black-box-type rovibrational method in comparison with a tailor-made
  approach: {C}ase study for the methane–water dimer},\ }\href
  {https://doi.org/10.1063/5.0054512} {\bibfield  {journal} {\bibinfo
  {journal} {J. Chem. Phys.}\ }\textbf {\bibinfo {volume} {154}},\ \bibinfo
  {pages} {224302} (\bibinfo {year} {2021}{\natexlab{b}})}\BibitemShut
  {NoStop}%
\bibitem [{\citenamefont {Sarka}\ \emph {et~al.}(2016)\citenamefont {Sarka},
  \citenamefont {Császár}, \citenamefont {Althorpe}, \citenamefont {Wales},\
  and\ \citenamefont {Mátyus}}]{SaCsAlWaMa16}%
  \BibitemOpen
  \bibfield  {author} {\bibinfo {author} {\bibfnamefont {J.}~\bibnamefont
  {Sarka}}, \bibinfo {author} {\bibfnamefont {A.~G.}\ \bibnamefont
  {Császár}}, \bibinfo {author} {\bibfnamefont {S.~C.}\ \bibnamefont
  {Althorpe}}, \bibinfo {author} {\bibfnamefont {D.~J.}\ \bibnamefont
  {Wales}},\ and\ \bibinfo {author} {\bibfnamefont {E.}~\bibnamefont
  {Mátyus}},\ }\bibfield  {title} {\bibinfo {title} {Rovibrational transitions
  of the methane-water dimer from intermolecular quantum dynamical
  computations},\ }\href {https://doi.org/10.1039/C6CP03062A} {\bibfield
  {journal} {\bibinfo  {journal} {Phys. Chem. Chem. Phys.}\ }\textbf {\bibinfo
  {volume} {18}},\ \bibinfo {pages} {22816} (\bibinfo {year}
  {2016})}\BibitemShut {NoStop}%
\bibitem [{\citenamefont {Sarka}\ \emph {et~al.}(2017)\citenamefont {Sarka},
  \citenamefont {Császár},\ and\ \citenamefont {Mátyus}}]{SaCsMa17}%
  \BibitemOpen
  \bibfield  {author} {\bibinfo {author} {\bibfnamefont {J.}~\bibnamefont
  {Sarka}}, \bibinfo {author} {\bibfnamefont {A.~G.}\ \bibnamefont
  {Császár}},\ and\ \bibinfo {author} {\bibfnamefont {E.}~\bibnamefont
  {Mátyus}},\ }\bibfield  {title} {\bibinfo {title} {Rovibrational quantum
  dynamical computations for deuterated isotopologues of the methane–water
  dimer},\ }\href {https://doi.org/10.1039/C7CP02061A} {\bibfield  {journal}
  {\bibinfo  {journal} {Phys. Chem. Chem. Phys.}\ }\textbf {\bibinfo {volume}
  {19}},\ \bibinfo {pages} {15335} (\bibinfo {year} {2017})}\BibitemShut
  {NoStop}%
\bibitem [{\citenamefont {Ferenc}\ and\ \citenamefont
  {Mátyus}(2019)}]{FeMa19}%
  \BibitemOpen
  \bibfield  {author} {\bibinfo {author} {\bibfnamefont {D.}~\bibnamefont
  {Ferenc}}\ and\ \bibinfo {author} {\bibfnamefont {E.}~\bibnamefont
  {Mátyus}},\ }\bibfield  {title} {\bibinfo {title} {Bound and unbound
  rovibrational states of the methane-argon dimer},\ }\href
  {https://doi.org/10.1080/00268976.2018.1547430} {\bibfield  {journal}
  {\bibinfo  {journal} {Mol. Phys.}\ }\textbf {\bibinfo {volume} {117}},\
  \bibinfo {pages} {1694} (\bibinfo {year} {2019})}\BibitemShut {NoStop}%
\bibitem [{\citenamefont {Ritz}(1909)}]{Ri1909}%
  \BibitemOpen
  \bibfield  {author} {\bibinfo {author} {\bibfnamefont {W.}~\bibnamefont
  {Ritz}},\ }\bibfield  {title} {\bibinfo {title} {{Über eine neue {M}ethode
  zur {L}ösung gewisser {V}ariationsprobleme der mathematischen {P}hysik.}},\
  }\href {http://eudml.org/doc/149295} {\bibfield  {journal} {\bibinfo
  {journal} {J. rein. angew. Math.}\ }\textbf {\bibinfo {volume} {135}},\
  \bibinfo {pages} {1} (\bibinfo {year} {1909})}\BibitemShut {NoStop}%
\bibitem [{\citenamefont {Bunker}\ and\ \citenamefont {Jensen}(1998)}]{BuJe98}%
  \BibitemOpen
  \bibfield  {author} {\bibinfo {author} {\bibfnamefont {P.~R.}\ \bibnamefont
  {Bunker}}\ and\ \bibinfo {author} {\bibfnamefont {P.}~\bibnamefont
  {Jensen}},\ }\href@noop {} {\emph {\bibinfo {title} {Molecular {S}ymmetry and
  {S}pectroscopy, 2nd Edition}}}\ (\bibinfo  {publisher} {NRC Research Press},\
  \bibinfo {address} {Ottawa},\ \bibinfo {year} {1998})\BibitemShut {NoStop}%
\bibitem [{\citenamefont {Wang}(1929)}]{Wa29}%
  \BibitemOpen
  \bibfield  {author} {\bibinfo {author} {\bibfnamefont {S.~C.}\ \bibnamefont
  {Wang}},\ }\bibfield  {title} {\bibinfo {title} {On the asymmetrical top in
  quantum mechanics},\ }\href {https://doi.org/10.1103/PhysRev.34.243}
  {\bibfield  {journal} {\bibinfo  {journal} {Phys. Rev.}\ }\textbf {\bibinfo
  {volume} {34}},\ \bibinfo {pages} {243} (\bibinfo {year} {1929})}\BibitemShut
  {NoStop}%
\bibitem [{\citenamefont {Golub}\ and\ \citenamefont {Welsch}(1969)}]{golub}%
  \BibitemOpen
  \bibfield  {author} {\bibinfo {author} {\bibfnamefont {G.~H.}\ \bibnamefont
  {Golub}}\ and\ \bibinfo {author} {\bibfnamefont {J.~H.}\ \bibnamefont
  {Welsch}},\ }\bibfield  {title} {\bibinfo {title} {Calculation of {G}auss
  quadrature rules},\ }\href@noop {} {\bibfield  {journal} {\bibinfo  {journal}
  {Math. Comp.}\ }\textbf {\bibinfo {volume} {23}},\ \bibinfo {pages} {221}
  (\bibinfo {year} {1969})}\BibitemShut {NoStop}%
\bibitem [{\citenamefont {Szalay}(1996)}]{Szalay96}%
  \BibitemOpen
  \bibfield  {author} {\bibinfo {author} {\bibfnamefont {V.}~\bibnamefont
  {Szalay}},\ }\bibfield  {title} {\bibinfo {title} {The generalized discrete
  variable representation. {A}n optimal design},\ }\href
  {https://doi.org/https://doi.org/10.1063/1.471986} {\bibfield  {journal}
  {\bibinfo  {journal} {J. Chem. Phys.}\ }\textbf {\bibinfo {volume} {105}},\
  \bibinfo {pages} {6940} (\bibinfo {year} {1996})}\BibitemShut {NoStop}%
\bibitem [{\citenamefont {Bacic}\ and\ \citenamefont {Light}(1986)}]{BaLi86}%
  \BibitemOpen
  \bibfield  {author} {\bibinfo {author} {\bibfnamefont {Z.}~\bibnamefont
  {Bacic}}\ and\ \bibinfo {author} {\bibfnamefont {J.~C.}\ \bibnamefont
  {Light}},\ }\bibfield  {title} {\bibinfo {title} {Highly excited vibrational
  levels of ‘‘floppy’’ triatomic molecules: A discrete variable
  representation—{D}istributed {G}aussian basis approach},\ }\href
  {https://doi.org/10.1063/1.451824} {\bibfield  {journal} {\bibinfo  {journal}
  {J. Chem. Phys.}\ }\textbf {\bibinfo {volume} {85}},\ \bibinfo {pages} {4594}
  (\bibinfo {year} {1986})}\BibitemShut {NoStop}%
\bibitem [{\citenamefont {Light}\ and\ \citenamefont
  {Carrington~Jr.}(2000)}]{LiCa00}%
  \BibitemOpen
  \bibfield  {author} {\bibinfo {author} {\bibfnamefont {J.~C.}\ \bibnamefont
  {Light}}\ and\ \bibinfo {author} {\bibfnamefont {T.}~\bibnamefont
  {Carrington~Jr.}},\ }\bibinfo {title} {Discrete-{V}ariable {R}epresentations
  and their {U}tilization},\ in\ \href
  {https://doi.org/10.1002/9780470141731.ch4} {\emph {\bibinfo {booktitle}
  {Adv. Chem. Phys.}}}\ (\bibinfo  {publisher} {John Wiley \& Sons, Ltd},\
  \bibinfo {year} {2000})\ Chap.~\bibinfo {chapter} {14}, pp.\ \bibinfo {pages}
  {263--310}\BibitemShut {NoStop}%
\bibitem [{\citenamefont {Wang}\ and\ \citenamefont
  {Carrington~Jr}(2003{\natexlab{a}})}]{WaCa03HF}%
  \BibitemOpen
  \bibfield  {author} {\bibinfo {author} {\bibfnamefont {X.-G.}\ \bibnamefont
  {Wang}}\ and\ \bibinfo {author} {\bibfnamefont {T.}~\bibnamefont
  {Carrington~Jr}},\ }\bibfield  {title} {\bibinfo {title} {Using {L}ebedev
  grids, sine spherical harmonics, and monomer contracted basis functions to
  calculate bending energy levels of {HF} trimer},\ }\href
  {https://doi.org/https://doi.org/10.1142/S0219633603000720} {\bibfield
  {journal} {\bibinfo  {journal} {J. Chem. Theo. Comp.}\ }\textbf {\bibinfo
  {volume} {2}},\ \bibinfo {pages} {599} (\bibinfo {year}
  {2003}{\natexlab{a}})}\BibitemShut {NoStop}%
\bibitem [{\citenamefont {Wang}\ and\ \citenamefont
  {T.~Carrington}(2021)}]{WaCa21}%
  \BibitemOpen
  \bibfield  {author} {\bibinfo {author} {\bibfnamefont {X.-G.}\ \bibnamefont
  {Wang}}\ and\ \bibinfo {author} {\bibfnamefont {J.}~\bibnamefont
  {T.~Carrington}},\ }\bibfield  {title} {\bibinfo {title} {Using nondirect
  product {W}igner {$D$} basis functions and the {S}ymmetry {A}dapted {L}anczos
  algorithm to compute the ro-vibrational spectrum of {CH}$_4$--{H}$_2${O}},\
  }\href {https://doi.org/10.1063/5.0044010} {\bibfield  {journal} {\bibinfo
  {journal} {J. Chem. Phys.}\ }\textbf {\bibinfo {volume} {154}},\ \bibinfo
  {pages} {124112} (\bibinfo {year} {2021})}\BibitemShut {NoStop}%
\bibitem [{\citenamefont {Schiffel}\ and\ \citenamefont
  {Manthe}(2010)}]{ScMa10}%
  \BibitemOpen
  \bibfield  {author} {\bibinfo {author} {\bibfnamefont {G.}~\bibnamefont
  {Schiffel}}\ and\ \bibinfo {author} {\bibfnamefont {U.}~\bibnamefont
  {Manthe}},\ }\bibfield  {title} {\bibinfo {title} {On direct product based
  discrete variable representations for angular coordinates and the treatment
  of singular terms in the kinetic energy operator},\ }\href
  {https://doi.org/10.1016/j.chemphys.2010.07.006} {\bibfield  {journal}
  {\bibinfo  {journal} {Chem. Phys.}\ }\textbf {\bibinfo {volume} {374}},\
  \bibinfo {pages} {118} (\bibinfo {year} {2010})}\BibitemShut {NoStop}%
\bibitem [{\citenamefont {Bramley}\ and\ \citenamefont
  {Carrington}(1993)}]{BraCa93}%
  \BibitemOpen
  \bibfield  {author} {\bibinfo {author} {\bibfnamefont {M.~J.}\ \bibnamefont
  {Bramley}}\ and\ \bibinfo {author} {\bibfnamefont {T.}~\bibnamefont
  {Carrington}},\ }\bibfield  {title} {\bibinfo {title} {A general discrete
  variable method to calculate vibrational energy levels of three‐ and
  four‐atom molecules},\ }\href {https://doi.org/10.1063/1.465576} {\bibfield
   {journal} {\bibinfo  {journal} {J. Chem. Phys.}\ }\textbf {\bibinfo {volume}
  {99}},\ \bibinfo {pages} {8519} (\bibinfo {year} {1993})}\BibitemShut
  {NoStop}%
\bibitem [{\citenamefont {Bramley}\ and\ \citenamefont
  {Carrington~Jr}(1994)}]{BraCa94}%
  \BibitemOpen
  \bibfield  {author} {\bibinfo {author} {\bibfnamefont {M.~J.}\ \bibnamefont
  {Bramley}}\ and\ \bibinfo {author} {\bibfnamefont {T.}~\bibnamefont
  {Carrington~Jr}},\ }\bibfield  {title} {\bibinfo {title} {Calculation of
  triatomic vibrational eigenstates: {P}roduct or contracted basis sets,
  {L}anczos or conventional eigensolvers? {W}hat is the most efficient
  combination?},\ }\href {https://doi.org/10.1063/1.468110} {\bibfield
  {journal} {\bibinfo  {journal} {J. Chem. Phys.}\ }\textbf {\bibinfo {volume}
  {101}},\ \bibinfo {pages} {8494} (\bibinfo {year} {1994})}\BibitemShut
  {NoStop}%
\bibitem [{\citenamefont {Roy}\ and\ \citenamefont
  {Carrington~Jr}(1996)}]{RoCa96}%
  \BibitemOpen
  \bibfield  {author} {\bibinfo {author} {\bibfnamefont {P.-N.}\ \bibnamefont
  {Roy}}\ and\ \bibinfo {author} {\bibfnamefont {T.}~\bibnamefont
  {Carrington~Jr}},\ }\bibfield  {title} {\bibinfo {title} {A direct-operation
  {L}anczos approach for calculating energy levels},\ }\href
  {https://doi.org/https://doi.org/10.1016/0009-2614(96)00505-2} {\bibfield
  {journal} {\bibinfo  {journal} {Chem. Phys. Lett.}\ }\textbf {\bibinfo
  {volume} {257}},\ \bibinfo {pages} {98} (\bibinfo {year} {1996})}\BibitemShut
  {NoStop}%
\bibitem [{\citenamefont {Wang}\ and\ \citenamefont
  {Carrington}(2001)}]{WaCa01}%
  \BibitemOpen
  \bibfield  {author} {\bibinfo {author} {\bibfnamefont {X.-G.}\ \bibnamefont
  {Wang}}\ and\ \bibinfo {author} {\bibfnamefont {T.}~\bibnamefont
  {Carrington}},\ }\bibfield  {title} {\bibinfo {title} {The utility of
  constraining basis function indices when using the {L}anczos algorithm to
  calculate vibrational energy levels},\ }\href
  {https://doi.org/https://doi.org/10.1021/jp003792s} {\bibfield  {journal}
  {\bibinfo  {journal} {J. Phys. Chem. A}\ }\textbf {\bibinfo {volume} {105}},\
  \bibinfo {pages} {2575} (\bibinfo {year} {2001})}\BibitemShut {NoStop}%
\bibitem [{\citenamefont {Wang}\ and\ \citenamefont
  {Carrington~Jr}(2003{\natexlab{b}})}]{WaCa03}%
  \BibitemOpen
  \bibfield  {author} {\bibinfo {author} {\bibfnamefont {X.-G.}\ \bibnamefont
  {Wang}}\ and\ \bibinfo {author} {\bibfnamefont {T.}~\bibnamefont
  {Carrington~Jr}},\ }\bibfield  {title} {\bibinfo {title} {A finite basis
  representation {L}anczos calculation of the bend energy levels of methane},\
  }\href {https://doi.org/10.1063/1.1554735} {\bibfield  {journal} {\bibinfo
  {journal} {J. Chem. Phys.}\ }\textbf {\bibinfo {volume} {118}},\ \bibinfo
  {pages} {6946} (\bibinfo {year} {2003}{\natexlab{b}})}\BibitemShut {NoStop}%
\bibitem [{\citenamefont {Carrington~Jr}\ and\ \citenamefont
  {Wang}(2011)}]{CaWa11}%
  \BibitemOpen
  \bibfield  {author} {\bibinfo {author} {\bibfnamefont {T.}~\bibnamefont
  {Carrington~Jr}}\ and\ \bibinfo {author} {\bibfnamefont {X.-G.}\ \bibnamefont
  {Wang}},\ }\bibfield  {title} {\bibinfo {title} {Computing ro-vibrational
  spectra of van der {W}aals molecules},\ }\href
  {https://doi.org/10.1002/wcms.73} {\bibfield  {journal} {\bibinfo  {journal}
  {Wiley Interdis. Rev.: Comp. Mol. Sci.}\ }\textbf {\bibinfo {volume} {1}},\
  \bibinfo {pages} {952} (\bibinfo {year} {2011})}\BibitemShut {NoStop}%
\bibitem [{\citenamefont {Lanczos}(1950)}]{lanczos}%
  \BibitemOpen
  \bibfield  {author} {\bibinfo {author} {\bibfnamefont {C.}~\bibnamefont
  {Lanczos}},\ }\bibfield  {title} {\bibinfo {title} {An iteration method for
  the solution of the eigenvalue problem of linear differential and integral
  operators},\ }\href {https://doi.org/10.6028/jres.045.026} {\bibfield
  {journal} {\bibinfo  {journal} {J. Res. Natl. Bur. Stand.}\ }\textbf
  {\bibinfo {volume} {45}},\ \bibinfo {pages} {255} (\bibinfo {year}
  {1950})}\BibitemShut {NoStop}%
\bibitem [{\citenamefont {Bramley}\ and\ \citenamefont {{Carrington,
  Jr.}}(1994)}]{BrCa94}%
  \BibitemOpen
  \bibfield  {author} {\bibinfo {author} {\bibfnamefont {M.~J.}\ \bibnamefont
  {Bramley}}\ and\ \bibinfo {author} {\bibfnamefont {T.}~\bibnamefont
  {{Carrington, Jr.}}},\ }\bibfield  {title} {\bibinfo {title} {Calculation of
  triatomic vibrational eigenstates: Product or contracted basis sets,
  {L}anczos or conventional eigensolvers? {W}hat is the most efficient
  combination?},\ }\href {https://doi.org/https://doi.org/10.1063/1.468110}
  {\bibfield  {journal} {\bibinfo  {journal} {J. Chem. Phys.}\ }\textbf
  {\bibinfo {volume} {101}},\ \bibinfo {pages} {8494} (\bibinfo {year}
  {1994})}\BibitemShut {NoStop}%
\bibitem [{\citenamefont {Mátyus}\ \emph {et~al.}(2009)\citenamefont
  {Mátyus}, \citenamefont {Šimunek},\ and\ \citenamefont
  {Császár}}]{MaSiCs09}%
  \BibitemOpen
  \bibfield  {author} {\bibinfo {author} {\bibfnamefont {E.}~\bibnamefont
  {Mátyus}}, \bibinfo {author} {\bibfnamefont {J.}~\bibnamefont {Šimunek}},\
  and\ \bibinfo {author} {\bibfnamefont {A.~G.}\ \bibnamefont {Császár}},\
  }\bibfield  {title} {\bibinfo {title} {On the variational computation of a
  large number of vibrational energy levels and wave functions for medium-sized
  molecules},\ }\href {https://doi.org/10.1063/1.3187528} {\bibfield  {journal}
  {\bibinfo  {journal} {J. Chem. Phys.}\ }\textbf {\bibinfo {volume} {131}},\
  \bibinfo {pages} {074106} (\bibinfo {year} {2009})}\BibitemShut {NoStop}%
\bibitem [{\citenamefont {Fábri}\ \emph {et~al.}(2014)\citenamefont {Fábri},
  \citenamefont {Sarka},\ and\ \citenamefont {Császár}}]{FaSaCs14}%
  \BibitemOpen
  \bibfield  {author} {\bibinfo {author} {\bibfnamefont {C.}~\bibnamefont
  {Fábri}}, \bibinfo {author} {\bibfnamefont {J.}~\bibnamefont {Sarka}},\ and\
  \bibinfo {author} {\bibfnamefont {A.~G.}\ \bibnamefont {Császár}},\
  }\bibfield  {title} {\bibinfo {title} {Communication: Rigidity of the
  molecular ion {H}$^+_5$},\ }\href
  {https://doi.org/https://doi.org/10.1063/1.4864360} {\bibfield  {journal}
  {\bibinfo  {journal} {J. Chem. Phys.}\ }\textbf {\bibinfo {volume} {140}},\
  \bibinfo {pages} {051101} (\bibinfo {year} {2014})}\BibitemShut {NoStop}%
\bibitem [{\citenamefont {Avila}\ and\ \citenamefont
  {Carrington}(2009)}]{AvCa09}%
  \BibitemOpen
  \bibfield  {author} {\bibinfo {author} {\bibfnamefont {G.}~\bibnamefont
  {Avila}}\ and\ \bibinfo {author} {\bibfnamefont {T.}~\bibnamefont
  {Carrington}},\ }\bibfield  {title} {\bibinfo {title} {Nonproduct quadrature
  grids for solving the vibrational {S}chrödinger equation},\ }\href
  {https://doi.org/10.1063/1.3246593} {\bibfield  {journal} {\bibinfo
  {journal} {J. Chem. Phys.}\ }\textbf {\bibinfo {volume} {131}},\ \bibinfo
  {pages} {174103} (\bibinfo {year} {2009})}\BibitemShut {NoStop}%
\bibitem [{\citenamefont {Avila}\ and\ \citenamefont
  {Carrington}(2011{\natexlab{a}})}]{AvCa11}%
  \BibitemOpen
  \bibfield  {author} {\bibinfo {author} {\bibfnamefont {G.}~\bibnamefont
  {Avila}}\ and\ \bibinfo {author} {\bibfnamefont {T.}~\bibnamefont
  {Carrington}},\ }\bibfield  {title} {\bibinfo {title} {Using nonproduct
  quadrature grids to solve the vibrational {S}chrödinger equation in {12D}},\
  }\href {https://doi.org/10.1063/1.3549817} {\bibfield  {journal} {\bibinfo
  {journal} {J. Chem. Phys.}\ }\textbf {\bibinfo {volume} {134}},\ \bibinfo
  {pages} {054126} (\bibinfo {year} {2011}{\natexlab{a}})}\BibitemShut
  {NoStop}%
\bibitem [{\citenamefont {Avila}\ and\ \citenamefont
  {Carrington}(2011{\natexlab{b}})}]{AvCa11b}%
  \BibitemOpen
  \bibfield  {author} {\bibinfo {author} {\bibfnamefont {G.}~\bibnamefont
  {Avila}}\ and\ \bibinfo {author} {\bibfnamefont {T.}~\bibnamefont
  {Carrington}},\ }\bibfield  {title} {\bibinfo {title} {Using a pruned basis,
  a non-product quadrature grid, and the exact {W}atson normal-coordinate
  kinetic energy operator to solve the vibrational {S}chrödinger equation for
  {C}$_2${H}$_4$},\ }\href {https://doi.org/10.1063/1.3617249} {\bibfield
  {journal} {\bibinfo  {journal} {J. Chem. Phys.}\ }\textbf {\bibinfo {volume}
  {135}},\ \bibinfo {pages} {064101} (\bibinfo {year}
  {2011}{\natexlab{b}})}\BibitemShut {NoStop}%
\bibitem [{\citenamefont {Lauvergnat}\ and\ \citenamefont
  {Nauts}(2014)}]{LaNa14}%
  \BibitemOpen
  \bibfield  {author} {\bibinfo {author} {\bibfnamefont {D.}~\bibnamefont
  {Lauvergnat}}\ and\ \bibinfo {author} {\bibfnamefont {A.}~\bibnamefont
  {Nauts}},\ }\bibfield  {title} {\bibinfo {title} {Quantum dynamics with
  sparse grids: {A} combination of {S}molyak scheme and cubature. {A}pplication
  to methanol in full dimensionality},\ }\href
  {https://doi.org/10.1016/j.saa.2013.05.068} {\bibfield  {journal} {\bibinfo
  {journal} {Spectrochim. Acta}\ }\textbf {\bibinfo {volume} {119}},\ \bibinfo
  {pages} {18} (\bibinfo {year} {2014})}\BibitemShut {NoStop}%
\bibitem [{\citenamefont {Zhang}\ \emph {et~al.}(1991)\citenamefont {Zhang},
  \citenamefont {Klippenstein},\ and\ \citenamefont {Marcus}}]{ZhKlMa91}%
  \BibitemOpen
  \bibfield  {author} {\bibinfo {author} {\bibfnamefont {Y.}~\bibnamefont
  {Zhang}}, \bibinfo {author} {\bibfnamefont {S.~J.}\ \bibnamefont
  {Klippenstein}},\ and\ \bibinfo {author} {\bibfnamefont {R.~A.}\ \bibnamefont
  {Marcus}},\ }\bibfield  {title} {\bibinfo {title} {Intramolecular dynamics.
  {I.} {C}urvilinear normal modes, local modes, molecular anharmonic
  {H}amiltonian, and application to benzene},\ }\href
  {https://doi.org/10.1063/1.460216} {\bibfield  {journal} {\bibinfo  {journal}
  {J. Chem. Phys.}\ }\textbf {\bibinfo {volume} {94}},\ \bibinfo {pages} {7319}
  (\bibinfo {year} {1991})}\BibitemShut {NoStop}%
\bibitem [{\citenamefont {Castro}\ \emph {et~al.}(2017)\citenamefont {Castro},
  \citenamefont {Avila}, \citenamefont {Manzhos}, \citenamefont {Agarwal},
  \citenamefont {Schaefer},\ and\ \citenamefont {{Carrington,
  Jr}}}]{CaAvMaAgScCa16}%
  \BibitemOpen
  \bibfield  {author} {\bibinfo {author} {\bibfnamefont {E.}~\bibnamefont
  {Castro}}, \bibinfo {author} {\bibfnamefont {G.}~\bibnamefont {Avila}},
  \bibinfo {author} {\bibfnamefont {S.}~\bibnamefont {Manzhos}}, \bibinfo
  {author} {\bibfnamefont {J.}~\bibnamefont {Agarwal}}, \bibinfo {author}
  {\bibfnamefont {H.~F.}\ \bibnamefont {Schaefer}},\ and\ \bibinfo {author}
  {\bibfnamefont {T.}~\bibnamefont {{Carrington, Jr}}},\ }\bibfield  {title}
  {\bibinfo {title} {Applying a {S}molyak collocation method to {C}l$_2${CO}},\
  }\href {https://doi.org/10.1080/00268976.2016.1271153} {\bibfield  {journal}
  {\bibinfo  {journal} {Mol. Phys.}\ }\textbf {\bibinfo {volume} {115}},\
  \bibinfo {pages} {1775} (\bibinfo {year} {2017})}\BibitemShut {NoStop}%
\bibitem [{\citenamefont {Papou{\v{s}}ek}\ \emph {et~al.}(1973)\citenamefont
  {Papou{\v{s}}ek}, \citenamefont {Stone},\ and\ \citenamefont
  {{\v{S}}pirko}}]{PaStSp73}%
  \BibitemOpen
  \bibfield  {author} {\bibinfo {author} {\bibfnamefont {D.}~\bibnamefont
  {Papou{\v{s}}ek}}, \bibinfo {author} {\bibfnamefont {J.}~\bibnamefont
  {Stone}},\ and\ \bibinfo {author} {\bibfnamefont {V.}~\bibnamefont
  {{\v{S}}pirko}},\ }\bibfield  {title} {\bibinfo {title}
  {Vibration-inversion-rotation spectra of ammonia. {A}
  vibration-inversion-rotation {H}amiltonian for {NH}$_3$},\ }\href
  {https://doi.org/https://doi.org/10.1016/0022-2852(73)90132-X} {\bibfield
  {journal} {\bibinfo  {journal} {J. Mol. Spectrosc.}\ }\textbf {\bibinfo
  {volume} {48}},\ \bibinfo {pages} {17} (\bibinfo {year} {1973})}\BibitemShut
  {NoStop}%
\bibitem [{\citenamefont {Miller}\ \emph {et~al.}(1980)\citenamefont {Miller},
  \citenamefont {Handy},\ and\ \citenamefont {Adams}}]{MiHaAd80}%
  \BibitemOpen
  \bibfield  {author} {\bibinfo {author} {\bibfnamefont {W.~H.}\ \bibnamefont
  {Miller}}, \bibinfo {author} {\bibfnamefont {N.~C.}\ \bibnamefont {Handy}},\
  and\ \bibinfo {author} {\bibfnamefont {J.~E.}\ \bibnamefont {Adams}},\
  }\bibfield  {title} {\bibinfo {title} {Reaction path {H}amiltonian for
  polyatomic molecules},\ }\href {https://doi.org/10.1063/1.438959} {\bibfield
  {journal} {\bibinfo  {journal} {J. Chem. Phys.}\ }\textbf {\bibinfo {volume}
  {72}},\ \bibinfo {pages} {99} (\bibinfo {year} {1980})}\BibitemShut {NoStop}%
\bibitem [{\citenamefont {Bowman}\ \emph {et~al.}(2007)\citenamefont {Bowman},
  \citenamefont {Huang}, \citenamefont {Handy},\ and\ \citenamefont
  {Carter}}]{BoHuHaNoCa07}%
  \BibitemOpen
  \bibfield  {author} {\bibinfo {author} {\bibfnamefont {J.~M.}\ \bibnamefont
  {Bowman}}, \bibinfo {author} {\bibfnamefont {X.}~\bibnamefont {Huang}},
  \bibinfo {author} {\bibfnamefont {N.~C.}\ \bibnamefont {Handy}},\ and\
  \bibinfo {author} {\bibfnamefont {S.}~\bibnamefont {Carter}},\ }\bibfield
  {title} {\bibinfo {title} {Vibrational levels of methanol calculated by the
  reaction path version of {MULTIMODE,} using an ab initio, full-dimensional
  potential},\ }\href {https://doi.org/https://doi.org/10.1021/jp070398m}
  {\bibfield  {journal} {\bibinfo  {journal} {J. Phys. Chem. A}\ }\textbf
  {\bibinfo {volume} {111}},\ \bibinfo {pages} {7317} (\bibinfo {year}
  {2007})}\BibitemShut {NoStop}%
\bibitem [{\citenamefont {Carrington}\ and\ \citenamefont
  {Miller}(1984)}]{CaMi84}%
  \BibitemOpen
  \bibfield  {author} {\bibinfo {author} {\bibfnamefont {T.}~\bibnamefont
  {Carrington}}\ and\ \bibinfo {author} {\bibfnamefont {W.~H.}\ \bibnamefont
  {Miller}},\ }\bibfield  {title} {\bibinfo {title} {Reaction surface
  hamiltonian for the dynamics of reactions in polyatomic systems},\ }\href
  {https://doi.org/10.1063/1.448187} {\bibfield  {journal} {\bibinfo  {journal}
  {J. Chem. Phys.}\ }\textbf {\bibinfo {volume} {81}},\ \bibinfo {pages} {3942}
  (\bibinfo {year} {1984})}\BibitemShut {NoStop}%
\bibitem [{\citenamefont {Carrington~Jr}\ and\ \citenamefont
  {Miller}(1986)}]{CaMi86}%
  \BibitemOpen
  \bibfield  {author} {\bibinfo {author} {\bibfnamefont {T.}~\bibnamefont
  {Carrington~Jr}}\ and\ \bibinfo {author} {\bibfnamefont {W.~H.}\ \bibnamefont
  {Miller}},\ }\bibfield  {title} {\bibinfo {title} {Reaction surface
  description of intramolecular hydrogen atom transfer in malonaldehyde},\
  }\href {https://doi.org/10.1063/1.450058} {\bibfield  {journal} {\bibinfo
  {journal} {J. Chem. Phys.}\ }\textbf {\bibinfo {volume} {84}},\ \bibinfo
  {pages} {4364} (\bibinfo {year} {1986})}\BibitemShut {NoStop}%
\bibitem [{\citenamefont {Shida}\ \emph {et~al.}(1991)\citenamefont {Shida},
  \citenamefont {Barbara},\ and\ \citenamefont {Alml{\"o}f}}]{ShBaAl91}%
  \BibitemOpen
  \bibfield  {author} {\bibinfo {author} {\bibfnamefont {N.}~\bibnamefont
  {Shida}}, \bibinfo {author} {\bibfnamefont {P.~F.}\ \bibnamefont {Barbara}},\
  and\ \bibinfo {author} {\bibfnamefont {J.}~\bibnamefont {Alml{\"o}f}},\
  }\bibfield  {title} {\bibinfo {title} {A reaction surface hamiltonian
  treatment of the double proton transfer of formic acid dimer},\ }\href
  {https://doi.org/10.1063/1.459734} {\bibfield  {journal} {\bibinfo  {journal}
  {J. Chem. Phys.}\ }\textbf {\bibinfo {volume} {94}},\ \bibinfo {pages} {3633}
  (\bibinfo {year} {1991})}\BibitemShut {NoStop}%
\bibitem [{\citenamefont {Whitehead}\ and\ \citenamefont
  {Handy}(1975)}]{WhHa75}%
  \BibitemOpen
  \bibfield  {author} {\bibinfo {author} {\bibfnamefont {R.~J.}\ \bibnamefont
  {Whitehead}}\ and\ \bibinfo {author} {\bibfnamefont {N.~C.}\ \bibnamefont
  {Handy}},\ }\bibfield  {title} {\bibinfo {title} {Variational calculation of
  vibration-rotation energy levels for triatomic molecules},\ }\href
  {https://doi.org/https://doi.org/10.1016/0022-2852(75)90274-X} {\bibfield
  {journal} {\bibinfo  {journal} {J. Mol. Spectrosc.}\ }\textbf {\bibinfo
  {volume} {55}},\ \bibinfo {pages} {356} (\bibinfo {year} {1975})}\BibitemShut
  {NoStop}%
\bibitem [{\citenamefont {Bowman}\ \emph {et~al.}(2003)\citenamefont {Bowman},
  \citenamefont {Carter},\ and\ \citenamefont {Huang}}]{BoCaHu03}%
  \BibitemOpen
  \bibfield  {author} {\bibinfo {author} {\bibfnamefont {J.~M.}\ \bibnamefont
  {Bowman}}, \bibinfo {author} {\bibfnamefont {S.}~\bibnamefont {Carter}},\
  and\ \bibinfo {author} {\bibfnamefont {X.}~\bibnamefont {Huang}},\ }\bibfield
   {title} {\bibinfo {title} {Multimode: {A} code to calculate rovibrational
  energies of polyatomic molecules},\ }\href
  {https://doi.org/10.1080/0144235031000124163} {\bibfield  {journal} {\bibinfo
   {journal} {Int. Rev. Phys. Chem.}\ }\textbf {\bibinfo {volume} {22}},\
  \bibinfo {pages} {533} (\bibinfo {year} {2003})}\BibitemShut {NoStop}%
\bibitem [{\citenamefont {Halverson}\ and\ \citenamefont
  {Poirier}(2015{\natexlab{a}})}]{HaPo15}%
  \BibitemOpen
  \bibfield  {author} {\bibinfo {author} {\bibfnamefont {T.}~\bibnamefont
  {Halverson}}\ and\ \bibinfo {author} {\bibfnamefont {B.}~\bibnamefont
  {Poirier}},\ }\bibfield  {title} {\bibinfo {title} {Large scale exact quantum
  dynamics calculations: {T}en thousand quantum states of acetonitrile},\
  }\href {https://doi.org/10.1016/j.cplett.2015.02.004} {\bibfield  {journal}
  {\bibinfo  {journal} {Chem. Phys. Lett.}\ }\textbf {\bibinfo {volume}
  {624}},\ \bibinfo {pages} {37} (\bibinfo {year}
  {2015}{\natexlab{a}})}\BibitemShut {NoStop}%
\bibitem [{\citenamefont {Halverson}\ and\ \citenamefont
  {Poirier}(2015{\natexlab{b}})}]{HaPo15b}%
  \BibitemOpen
  \bibfield  {author} {\bibinfo {author} {\bibfnamefont {T.}~\bibnamefont
  {Halverson}}\ and\ \bibinfo {author} {\bibfnamefont {B.}~\bibnamefont
  {Poirier}},\ }\bibfield  {title} {\bibinfo {title} {One {M}illion {Q}uantum
  {S}tates of {B}enzene},\ }\href {https://doi.org/10.1021/acs.jpca.5b07868}
  {\bibfield  {journal} {\bibinfo  {journal} {J. Phys. Chem. A}\ }\textbf
  {\bibinfo {volume} {119}},\ \bibinfo {pages} {12417} (\bibinfo {year}
  {2015}{\natexlab{b}})}\BibitemShut {NoStop}%
\bibitem [{\citenamefont {Sarka}\ and\ \citenamefont {Poirier}(2021)}]{SaPo21}%
  \BibitemOpen
  \bibfield  {author} {\bibinfo {author} {\bibfnamefont {J.}~\bibnamefont
  {Sarka}}\ and\ \bibinfo {author} {\bibfnamefont {B.}~\bibnamefont
  {Poirier}},\ }\bibfield  {title} {\bibinfo {title} {Hitting the trifecta: How
  to simultaneously push the limits of {S}chrödinger solution with respect to
  system size, convergence accuracy, and number of computed states},\ }\href
  {https://doi.org/10.1021/acs.jctc.1c00824} {\bibfield  {journal} {\bibinfo
  {journal} {J. Chem. Theo. Comp.}\ }\textbf {\bibinfo {volume} {17}},\
  \bibinfo {pages} {7732} (\bibinfo {year} {2021})}\BibitemShut {NoStop}%
\bibitem [{\citenamefont {Chen}\ and\ \citenamefont
  {Lauvergnat}(2021)}]{ChLa21}%
  \BibitemOpen
  \bibfield  {author} {\bibinfo {author} {\bibfnamefont {A.}~\bibnamefont
  {Chen}}\ and\ \bibinfo {author} {\bibfnamefont {D.}~\bibnamefont
  {Lauvergnat}},\ }\bibfield  {title} {\bibinfo {title} {El{V}ib{R}ot-{MPI}:
  parallel quantum dynamics with {S}molyak algorithm for general molecular
  simulation},\ }\bibfield  {journal} {\bibinfo  {journal} {arXiv preprint
  arXiv:2111.13655}\ }\href {https://doi.org/10.48550/arXiv.2111.13655}
  {10.48550/arXiv.2111.13655} (\bibinfo {year} {2021})\BibitemShut {NoStop}%
\bibitem [{\citenamefont {Avila}\ and\ \citenamefont
  {Carrington~Jr}(2012)}]{AvCa12}%
  \BibitemOpen
  \bibfield  {author} {\bibinfo {author} {\bibfnamefont {G.}~\bibnamefont
  {Avila}}\ and\ \bibinfo {author} {\bibfnamefont {T.}~\bibnamefont
  {Carrington~Jr}},\ }\bibfield  {title} {\bibinfo {title} {Solving the
  vibrational {S}chr{\"o}dinger equation using bases pruned to include strongly
  coupled functions and compatible quadratures},\ }\href
  {https://doi.org/10.1063/1.4764099} {\bibfield  {journal} {\bibinfo
  {journal} {J. Chem. Phys.}\ }\textbf {\bibinfo {volume} {137}},\ \bibinfo
  {pages} {174108} (\bibinfo {year} {2012})}\BibitemShut {NoStop}%
\bibitem [{\citenamefont {Heiss}\ and\ \citenamefont
  {Winschel}(2008)}]{HeWi08}%
  \BibitemOpen
  \bibfield  {author} {\bibinfo {author} {\bibfnamefont {F.}~\bibnamefont
  {Heiss}}\ and\ \bibinfo {author} {\bibfnamefont {V.}~\bibnamefont
  {Winschel}},\ }\bibfield  {title} {\bibinfo {title} {Likelihood approximation
  by numerical integration on sparse grids},\ }\href
  {https://doi.org/0.1016/j.jeconom.2007.12.004} {\bibfield  {journal}
  {\bibinfo  {journal} {J. Econometrics}\ }\textbf {\bibinfo {volume} {144}},\
  \bibinfo {pages} {62} (\bibinfo {year} {2008})}\BibitemShut {NoStop}%
\bibitem [{\citenamefont {Nauts}\ and\ \citenamefont
  {Lauvergnat}(2018)}]{NaLa18}%
  \BibitemOpen
  \bibfield  {author} {\bibinfo {author} {\bibfnamefont {A.}~\bibnamefont
  {Nauts}}\ and\ \bibinfo {author} {\bibfnamefont {D.}~\bibnamefont
  {Lauvergnat}},\ }\bibfield  {title} {\bibinfo {title} {Numerical on-the-fly
  implementation of the action of the kinetic energy operator on a vibrational
  wave function: application to methanol},\ }\href
  {https://doi.org/https://doi.org/10.1080/00268976.2018.1473652} {\bibfield
  {journal} {\bibinfo  {journal} {Mol. Phys.}\ }\textbf {\bibinfo {volume}
  {116}},\ \bibinfo {pages} {3701} (\bibinfo {year} {2018})}\BibitemShut
  {NoStop}%
\bibitem [{\citenamefont {Beck}\ \emph {et~al.}(2000)\citenamefont {Beck},
  \citenamefont {J{\"a}ckle}, \citenamefont {Worth},\ and\ \citenamefont
  {Meyer}}]{BeMiJa2000}%
  \BibitemOpen
  \bibfield  {author} {\bibinfo {author} {\bibfnamefont {M.~H.}\ \bibnamefont
  {Beck}}, \bibinfo {author} {\bibfnamefont {A.}~\bibnamefont {J{\"a}ckle}},
  \bibinfo {author} {\bibfnamefont {G.~A.}\ \bibnamefont {Worth}},\ and\
  \bibinfo {author} {\bibfnamefont {H.-D.}\ \bibnamefont {Meyer}},\ }\bibfield
  {title} {\bibinfo {title} {The multiconfiguration time-dependent {H}artree
  ({MCTDH}) method: a highly efficient algorithm for propagating wavepackets},\
  }\href {https://doi.org/10.1016/S0370-1573(99)00047-2} {\bibfield  {journal}
  {\bibinfo  {journal} {Phys. Rep.}\ }\textbf {\bibinfo {volume} {324}},\
  \bibinfo {pages} {1} (\bibinfo {year} {2000})}\BibitemShut {NoStop}%
\bibitem [{\citenamefont {Ba{\v{c}}i{\'c}}\ and\ \citenamefont
  {Light}(1989)}]{BaLi89}%
  \BibitemOpen
  \bibfield  {author} {\bibinfo {author} {\bibfnamefont {Z.}~\bibnamefont
  {Ba{\v{c}}i{\'c}}}\ and\ \bibinfo {author} {\bibfnamefont {J.~C.}\
  \bibnamefont {Light}},\ }\bibfield  {title} {\bibinfo {title} {Theoretical
  methods for rovibrational states of floppy molecules},\ }\href
  {https://doi.org/10.1146/annurev.pc.40.100189.002345} {\bibfield  {journal}
  {\bibinfo  {journal} {Ann. Rev. Phys. Chem.}\ }\textbf {\bibinfo {volume}
  {40}},\ \bibinfo {pages} {469} (\bibinfo {year} {1989})}\BibitemShut
  {NoStop}%
\bibitem [{\citenamefont {Henderson}\ and\ \citenamefont
  {Tennyson}(1990)}]{HeTe90}%
  \BibitemOpen
  \bibfield  {author} {\bibinfo {author} {\bibfnamefont {J.~R.}\ \bibnamefont
  {Henderson}}\ and\ \bibinfo {author} {\bibfnamefont {J.}~\bibnamefont
  {Tennyson}},\ }\bibfield  {title} {\bibinfo {title} {All the vibrational
  bound states of {H}$^+_3$},\ }\href
  {https://doi.org/10.1016/0009-2614(90)80066-M} {\bibfield  {journal}
  {\bibinfo  {journal} {Chem. Phys. Lett.}\ }\textbf {\bibinfo {volume}
  {173}},\ \bibinfo {pages} {133} (\bibinfo {year} {1990})}\BibitemShut
  {NoStop}%
\bibitem [{\citenamefont {Wang}\ and\ \citenamefont
  {Carrington~Jr}(2004)}]{WaCa04}%
  \BibitemOpen
  \bibfield  {author} {\bibinfo {author} {\bibfnamefont {X.-G.}\ \bibnamefont
  {Wang}}\ and\ \bibinfo {author} {\bibfnamefont {T.}~\bibnamefont
  {Carrington~Jr}},\ }\bibfield  {title} {\bibinfo {title} {Contracted basis
  lanczos methods for computing numerically exact rovibrational levels of
  methane},\ }\href {https://doi.org/10.1063/1.1767093} {\bibfield  {journal}
  {\bibinfo  {journal} {J. Chem. Phys.}\ }\textbf {\bibinfo {volume} {121}},\
  \bibinfo {pages} {2937} (\bibinfo {year} {2004})}\BibitemShut {NoStop}%
\bibitem [{\citenamefont {Wang}\ and\ \citenamefont
  {Carrington~Jr}(2018)}]{WaCa18}%
  \BibitemOpen
  \bibfield  {author} {\bibinfo {author} {\bibfnamefont {X.-G.}\ \bibnamefont
  {Wang}}\ and\ \bibinfo {author} {\bibfnamefont {T.}~\bibnamefont
  {Carrington~Jr}},\ }\bibfield  {title} {\bibinfo {title} {Using monomer
  vibrational wavefunctions to compute numerically exact ({12D}) rovibrational
  levels of water dimer},\ }\href {https://doi.org/10.1063/1.5020426}
  {\bibfield  {journal} {\bibinfo  {journal} {J. Chem. Phys.}\ }\textbf
  {\bibinfo {volume} {148}},\ \bibinfo {pages} {074108} (\bibinfo {year}
  {2018})}\BibitemShut {NoStop}%
\bibitem [{\citenamefont {Felker}\ and\ \citenamefont
  {Bačić}(2019)}]{FeBa19}%
  \BibitemOpen
  \bibfield  {author} {\bibinfo {author} {\bibfnamefont {P.~M.}\ \bibnamefont
  {Felker}}\ and\ \bibinfo {author} {\bibfnamefont {Z.}~\bibnamefont
  {Bačić}},\ }\bibfield  {title} {\bibinfo {title} {Weakly bound molecular
  dimers: Intramolecular vibrational fundamentals, overtones, and tunneling
  splittings from full-dimensional quantum calculations using compact
  contracted bases of intramolecular and low-energy rigid-monomer
  intermolecular eigenstates},\ }\href {https://doi.org/10.1063/1.5111131}
  {\bibfield  {journal} {\bibinfo  {journal} {J. Chem. Phys.}\ }\textbf
  {\bibinfo {volume} {151}},\ \bibinfo {pages} {024305} (\bibinfo {year}
  {2019})}\BibitemShut {NoStop}%
\bibitem [{\citenamefont {Felker}\ and\ \citenamefont
  {Bačić}(2020)}]{FeBa20}%
  \BibitemOpen
  \bibfield  {author} {\bibinfo {author} {\bibfnamefont {P.~M.}\ \bibnamefont
  {Felker}}\ and\ \bibinfo {author} {\bibfnamefont {Z.}~\bibnamefont
  {Bačić}},\ }\bibfield  {title} {\bibinfo {title} {{H}$_2${O}--{CO} and
  {D}$_2${O}--{CO} complexes: Intra- and intermolecular rovibrational states
  from full-dimensional and fully coupled quantum calculations},\ }\href
  {https://doi.org/10.1063/5.0020566} {\bibfield  {journal} {\bibinfo
  {journal} {J. Chem. Phys.}\ }\textbf {\bibinfo {volume} {153}},\ \bibinfo
  {pages} {074107} (\bibinfo {year} {2020})}\BibitemShut {NoStop}%
\bibitem [{\citenamefont {Liu}\ \emph {et~al.}(2021)\citenamefont {Liu},
  \citenamefont {Li}, \citenamefont {Felker},\ and\ \citenamefont
  {Bačić}}]{LiLiFeBa21}%
  \BibitemOpen
  \bibfield  {author} {\bibinfo {author} {\bibfnamefont {Y.}~\bibnamefont
  {Liu}}, \bibinfo {author} {\bibfnamefont {J.}~\bibnamefont {Li}}, \bibinfo
  {author} {\bibfnamefont {P.~M.}\ \bibnamefont {Felker}},\ and\ \bibinfo
  {author} {\bibfnamefont {Z.}~\bibnamefont {Bačić}},\ }\bibfield  {title}
  {\bibinfo {title} {{HC}l--{H}$_2${O} dimer: an accurate full-dimensional
  potential energy surface and fully coupled quantum calculations of intra- and
  intermolecular vibrational states and frequency shifts},\ }\href
  {https://doi.org/10.1039/D1CP00865J} {\bibfield  {journal} {\bibinfo
  {journal} {Phys. Chem. Chem. Phys.}\ }\textbf {\bibinfo {volume} {23}},\
  \bibinfo {pages} {7101} (\bibinfo {year} {2021})}\BibitemShut {NoStop}%
\bibitem [{\citenamefont {J{\"a}ckle}\ and\ \citenamefont
  {Meyer}(1996)}]{JaMe96}%
  \BibitemOpen
  \bibfield  {author} {\bibinfo {author} {\bibfnamefont {A.}~\bibnamefont
  {J{\"a}ckle}}\ and\ \bibinfo {author} {\bibfnamefont {H.-D.}\ \bibnamefont
  {Meyer}},\ }\bibfield  {title} {\bibinfo {title} {Product representation of
  potential energy surfaces},\ }\href {https://doi.org/10.1063/1.471513}
  {\bibfield  {journal} {\bibinfo  {journal} {J. Chem. Phys.}\ }\textbf
  {\bibinfo {volume} {104}},\ \bibinfo {pages} {7974} (\bibinfo {year}
  {1996})}\BibitemShut {NoStop}%
\bibitem [{\citenamefont {Thomas}\ and\ \citenamefont
  {Carrington~Jr}(2017)}]{ThoCarr17}%
  \BibitemOpen
  \bibfield  {author} {\bibinfo {author} {\bibfnamefont {P.~S.}\ \bibnamefont
  {Thomas}}\ and\ \bibinfo {author} {\bibfnamefont {T.}~\bibnamefont
  {Carrington~Jr}},\ }\bibfield  {title} {\bibinfo {title} {An intertwined
  method for making low-rank, sum-of-product basis functions that makes it
  possible to compute vibrational spectra of molecules with more than 10
  atoms},\ }\href {https://doi.org/10.1063/1.4983695} {\bibfield  {journal}
  {\bibinfo  {journal} {J. Chem. Phys.}\ }\textbf {\bibinfo {volume} {146}},\
  \bibinfo {pages} {204110} (\bibinfo {year} {2017})}\BibitemShut {NoStop}%
\bibitem [{\citenamefont {Thomas}\ \emph {et~al.}(2018)\citenamefont {Thomas},
  \citenamefont {Carrington~Jr}, \citenamefont {Agarwal},\ and\ \citenamefont
  {Schaefer~III}}]{ThoCarr18}%
  \BibitemOpen
  \bibfield  {author} {\bibinfo {author} {\bibfnamefont {P.~S.}\ \bibnamefont
  {Thomas}}, \bibinfo {author} {\bibfnamefont {T.}~\bibnamefont
  {Carrington~Jr}}, \bibinfo {author} {\bibfnamefont {J.}~\bibnamefont
  {Agarwal}},\ and\ \bibinfo {author} {\bibfnamefont {H.~F.}\ \bibnamefont
  {Schaefer~III}},\ }\bibfield  {title} {\bibinfo {title} {Using an iterative
  eigensolver and intertwined rank reduction to compute vibrational spectra of
  molecules with more than a dozen atoms: {U}racil and naphthalene},\ }\href
  {https://doi.org/10.1063/1.5039147} {\bibfield  {journal} {\bibinfo
  {journal} {J. Chem. Phys.}\ }\textbf {\bibinfo {volume} {149}},\ \bibinfo
  {pages} {064108} (\bibinfo {year} {2018})}\BibitemShut {NoStop}%
\bibitem [{\citenamefont {Carter}\ \emph {et~al.}(1997)\citenamefont {Carter},
  \citenamefont {Culik},\ and\ \citenamefont {Bowman}}]{CaCuBo97}%
  \BibitemOpen
  \bibfield  {author} {\bibinfo {author} {\bibfnamefont {S.}~\bibnamefont
  {Carter}}, \bibinfo {author} {\bibfnamefont {S.~J.}\ \bibnamefont {Culik}},\
  and\ \bibinfo {author} {\bibfnamefont {J.~M.}\ \bibnamefont {Bowman}},\
  }\bibfield  {title} {\bibinfo {title} {Vibrational self-consistent field
  method for many-mode systems: A new approach and application to the
  vibrations of {CO} adsorbed on {Cu}(100)},\ }\href
  {https://doi.org/10.1063/1.474210} {\bibfield  {journal} {\bibinfo  {journal}
  {J. Chem. Phys.}\ }\textbf {\bibinfo {volume} {107}},\ \bibinfo {pages}
  {10458} (\bibinfo {year} {1997})}\BibitemShut {NoStop}%
\bibitem [{\citenamefont {Carter}\ \emph {et~al.}(1998)\citenamefont {Carter},
  \citenamefont {Bowman},\ and\ \citenamefont {Handy}}]{CaBoHa98}%
  \BibitemOpen
  \bibfield  {author} {\bibinfo {author} {\bibfnamefont {S.}~\bibnamefont
  {Carter}}, \bibinfo {author} {\bibfnamefont {J.~M.}\ \bibnamefont {Bowman}},\
  and\ \bibinfo {author} {\bibfnamefont {N.~C.}\ \bibnamefont {Handy}},\
  }\bibfield  {title} {\bibinfo {title} {Extensions and tests of
  “multimode”: a code to obtain accurate vibration/rotation energies of
  many-mode molecules},\ }\href {https://doi.org/10.1007/s002140050379}
  {\bibfield  {journal} {\bibinfo  {journal} {Theor. Chem. Acc.}\ }\textbf
  {\bibinfo {volume} {100}},\ \bibinfo {pages} {191} (\bibinfo {year}
  {1998})}\BibitemShut {NoStop}%
\bibitem [{\citenamefont {Ziegler}\ and\ \citenamefont
  {Rauhut}(2019)}]{ZiRa19}%
  \BibitemOpen
  \bibfield  {author} {\bibinfo {author} {\bibfnamefont {B.}~\bibnamefont
  {Ziegler}}\ and\ \bibinfo {author} {\bibfnamefont {G.}~\bibnamefont
  {Rauhut}},\ }\bibfield  {title} {\bibinfo {title} {Accurate vibrational
  configuration interaction calculations on diborane and its isotopologues},\
  }\href {https://doi.org/10.1021/acs.jpca.9b01604} {\bibfield  {journal}
  {\bibinfo  {journal} {J. Phys. Chem. A}\ }\textbf {\bibinfo {volume} {123}},\
  \bibinfo {pages} {3367} (\bibinfo {year} {2019})}\BibitemShut {NoStop}%
\bibitem [{\citenamefont {Wang}\ \emph {et~al.}(2015)\citenamefont {Wang},
  \citenamefont {Carter},\ and\ \citenamefont {Bowman}}]{WaCaBo15}%
  \BibitemOpen
  \bibfield  {author} {\bibinfo {author} {\bibfnamefont {X.}~\bibnamefont
  {Wang}}, \bibinfo {author} {\bibfnamefont {S.}~\bibnamefont {Carter}},\ and\
  \bibinfo {author} {\bibfnamefont {J.~M.}\ \bibnamefont {Bowman}},\ }\bibfield
   {title} {\bibinfo {title} {Pruning the {H}amiltonian matrix in {MULTIMODE}:
  test for {C}$_2${H}$_4$ and application to {CH}$_3${NO}$_2$ using a new ab
  initio potential energy surface},\ }\href
  {https://doi.org/https://doi.org/10.1021/acs.jpca.5b09816} {\bibfield
  {journal} {\bibinfo  {journal} {J. Phys. Chem. A}\ }\textbf {\bibinfo
  {volume} {119}},\ \bibinfo {pages} {11632} (\bibinfo {year}
  {2015})}\BibitemShut {NoStop}%
\bibitem [{\citenamefont {Rabitz}\ and\ \citenamefont {Alis}(1999)}]{RaAl99}%
  \BibitemOpen
  \bibfield  {author} {\bibinfo {author} {\bibfnamefont {H.}~\bibnamefont
  {Rabitz}}\ and\ \bibinfo {author} {\bibfnamefont {O.~F.}\ \bibnamefont
  {Alis}},\ }\bibfield  {title} {\bibinfo {title} {General foundations of
  high-dimensional model representations},\ }\href
  {https://doi.org/10.1023/A:1019188517934} {\bibfield  {journal} {\bibinfo
  {journal} {J. Math. Chem.}\ }\textbf {\bibinfo {volume} {25}},\ \bibinfo
  {pages} {197} (\bibinfo {year} {1999})}\BibitemShut {NoStop}%
\bibitem [{\citenamefont {Manzhos}\ and\ \citenamefont
  {Carrington}(2006)}]{MaCa06}%
  \BibitemOpen
  \bibfield  {author} {\bibinfo {author} {\bibfnamefont {S.}~\bibnamefont
  {Manzhos}}\ and\ \bibinfo {author} {\bibfnamefont {T.}~\bibnamefont
  {Carrington}},\ }\bibfield  {title} {\bibinfo {title} {A random-sampling high
  dimensional model representation neural network for building potential energy
  surfaces},\ }\href {https://doi.org/10.1063/1.2336223} {\bibfield  {journal}
  {\bibinfo  {journal} {J. Chem. Phys.}\ }\textbf {\bibinfo {volume} {125}},\
  \bibinfo {pages} {084109} (\bibinfo {year} {2006})}\BibitemShut {NoStop}%
\bibitem [{\citenamefont {Manzhos}\ and\ \citenamefont
  {Carrington}(2007)}]{MaCa07}%
  \BibitemOpen
  \bibfield  {author} {\bibinfo {author} {\bibfnamefont {S.}~\bibnamefont
  {Manzhos}}\ and\ \bibinfo {author} {\bibfnamefont {T.}~\bibnamefont
  {Carrington}},\ }\bibfield  {title} {\bibinfo {title} {Using redundant
  coordinates to represent potential energy surfaces with lower-dimensional
  functions},\ }\href {https://doi.org/10.1063/1.2746846} {\bibfield  {journal}
  {\bibinfo  {journal} {J. Chem. Phys.}\ }\textbf {\bibinfo {volume} {127}},\
  \bibinfo {pages} {014103} (\bibinfo {year} {2007})}\BibitemShut {NoStop}%
\bibitem [{\citenamefont {Manzhos}\ \emph {et~al.}(2011)\citenamefont
  {Manzhos}, \citenamefont {Yamashita},\ and\ \citenamefont
  {Carrington}}]{MaYaCa11}%
  \BibitemOpen
  \bibfield  {author} {\bibinfo {author} {\bibfnamefont {S.}~\bibnamefont
  {Manzhos}}, \bibinfo {author} {\bibfnamefont {K.}~\bibnamefont {Yamashita}},\
  and\ \bibinfo {author} {\bibfnamefont {T.}~\bibnamefont {Carrington}},\
  }\bibfield  {title} {\bibinfo {title} {On the advantages of a rectangular
  matrix collocation equation for computing vibrational spectra from small
  basis sets},\ }\href
  {https://doi.org/https://doi.org/10.1016/j.cplett.2011.06.040} {\bibfield
  {journal} {\bibinfo  {journal} {Chem. Phys. Lett.}\ }\textbf {\bibinfo
  {volume} {511}},\ \bibinfo {pages} {434} (\bibinfo {year}
  {2011})}\BibitemShut {NoStop}%
\bibitem [{\citenamefont {Avila}\ and\ \citenamefont
  {Carrington}(2013)}]{AvCa13}%
  \BibitemOpen
  \bibfield  {author} {\bibinfo {author} {\bibfnamefont {G.}~\bibnamefont
  {Avila}}\ and\ \bibinfo {author} {\bibfnamefont {T.}~\bibnamefont
  {Carrington}},\ }\bibfield  {title} {\bibinfo {title} {Solving the
  {S}chroedinger equation using {S}molyak interpolants},\ }\href
  {https://doi.org/10.1063/1.4821348} {\bibfield  {journal} {\bibinfo
  {journal} {J. Chem. Phys.}\ }\textbf {\bibinfo {volume} {139}},\ \bibinfo
  {pages} {134114} (\bibinfo {year} {2013})}\BibitemShut {NoStop}%
\bibitem [{\citenamefont {Avila}\ and\ \citenamefont
  {Carrington}(2015)}]{AvCa15}%
  \BibitemOpen
  \bibfield  {author} {\bibinfo {author} {\bibfnamefont {G.}~\bibnamefont
  {Avila}}\ and\ \bibinfo {author} {\bibfnamefont {T.}~\bibnamefont
  {Carrington}},\ }\bibfield  {title} {\bibinfo {title} {A multi-dimensional
  {S}molyak collocation method in curvilinear coordinates for computing
  vibrational spectra},\ }\href {https://doi.org/10.1063/1.4936294} {\bibfield
  {journal} {\bibinfo  {journal} {J. Chem. Phys.}\ }\textbf {\bibinfo {volume}
  {143}},\ \bibinfo {pages} {214108} (\bibinfo {year} {2015})}\BibitemShut
  {NoStop}%
\bibitem [{\citenamefont {Avila}\ and\ \citenamefont
  {Carrington}(2017)}]{AvCa17}%
  \BibitemOpen
  \bibfield  {author} {\bibinfo {author} {\bibfnamefont {G.}~\bibnamefont
  {Avila}}\ and\ \bibinfo {author} {\bibfnamefont {T.}~\bibnamefont
  {Carrington}},\ }\bibfield  {title} {\bibinfo {title} {Reducing the cost of
  using collocation to compute vibrational energy levels: Results for
  {CH}$_2${NH}},\ }\href {https://doi.org/10.1063/1.4994920} {\bibfield
  {journal} {\bibinfo  {journal} {J. Chem. Phys.}\ }\textbf {\bibinfo {volume}
  {147}},\ \bibinfo {pages} {064103} (\bibinfo {year} {2017})}\BibitemShut
  {NoStop}%
\bibitem [{\citenamefont {Wodraszka}\ and\ \citenamefont
  {Carrington~Jr}(2019)}]{WoCa19}%
  \BibitemOpen
  \bibfield  {author} {\bibinfo {author} {\bibfnamefont {R.}~\bibnamefont
  {Wodraszka}}\ and\ \bibinfo {author} {\bibfnamefont {T.}~\bibnamefont
  {Carrington~Jr}},\ }\bibfield  {title} {\bibinfo {title} {A pruned
  collocation-based multiconfiguration time-dependent {H}artree approach using
  a {S}molyak grid for solving the {S}chr{\"o}dinger equation with a general
  potential energy surface},\ }\href {https://doi.org/10.1063/1.5093317}
  {\bibfield  {journal} {\bibinfo  {journal} {J. Chem. Phys.}\ }\textbf
  {\bibinfo {volume} {150}},\ \bibinfo {pages} {154108} (\bibinfo {year}
  {2019})}\BibitemShut {NoStop}%
\bibitem [{\citenamefont {Wodraszka}\ and\ \citenamefont
  {Carrington}(2021)}]{WoCa21}%
  \BibitemOpen
  \bibfield  {author} {\bibinfo {author} {\bibfnamefont {R.}~\bibnamefont
  {Wodraszka}}\ and\ \bibinfo {author} {\bibfnamefont {T.}~\bibnamefont
  {Carrington}},\ }\bibfield  {title} {\bibinfo {title} {A rectangular
  collocation multi-configuration time-dependent {H}artree ({MCTDH}) approach
  with time-independent points for calculations on general potential energy
  surfaces},\ }\href {https://doi.org/10.1063/5.0046425} {\bibfield  {journal}
  {\bibinfo  {journal} {J. Chem. Phys.}\ }\textbf {\bibinfo {volume} {154}},\
  \bibinfo {pages} {114107} (\bibinfo {year} {2021})}\BibitemShut {NoStop}%
\bibitem [{\citenamefont {Carrington}(2021)}]{Ca21}%
  \BibitemOpen
  \bibfield  {author} {\bibinfo {author} {\bibfnamefont {T.}~\bibnamefont
  {Carrington}},\ }\bibfield  {title} {\bibinfo {title} {Using collocation to
  study the vibrational dynamics of molecules},\ }\href
  {https://doi.org/https://doi.org/10.1016/j.saa.2020.119158} {\bibfield
  {journal} {\bibinfo  {journal} {Spectrochim. Acta}\ }\textbf {\bibinfo
  {volume} {248}},\ \bibinfo {pages} {119158} (\bibinfo {year}
  {2021})}\BibitemShut {NoStop}%
\bibitem [{\citenamefont {Ollitrault}\ \emph {et~al.}(2020)\citenamefont
  {Ollitrault}, \citenamefont {Baiardi}, \citenamefont {Reiher},\ and\
  \citenamefont {Tavernelli}}]{OlBaReTa20}%
  \BibitemOpen
  \bibfield  {author} {\bibinfo {author} {\bibfnamefont {P.~J.}\ \bibnamefont
  {Ollitrault}}, \bibinfo {author} {\bibfnamefont {A.}~\bibnamefont {Baiardi}},
  \bibinfo {author} {\bibfnamefont {M.}~\bibnamefont {Reiher}},\ and\ \bibinfo
  {author} {\bibfnamefont {I.}~\bibnamefont {Tavernelli}},\ }\bibfield  {title}
  {\bibinfo {title} {Hardware efficient quantum algorithms for vibrational
  structure calculations},\ }\href {https://doi.org/10.1039/D0SC01908A}
  {\bibfield  {journal} {\bibinfo  {journal} {Chem. Sci.}\ }\textbf {\bibinfo
  {volume} {11}},\ \bibinfo {pages} {6842} (\bibinfo {year}
  {2020})}\BibitemShut {NoStop}%
\bibitem [{\citenamefont {Christiansen}(2004)}]{Chr04}%
  \BibitemOpen
  \bibfield  {author} {\bibinfo {author} {\bibfnamefont {O.}~\bibnamefont
  {Christiansen}},\ }\bibfield  {title} {\bibinfo {title} {A second
  quantization formulation of multimode dynamics},\ }\href
  {https://doi.org/10.1063/1.1637578} {\bibfield  {journal} {\bibinfo
  {journal} {J. Chem. Phys.}\ }\textbf {\bibinfo {volume} {120}},\ \bibinfo
  {pages} {2140} (\bibinfo {year} {2004})}\BibitemShut {NoStop}%
\bibitem [{\citenamefont {Su}\ \emph {et~al.}(2021)\citenamefont {Su},
  \citenamefont {Berry}, \citenamefont {Wiebe}, \citenamefont {Rubin},\ and\
  \citenamefont {Babbush}}]{SuBeWiRuBa21}%
  \BibitemOpen
  \bibfield  {author} {\bibinfo {author} {\bibfnamefont {Y.}~\bibnamefont
  {Su}}, \bibinfo {author} {\bibfnamefont {D.~W.}\ \bibnamefont {Berry}},
  \bibinfo {author} {\bibfnamefont {N.}~\bibnamefont {Wiebe}}, \bibinfo
  {author} {\bibfnamefont {N.}~\bibnamefont {Rubin}},\ and\ \bibinfo {author}
  {\bibfnamefont {R.}~\bibnamefont {Babbush}},\ }\bibfield  {title} {\bibinfo
  {title} {Fault-tolerant quantum simulations of chemistry in first
  quantization},\ }\href {https://doi.org/10.1103/PRXQuantum.2.040332}
  {\bibfield  {journal} {\bibinfo  {journal} {PRX Quant.}\ }\textbf {\bibinfo
  {volume} {2}},\ \bibinfo {pages} {040332} (\bibinfo {year}
  {2021})}\BibitemShut {NoStop}%
\bibitem [{\citenamefont {M\'atyus}\ \emph
  {et~al.}(2011{\natexlab{a}})\citenamefont {M\'atyus}, \citenamefont {Hutter},
  \citenamefont {M\"uller-Herold},\ and\ \citenamefont {Reiher}}]{MaHuMuRe11a}%
  \BibitemOpen
  \bibfield  {author} {\bibinfo {author} {\bibfnamefont {E.}~\bibnamefont
  {M\'atyus}}, \bibinfo {author} {\bibfnamefont {J.}~\bibnamefont {Hutter}},
  \bibinfo {author} {\bibfnamefont {U.}~\bibnamefont {M\"uller-Herold}},\ and\
  \bibinfo {author} {\bibfnamefont {M.}~\bibnamefont {Reiher}},\ }\bibfield
  {title} {\bibinfo {title} {On the emergence of molecular structure},\ }\href
  {https://doi.org/10.1103/PhysRevA.83.052512} {\bibfield  {journal} {\bibinfo
  {journal} {Phys. Rev. A}\ }\textbf {\bibinfo {volume} {83}},\ \bibinfo
  {pages} {052512} (\bibinfo {year} {2011}{\natexlab{a}})}\BibitemShut
  {NoStop}%
\bibitem [{\citenamefont {M\'atyus}\ \emph
  {et~al.}(2011{\natexlab{b}})\citenamefont {M\'atyus}, \citenamefont {Hutter},
  \citenamefont {M\"uller-Herold},\ and\ \citenamefont {Reiher}}]{MaHuMuRe11b}%
  \BibitemOpen
  \bibfield  {author} {\bibinfo {author} {\bibfnamefont {E.}~\bibnamefont
  {M\'atyus}}, \bibinfo {author} {\bibfnamefont {J.}~\bibnamefont {Hutter}},
  \bibinfo {author} {\bibfnamefont {U.}~\bibnamefont {M\"uller-Herold}},\ and\
  \bibinfo {author} {\bibfnamefont {M.}~\bibnamefont {Reiher}},\ }\bibfield
  {title} {\bibinfo {title} {Extracting elements of molecular structure from
  the all-particle wave function},\ }\href {https://doi.org/10.1063/1.3662487}
  {\bibfield  {journal} {\bibinfo  {journal} {J. Chem. Phys.}\ }\textbf
  {\bibinfo {volume} {135}},\ \bibinfo {pages} {204302} (\bibinfo {year}
  {2011}{\natexlab{b}})}\BibitemShut {NoStop}%
\bibitem [{\citenamefont {M\'atyus}\ and\ \citenamefont
  {Reiher}(2012)}]{MaRe12}%
  \BibitemOpen
  \bibfield  {author} {\bibinfo {author} {\bibfnamefont {E.}~\bibnamefont
  {M\'atyus}}\ and\ \bibinfo {author} {\bibfnamefont {M.}~\bibnamefont
  {Reiher}},\ }\bibfield  {title} {\bibinfo {title} {Molecular structure
  calculations: a unified quantum mechanical description of electrons and
  nuclei using explicitly correlated gaussian functions and the global vector
  representation},\ }\href@noop {} {\bibfield  {journal} {\bibinfo  {journal}
  {J. Chem. Phys.}\ }\textbf {\bibinfo {volume} {137}},\ \bibinfo {pages}
  {024104} (\bibinfo {year} {2012})}\BibitemShut {NoStop}%
\bibitem [{\citenamefont {Kassal}\ \emph {et~al.}(2011)\citenamefont {Kassal},
  \citenamefont {Whitfield}, \citenamefont {Perdomo-Ortiz}, \citenamefont
  {Yung},\ and\ \citenamefont {Aspuru-Guzik}}]{KaWhPeYuAs11}%
  \BibitemOpen
  \bibfield  {author} {\bibinfo {author} {\bibfnamefont {I.}~\bibnamefont
  {Kassal}}, \bibinfo {author} {\bibfnamefont {J.~D.}\ \bibnamefont
  {Whitfield}}, \bibinfo {author} {\bibfnamefont {A.}~\bibnamefont
  {Perdomo-Ortiz}}, \bibinfo {author} {\bibfnamefont {M.-H.}\ \bibnamefont
  {Yung}},\ and\ \bibinfo {author} {\bibfnamefont {A.}~\bibnamefont
  {Aspuru-Guzik}},\ }\bibfield  {title} {\bibinfo {title} {Simulating chemistry
  using quantum computers},\ }\href
  {https://doi.org/10.1146/annurev-physchem-032210-103512} {\bibfield
  {journal} {\bibinfo  {journal} {Ann. Rev. Phys. Chem.}\ }\textbf {\bibinfo
  {volume} {62}},\ \bibinfo {pages} {185} (\bibinfo {year} {2011})}\BibitemShut
  {NoStop}%
\bibitem [{\citenamefont {Fábri}\ \emph {et~al.}(2011)\citenamefont {Fábri},
  \citenamefont {Mátyus}, \citenamefont {Furtenbacher}, \citenamefont {Nemes},
  \citenamefont {Mihály}, \citenamefont {Zoltáni},\ and\ \citenamefont
  {Császár}}]{FaMaFuNeMiZoCs11}%
  \BibitemOpen
  \bibfield  {author} {\bibinfo {author} {\bibfnamefont {C.}~\bibnamefont
  {Fábri}}, \bibinfo {author} {\bibfnamefont {E.}~\bibnamefont {Mátyus}},
  \bibinfo {author} {\bibfnamefont {T.}~\bibnamefont {Furtenbacher}}, \bibinfo
  {author} {\bibfnamefont {L.}~\bibnamefont {Nemes}}, \bibinfo {author}
  {\bibfnamefont {B.}~\bibnamefont {Mihály}}, \bibinfo {author} {\bibfnamefont
  {T.}~\bibnamefont {Zoltáni}},\ and\ \bibinfo {author} {\bibfnamefont
  {A.~G.}\ \bibnamefont {Császár}},\ }\bibfield  {title} {\bibinfo {title}
  {Variational quantum mechanical and active database approaches to the
  rotational-vibrational spectroscopy of ketene, {H}$_2${CCO}},\ }\href
  {https://doi.org/10.1063/1.3625404} {\bibfield  {journal} {\bibinfo
  {journal} {J. Chem. Phys.}\ }\textbf {\bibinfo {volume} {135}},\ \bibinfo
  {pages} {094307} (\bibinfo {year} {2011})}\BibitemShut {NoStop}%
\bibitem [{\citenamefont {Owens}\ \emph {et~al.}(2017)\citenamefont {Owens},
  \citenamefont {Zak}, \citenamefont {Chubb}, \citenamefont {Yurchenko},
  \citenamefont {Tennyson},\ and\ \citenamefont {Yachmenev}}]{OwZaChYuTeYa17}%
  \BibitemOpen
  \bibfield  {author} {\bibinfo {author} {\bibfnamefont {A.}~\bibnamefont
  {Owens}}, \bibinfo {author} {\bibfnamefont {E.}~\bibnamefont {Zak}}, \bibinfo
  {author} {\bibfnamefont {K.}~\bibnamefont {Chubb}}, \bibinfo {author}
  {\bibfnamefont {S.~N.}\ \bibnamefont {Yurchenko}}, \bibinfo {author}
  {\bibfnamefont {J.}~\bibnamefont {Tennyson}},\ and\ \bibinfo {author}
  {\bibfnamefont {A.}~\bibnamefont {Yachmenev}},\ }\bibfield  {title} {\bibinfo
  {title} {Simulating electric field interactions with polar molecules using
  spectroscopic databases},\ }\href {https://doi.org/10.1038/srep45068}
  {\bibfield  {journal} {\bibinfo  {journal} {Sci. Rep.}\ }\textbf {\bibinfo
  {volume} {7}},\ \bibinfo {pages} {45068} (\bibinfo {year}
  {2017})}\BibitemShut {NoStop}%
\bibitem [{\citenamefont {Szidarovszky}\ and\ \citenamefont
  {Yamanouchi}(2017)}]{SzYa17}%
  \BibitemOpen
  \bibfield  {author} {\bibinfo {author} {\bibfnamefont {T.}~\bibnamefont
  {Szidarovszky}}\ and\ \bibinfo {author} {\bibfnamefont {K.}~\bibnamefont
  {Yamanouchi}},\ }\bibfield  {title} {\bibinfo {title} {Full-dimensional
  simulation of the laser-induced alignment dynamics of {H}$_2${H}e$^+$},\
  }\href {https://doi.org/10.1080/00268976.2017.1297863} {\bibfield  {journal}
  {\bibinfo  {journal} {Mol. Phys.}\ }\textbf {\bibinfo {volume} {115}},\
  \bibinfo {pages} {1916} (\bibinfo {year} {2017})}\BibitemShut {NoStop}%
\bibitem [{\citenamefont {Wang}\ \emph {et~al.}(2011)\citenamefont {Wang},
  \citenamefont {Carrington~Jr}, \citenamefont {Dawes},\ and\ \citenamefont
  {Jasper}}]{WaCa11}%
  \BibitemOpen
  \bibfield  {author} {\bibinfo {author} {\bibfnamefont {X.-G.}\ \bibnamefont
  {Wang}}, \bibinfo {author} {\bibfnamefont {T.}~\bibnamefont {Carrington~Jr}},
  \bibinfo {author} {\bibfnamefont {R.}~\bibnamefont {Dawes}},\ and\ \bibinfo
  {author} {\bibfnamefont {A.~W.}\ \bibnamefont {Jasper}},\ }\bibfield  {title}
  {\bibinfo {title} {The vibration--rotation--tunneling spectrum of the polar
  and {T}-shaped-{N}-in isomers of ({NNO})$_2$},\ }\href
  {https://doi.org/10.1016/j.jms.2011.03.017} {\bibfield  {journal} {\bibinfo
  {journal} {J. Mol. Spectrosc.}\ }\textbf {\bibinfo {volume} {268}},\ \bibinfo
  {pages} {53} (\bibinfo {year} {2011})}\BibitemShut {NoStop}%
\bibitem [{\citenamefont {Mátyus}\ \emph {et~al.}(2010)\citenamefont
  {Mátyus}, \citenamefont {Fábri}, \citenamefont {Szidarovszky},
  \citenamefont {Czakó}, \citenamefont {Allen},\ and\ \citenamefont
  {Császár}}]{MaFaSzCzAlCs10}%
  \BibitemOpen
  \bibfield  {author} {\bibinfo {author} {\bibfnamefont {E.}~\bibnamefont
  {Mátyus}}, \bibinfo {author} {\bibfnamefont {C.}~\bibnamefont {Fábri}},
  \bibinfo {author} {\bibfnamefont {T.}~\bibnamefont {Szidarovszky}}, \bibinfo
  {author} {\bibfnamefont {G.}~\bibnamefont {Czakó}}, \bibinfo {author}
  {\bibfnamefont {W.~D.}\ \bibnamefont {Allen}},\ and\ \bibinfo {author}
  {\bibfnamefont {A.~G.}\ \bibnamefont {Császár}},\ }\bibfield  {title}
  {\bibinfo {title} {Assigning quantum labels to variationally computed
  rotational-vibrational eigenstates of polyatomic molecules},\ }\href
  {https://doi.org/10.1063/1.3451075} {\bibfield  {journal} {\bibinfo
  {journal} {J. Chem. Phys.}\ }\textbf {\bibinfo {volume} {133}},\ \bibinfo
  {pages} {034113} (\bibinfo {year} {2010})}\BibitemShut {NoStop}%
\bibitem [{\citenamefont {Metz}\ \emph {et~al.}(2019)\citenamefont {Metz},
  \citenamefont {Szalewicz}, \citenamefont {Sarka}, \citenamefont {Tóbiás},
  \citenamefont {Császár},\ and\ \citenamefont {Mátyus}}]{dimers}%
  \BibitemOpen
  \bibfield  {author} {\bibinfo {author} {\bibfnamefont {M.~P.}\ \bibnamefont
  {Metz}}, \bibinfo {author} {\bibfnamefont {K.}~\bibnamefont {Szalewicz}},
  \bibinfo {author} {\bibfnamefont {J.}~\bibnamefont {Sarka}}, \bibinfo
  {author} {\bibfnamefont {R.}~\bibnamefont {Tóbiás}}, \bibinfo {author}
  {\bibfnamefont {A.~G.}\ \bibnamefont {Császár}},\ and\ \bibinfo {author}
  {\bibfnamefont {E.}~\bibnamefont {Mátyus}},\ }\bibfield  {title} {\bibinfo
  {title} {Molecular dimers of methane clathrates: \emph{ab initio} potential
  energy surfaces and variational vibrational states},\ }\href
  {https://doi.org/10.1039/C9CP00993K} {\bibfield  {journal} {\bibinfo
  {journal} {Phys. Chem. Chem. Phys.}\ }\textbf {\bibinfo {volume} {21}},\
  \bibinfo {pages} {13504} (\bibinfo {year} {2019})}\BibitemShut {NoStop}%
\bibitem [{\citenamefont {Coles}\ \emph {et~al.}(2018)\citenamefont {Coles},
  \citenamefont {Owens}, \citenamefont {Küpper},\ and\ \citenamefont
  {Yachmenev}}]{CoOwKuYa18}%
  \BibitemOpen
  \bibfield  {author} {\bibinfo {author} {\bibfnamefont {P.~A.}\ \bibnamefont
  {Coles}}, \bibinfo {author} {\bibfnamefont {A.}~\bibnamefont {Owens}},
  \bibinfo {author} {\bibfnamefont {J.}~\bibnamefont {Küpper}},\ and\ \bibinfo
  {author} {\bibfnamefont {A.}~\bibnamefont {Yachmenev}},\ }\bibfield  {title}
  {\bibinfo {title} {A {H}yperfine-resolved rotation-vibration line list of
  ammonia ({NH}$_3$)},\ }\href {https://doi.org/10.3847/1538-4357/aaef7e}
  {\bibfield  {journal} {\bibinfo  {journal} {Astrophys. J.}\ }\textbf
  {\bibinfo {volume} {870}},\ \bibinfo {pages} {24} (\bibinfo {year}
  {2018})}\BibitemShut {NoStop}%
\bibitem [{\citenamefont {Yachmenev}\ \emph {et~al.}(2022)\citenamefont
  {Yachmenev}, \citenamefont {Yang}, \citenamefont {Zak}, \citenamefont
  {Yurchenko},\ and\ \citenamefont {Küpper}}]{YaYaZaYuKu22}%
  \BibitemOpen
  \bibfield  {author} {\bibinfo {author} {\bibfnamefont {A.}~\bibnamefont
  {Yachmenev}}, \bibinfo {author} {\bibfnamefont {G.}~\bibnamefont {Yang}},
  \bibinfo {author} {\bibfnamefont {E.}~\bibnamefont {Zak}}, \bibinfo {author}
  {\bibfnamefont {S.}~\bibnamefont {Yurchenko}},\ and\ \bibinfo {author}
  {\bibfnamefont {J.}~\bibnamefont {Küpper}},\ }\bibfield  {title} {\bibinfo
  {title} {The nuclear-spin-forbidden rovibrational transitions of water from
  first principles},\ }\href {https://doi.org/10.1063/5.0090771} {\bibfield
  {journal} {\bibinfo  {journal} {J. Chem. Phys.}\ }\textbf {\bibinfo {volume}
  {156}},\ \bibinfo {pages} {204307} (\bibinfo {year} {2022})}\BibitemShut
  {NoStop}%
\bibitem [{\citenamefont {Owens}\ and\ \citenamefont
  {Yachmenev}(2018)}]{OwYa18}%
  \BibitemOpen
  \bibfield  {author} {\bibinfo {author} {\bibfnamefont {A.}~\bibnamefont
  {Owens}}\ and\ \bibinfo {author} {\bibfnamefont {A.}~\bibnamefont
  {Yachmenev}},\ }\bibfield  {title} {\bibinfo {title} {Richmol: A general
  variational approach for rovibrational molecular dynamics in external
  electric fields},\ }\href {https://doi.org/10.1063/1.5023874} {\bibfield
  {journal} {\bibinfo  {journal} {J. Chem. Phys.}\ }\textbf {\bibinfo {volume}
  {148}},\ \bibinfo {pages} {124102} (\bibinfo {year} {2018})}\BibitemShut
  {NoStop}%
\bibitem [{\citenamefont {Erfort}\ \emph {et~al.}(2020)\citenamefont {Erfort},
  \citenamefont {Tsch{\"o}pe},\ and\ \citenamefont {Rauhut}}]{ErTsRa20}%
  \BibitemOpen
  \bibfield  {author} {\bibinfo {author} {\bibfnamefont {S.}~\bibnamefont
  {Erfort}}, \bibinfo {author} {\bibfnamefont {M.}~\bibnamefont
  {Tsch{\"o}pe}},\ and\ \bibinfo {author} {\bibfnamefont {G.}~\bibnamefont
  {Rauhut}},\ }\bibfield  {title} {\bibinfo {title} {Toward a fully automated
  calculation of rovibrational infrared intensities for semi-rigid polyatomic
  molecules},\ }\href {https://doi.org/10.1063/5.0011832} {\bibfield  {journal}
  {\bibinfo  {journal} {J. Chem. Phys.}\ }\textbf {\bibinfo {volume} {152}},\
  \bibinfo {pages} {244104} (\bibinfo {year} {2020})}\BibitemShut {NoStop}%
\bibitem [{\citenamefont {Erfort}\ \emph {et~al.}(2022)\citenamefont {Erfort},
  \citenamefont {Tschöpe},\ and\ \citenamefont {Rauhut}}]{ErTsRa22}%
  \BibitemOpen
  \bibfield  {author} {\bibinfo {author} {\bibfnamefont {S.}~\bibnamefont
  {Erfort}}, \bibinfo {author} {\bibfnamefont {M.}~\bibnamefont {Tschöpe}},\
  and\ \bibinfo {author} {\bibfnamefont {G.}~\bibnamefont {Rauhut}},\
  }\bibfield  {title} {\bibinfo {title} {Efficient and automated quantum
  chemical calculation of rovibrational nonresonant {R}aman spectra},\ }\href
  {https://doi.org/10.1063/5.0087359} {\bibfield  {journal} {\bibinfo
  {journal} {J. Chem. Phys.}\ }\textbf {\bibinfo {volume} {156}},\ \bibinfo
  {pages} {124102} (\bibinfo {year} {2022})}\BibitemShut {NoStop}%
\end{thebibliography}
%

\end{document}